\documentclass[prb,aps,reprint,twocolumn,showpacs]{revtex4-1}
\usepackage{amsmath,amssymb,amsfonts,float}
\usepackage[pdftex]{graphicx,color}    
\usepackage{epstopdf}
\usepackage{times,pifont}
 
\newcommand{\st}{\text{st}}
\newcommand{\tr}{\text{tr}}
\newcommand{\str}{\text{str}}
  
\newcommand{\be}{\text{e}}

\newsavebox{\dotdot}
\savebox{\dotdot}[3mm]{\shortstack{\circle*{0.8}\\ \\ \circle*{0.8}}}

\begin{document}
\title{ 
Quantum Entanglement and 
Topological Order in Hole-Doped Valence Bond Solid States
}
\author{Kazuki~Hasebe$^{1}$ and Keisuke~Totsuka$^2$}
\affiliation{
$^1$  Kagawa National College of 
Technology, Takuma-cho, Mitoyo, Kagawa 769-1192, Japan
\\
$^2$ Yukawa Institute for Theoretical Physics, 
Kyoto University, Kitashirakawa Oiwake-Cho, Kyoto 606-8502, Japan}

\begin{abstract}
We present a detailed analysis of topological properties of the valence bond solid (VBS) states 
doped with fermionic holes.  As concrete examples, we consider the supersymmetric extension of the SU(2)- and the SO(5) VBS states, dubbed UOSp(1$|$2) and UOSp(1$|$4) 
supersymmetric VBS states, respectively. 
Specifically, we investigate the string-order parameters and the entanglement spectra 
of these states to find that, even when the parent states (bosonic VBS states) do not  
support the string order, they recover it when holes are doped and the fermionic sector 
appears in the entanglement spectrum.  
These peculiar properties are discussed in light of the symmetry-protected topological order.  To this end, we characterize a few typical classes of symmetry-protected topological orders in terms of supermatrix-product states (SMPS).  From this, we see that the topological order in the bulk manifests itself in the transformation properties of the SMPS in question 
and thereby affects the structure of the entanglement spectrum.  
Then, we explicitly relate the existence of the string order and the structure of the entanglement 
spectrum to explain the recovery and the stabilization of the string order in 
the supersymmetric systems.   
\end{abstract}            

\maketitle


\section{Introduction}

The valence-bond solid (VBS) states had been originally introduced 
by Affleck, Kennedy, Lieb and Tasaki\cite{Affleck-K-L-T-87, *Affleck-K-L-T-88a} 
to build explicit model ground states which realize the properties of the generic integer-spin antiferromagnetic spin chains conjectured by Haldane.\cite{Haldane-83a,*Haldane-83b}  
Quite unexpectedly, on top of the properties already anticipated from other analyses 
(e.g. quantum-disordered ground state with short-range spin correlations, 
gapped triplet spin excitations, etc.), these states exhibit many striking features such as 
the emergent boundary excitations (edge states)\cite{Hagiwara-K-A-H-R-90} 
and the existence of hidden string order.\cite{denNijs-R-89,Tasaki-91}  
In the case of spin-1 systems, it has been argued\cite{Kennedy-T-92-PRB,*Kennedy-T-92-CMP}  
that the hidden topological (string) order is a consequence 
of the $\mathbb{Z}_2{\times} \mathbb{Z}_2$-symmetry breaking occurring 
in the system after applying the non-local unitary transformation.  
The idea of non-local hidden order and edge states has been to some extent 
generalized\cite{Hatsugai-92,Oshikawa-92,Totsuka-S-mpg-95,Nishiyama-T-H-S-95}  
to other values of integer-spin-$S$ although the hidden $\mathbb{Z}_2{\times} \mathbb{Z}_2$-symmetry is never broken\cite{Oshikawa-92} in the case of even-$S$. 
Through these studies, it has been recognized that there are some differences\cite{Oshikawa-92,Totsuka-S-mpg-95} in the ground-state properties according to the 
parity of $S$.  
Nevertheless, by analogy with the quantum-Hall systems\cite{Girvin-A-89}, 
the ground state of generic integer-spin antiferromagnetic chains, 
including the original VBS state and its higher-spin 
generalizations\cite{Arovas-A-H-88}, characterized by certain kinds of 
non-local correlations and emergent edge states have been called `topological' 
in a rough sense.  

Recent development in quantum-information-theoretic approaches to quantum many-body 
problems enables us to extract information on the bulk topological order from 
the entanglement properties of the {\em ground-state} 
wave function\cite{Levin-W-06,Kitaev-P-06,Li-H-08}.  
The topological states in one-dimensional  (1D) spin systems have been 
reconsidered\cite{Gu-W-09,Pollmann-T-B-O-10,*Pollmann-B-T-O-12} 
from the modern point of view and the precise meaning of 
the topological Haldane phase has been clarified.    
In these studies, the string order parameters and the edge states, which in general 
are not robust against small perturbations, are replaced by more robust objects 
(i.e. the structure of the entanglement spectrum or the structure of tensor-network).  
In particular, it has been shown in Ref.~\onlinecite{Pollmann-T-B-O-10,*Pollmann-B-T-O-12} that the existence of (at least one of) the discrete symmetries 
(time-reversal, link-inversion and $\mathbb{Z}_2 \times \mathbb{Z}_2 $ symmetry) divides  
all states of matter in 1D into two categories--topologically-non-trivial ones and the rest. 
Generic odd-integer-$S$ spin chains belong to the former while even-$S$ chains to 
the latter. The hallmark of the topological phase protected by the above discrete symmetries 
is that all entanglement levels are even-fold degenerate.   
In this formulation, the difference between odd-$S$ and even-$S$ is naturally understood 
in terms of the entanglement structure.   
It should also be mentioned that the topological phases of one-dimensional gapped spin systems have been 
classified by group cohomology, \cite{Chen-G-W-11,*Chen-G-W-11b,Schuch-G-C-11} 
and the detailed analyses based on the Lie group symmetries are reported 
in Ref.~\onlinecite{Duiv-Quella-12a,*Duiv-Quella-12b}.

In this paper, we present an exhaustive discussion about the effects of 
coexisting bosonic- and fermionic degrees of freedom 
on (symmetry-protected) topological phases in 1D.  
Clearly, this kind of questions is motivated in part by hole doping in the Haldane-gap 
systems.\cite{Zhang-A-89,Penc-S-95,xu2000hqs}   
In order to incorporate the coexisting bosons and fermions, 
for mathematical convenience, we use supersymmetry (SUSY)    
which relates bosons carrying integer spins and fermions with half-odd-integer spins.   
Several  ``topological phases'' with SUSY have been found so far in, 
e.g., quantum-Hall systems,\cite{hasebe2005PRL}  
VBS states,\cite{Arovas-H-Q-Z-09} and 
ultra-cold atom systems\cite{YuYang2008}.   
However, the precise characterization of these SUSY topological phases has not 
been obtained so far and it would be quite useful to investigate symmetry-protected 
topological order in model SUSY systems from the entanglement point of view. 

As the model SUSY states, we consider a class of supersymmetric VBS (SVBS) states 
defined by the Schwinger operator consisting of $2K$ bosons which represent 
the bosonic degrees of freedom at each site (e.g. localized integer spins) and $N$ fermions which correspond to doped fermionic holes (with $K$ and $N$ being integers).  
This class is interesting 
since it includes the SVBS states investigated in Refs.~\onlinecite{Arovas-H-Q-Z-09,Hasebe-T-11} 
as well as the SUSY-extension of the SO(5) VBS state and the Sp($N$) VBS state introduced 
respectively in Refs.~\onlinecite{Tu-Z-X-X-N-09} and \onlinecite{Schuricht-R-08}.   
The (S)VBS states are rare examples where we can study non-trivial topological properties 
even in 1D and most of the calculations can be done without relying on any approximation.  
Taking advantage of such properties of the SVBS states, 
we uncover the roles of SUSY in topological phases in 1D. 

The generalized hidden string order in the SVBS states\cite{Arovas-H-Q-Z-09} 
has been investigated already in the previous work\cite{Hasebe-T-11} by the authors. 
In contrast to what is known for the bosonic counterpart (the spin-$S$ VBS state\cite{Arovas-A-H-88}), the symptom of the non-trivial topological order has been observed 
in the analysis of the string order even for the even-integer superspin.    
To be more precise, even when the string order vanishes, it revives upon the hole doping; 
this might suggest the existence of topological order in the SVBS states {\em regardless} 
of the parity of bulk superspin $\mathcal{S}$.  
In order for the better understanding of this phenomenon, we first characterize 
symmetry-protected topological orders in SUSY systems in the language of entanglement. 
To this end, we use the supermatrix-product-state (SMPS) formalism 
to generalize the arguments of Ref.~\onlinecite{Pollmann-T-B-O-10,*Pollmann-B-T-O-12}  
and derive the relation between topological order in the bulk and the entanglement structure.   
The SMPS formalism further enables us to obtain the explicit relation between 
the entanglement spectrum and the string order parameters, and thereby to clarify 
why the hidden string order revives after doping. 

As has been emphasized in the previous work\cite{Hasebe-T-11}, 
in spite of its name, the SMPS formalism does not assume any particular form of SUSY.   
In fact, we do not need even postulate {\em exact} SUSY and the only prerequisite is that 
the local Hilbert space is made up of the bosonic part and the fermionic one.  
In view of the ability of (S)MPS in approximating any gapped states in 1D with arbitrary 
precision\cite{Verstraete-C-06,Hastings-06,*Hastings-area-law-07}, 
our results are applicable to a wider class of 1D systems 
with some kind of relation between bosons and fermions.  
 
The organization of this paper is as follows. 
In Sec. \ref{sect:basicsofsupersymmetry}, we introduce a class of 
UOSp($N{|}2K$)-invariant SVBS states ($2K$ being the number of boson species and 
$N$ for fermions) with arbitrary superspins using the Schwinger operator.  
We then construct the explicit SMPS representation 
for $(N,K)=(1,1)$ [UOSp(1$|$2)] and $(1,2)$ [UOSp(1$|$4)]  
and summarize several important properties of these states. 
As the first step toward the investigation of topological order, 
we explicitly evaluate the string order parameters in the above two types of 
SVBS states for different values of superspins in Sec. \ref{sec:stringorder}.  
There we find that the revival of the string order already observed for UOSp(1$|$2) 
in Ref.~\onlinecite{Hasebe-T-11} occurs in other SUSY cases as well. 
In Sec. \ref{sect:entanglementSVBS}, the entanglement spectrum of these SVBS states 
(in the limit of infinite-size systems) is derived and typical features of the spectrum are discussed. 
In order to understand the results obtained in the previous section 
and characterize symmetry-protected topological order in 1D SUSY systems,  
we generalize the argument of Ref.~\onlinecite{Pollmann-T-B-O-10,*Pollmann-B-T-O-12} to 
SUSY systems in Sec. \ref{sect:supersymmetryprotectedtopologicalorder} and 
relate the structure of the entanglement spectrum and the bulk topological order.  
Finally, the relationship between the degeneracy of the entanglement spectrum 
and non-vanishing string order parameters is clarified in Sec. \ref{sec:relation-string-top}
by using the (S)MPS formalism.  
Section \ref{sect:summary} is devoted to summary and discussions. 

\section{SVBS states and SMPS formalism}
\label{sect:basicsofsupersymmetry}

In this section, we briefly describe how the standard MPS formalism is generalized 
to the cases with SUSY.  
Let us begin with constructing the MPS of the spin-$M$ ($M$: integer) 
SU(2) valence-bond solid (VBS) state\cite{Arovas-A-H-88}  
starting from its representation in terms of 
the SU(2) Schwinger operators $\phi=({b^{1}}^{\dagger},{b^2}^{\dagger})^{\text{t}}$: 
\begin{equation}
\begin{split}
|\text{VBS}\rangle^{(M)} &=\prod_{j} 
( {b^1_j}^{\dagger} {b^2_{j+1}}^{\dagger}  - {b^2_j}^{\dagger} {b^1_{j+1}}^{\dagger})^M 
|\text{vac}\rangle \\
&=\prod_{j} (\phi^{\text{t}}_j \, i\sigma_2 \phi_{j+1})^M |\text{vac}\rangle, 
\end{split}
\end{equation}
where the metric [or, the SU(2) charge conjugation matrix] 
\begin{equation}
i\sigma_2=\begin{pmatrix} 0 & 1 \\
-1 & 0 \end{pmatrix}
\end{equation}
has been used to form a maximally-entangled (singlet) pair between the sites $j$ and $j+1$.  
Therefore, by construction, 
the VBS state is SU(2) invariant and represents a spin-isotropic state. 
 
\subsection{General idea}
\label{sec:general-idea}

The standard construction of the VBS-type of states\cite{Garcia-V-W-C-07} starts by preparing two 
auxiliary degrees of freedom on each site of the lattice.  
Then, the (bosonic) VBS state is constructed first by creating 
singlets between pairs of those auxiliary objects on adjacent sites and then by projecting 
the tensor-product of the two auxiliary objects on each site onto the desired physical Hilbert space.  

The SVBS states are introduced by including the states with one- or more fermionic holes 
into the above bosonic Hilbert space. 
Mathematically, we replace the usual Lie-group symmetry [e.g. SU(2)] with that of 
the super Lie group UOSp($N|2K$) corresponding to $2K$ bosonic degrees of freedom 
and $N$ fermionic ones [for a review of super Lie groups, see,  
for instance, Ref.~\onlinecite{SUSY-dictionary}, and for UOSp($N|2K$), Ref.~\onlinecite{hasebe-2011}]. 
Specifically, the SVBS states with UOSp($N | 2K$)-symmetry are defined as
\begin{equation}
|\text{SVBS}(N|2K)\rangle^{(M)} 
=\prod_{\langle i,j\rangle} 
(\psi^{\text{t}}_i \mathcal{R}_{N|2K} \psi_j)^M|\text{vac}\rangle \; ,
\label{typeNSVBS}
\end{equation}
where $\psi$ stands for the UOSp($N | 2K$) Schwinger operator  
\begin{equation}
\psi=({b^1}^{\dagger},{b^2}^{\dagger},\cdots,{b^{2K}}^{\dagger},
{f^1}^{\dagger},\cdots,{f^{N}}^{\dagger})^{\text{t}} \; .
\label{typeNSVBS2}
\end{equation}
The $2K$ bosons ${b^{\sigma}}^{\dagger}$ $(\sigma=1,2,\cdots,2K)$ and 
the $N$ fermions $f^{\mu}$ $(\mu=1,2,\cdots,N)$ satisfy the commutation relations   
$[b^{\sigma},{b^{\tau}}^{\dagger}]=\delta^{\sigma\tau}$, $\{{f^{\mu}},{f^{\nu}}^{\dagger}\}=\delta^{\mu\nu}$, 
$[b^{\sigma},f^{\mu}]=[b^{\sigma},{f^{\mu}}^{\dagger}]=0$. 
The matrix $\mathcal{R}_{N|2K}$ signifies the UOSp($N|2K$) invariant matrix: 
\begin{equation}
\mathcal{R}_{N|2K} =
\begin{pmatrix}
J_{2K} & 0 \\
0 & -1_N
\end{pmatrix}, 
\label{eqn:def-general-metric}
\end{equation}
where the USp($2K$)-invariant $2K {\times} 2K$  antisymmetric matrix $J_{2K}$ is defined using the Pauli matrix 
$\sigma_{2}$ as:  
\begin{equation}
J_{2K}= 
\begin{pmatrix}
i\sigma_2 &             &     &  {0}   \\
          &   i\sigma_2 &     &    \\ 
        &             & \ddots & \\
      0    &             &        & i\sigma_2 
\end{pmatrix}
\end{equation}
and $1_{N}$ denotes the $N$-dimensional identity matrix.   
By using the above equations, it is straightforward to show that 
the product of spinors $\psi^{\text{t}}_i \mathcal{R}_{N|2K} \psi_j$ is singlet 
under UOSp($N|2K$).  

As the number of fermion species $N$ corresponds 
to that of the SUSY in the system, hereafter we call the SVBS states 
defined by \eqref{typeNSVBS} and \eqref{typeNSVBS2} 
the UOSp($N|2K$) SVBS states. 
In this paper, we give the detailed discussions for the two $N=1$ cases, 
specifically $(K,N)=(1,1)$ and $(K,N)=(2,1)$,  
in which the following isomorphisms between the orthogonal groups and 
the unitary symplectic groups hold: 
$\text{SO(3)}\simeq \text{USp(2)}/\mathbb{Z}_2$ ($K=1$), 
$\text{SO(5)}\simeq \text{UOSp(4)}/\mathbb{Z}_2$ ($K=2$). 
For UOSp($N|2$) ($K=1$), the metric matrix is given by 
\begin{equation}
\mathcal{R}_{N|2}=
\begin{pmatrix}
i\sigma_2 & 0 \\
0 & -1_N
\end{pmatrix}, 
\label{osp2Nmetric}
\end{equation}
and for UOSp($N|4$) ($K=2$), by
\begin{equation}
\mathcal{R}_{N|4}=
\begin{pmatrix}
i\sigma_2 & 0 & 0  \\ 
0 & i\sigma_2 & 0  \\
0 & 0 & -1_N 
\end{pmatrix}.
\label{osp4Nmetric}
\end{equation}

The particle number at each site is related to the superspin $\mathcal{S}$ via 
\begin{equation}
2\mathcal{S}= 
\sum_{\alpha=1}^{2K+N}{\psi^{\alpha}}^{\dagger}\psi^{\alpha} 
=\sum_{\sigma=1}^{2K} {b^{\sigma}}^{\dagger} b^{\sigma}
+\sum_{\mu=1}^N {f^{\mu}}^{\dagger}f^{\mu}=zM, 
\label{eqn:superspin-to-M}
\end{equation}
where $z$ is the lattice-coordination number ($z=2$ in one dimension).  
Throughout this paper, we reserve the symbol $\mathcal{S}$ for superspin and 
use $S$ for the bosonic spin. 
Since $\sum_{\mu}{f^{\mu}}^{\dagger}f^{\mu}$ takes either 0 or 1, 
the possible values of SU(2) spin, which is equal to 
the half of the number of bosons at each site, are:  
\begin{align}
S&=\frac{1}{2}\sum_{\sigma=1}^{2K} {b^{\sigma}}^{\dagger}b^{\sigma} \nonumber\\
&=\frac{1}{2}zM,~\frac{1}{2}zM-\frac{1}{2},~\frac{1}{2}zM-1,\cdots,~\frac{1}{2}zM-\frac{1}{2}N.
\end{align}
(If $N\ge zM$, it is implied that the above sequence terminates at $S=0$).  
One may find that the inclusion of SUSY introduces, as well as the states 
with the spin magnitude $zM/2$ which exist already in the SU(2) case,  
those with spin smaller by $1/2$.   
In what follows, we consider the one-dimensional cases (i.e. $z=2$) unless 
otherwise stated.  

For the 1D chain $(z=2)$, the above sequence reads 
\begin{equation}
S=M,~M-\frac{1}{2},~M-1,\cdots,~M-\frac{1}{2}N, 
\end{equation}
and correspondingly the emergent edge spin takes the following values 
\begin{equation}
s=\frac{1}{2}M,~\frac{1}{2}M-\frac{1}{2},~~\frac{1}{2}M-1,\cdots,~\frac{1}{2}M-\frac{1}{2}N. 
\end{equation}
(again, if $N\ge M$, the above sequence is understood as to stop at $s=0$.) 
The dimension of the physical Hilbert space at each site constructed in this way is given by 
the sum of the one of each bosonic Hilbert space with a fixed boson number 
($2\mathcal{S}-n$): 
\begin{equation}
d_{\mathcal{S}}(N|2K) 
=\sum_{n=0}^{N}
\begin{pmatrix}
2K+2\mathcal{S} -n-1 \\ 2K -1
\end{pmatrix}
\; .
\label{eqn:dim-physical-space}
\end{equation} 
It should be noted here that the Schwinger-operator construction presented here 
does not cover all the possible VBS-type states with UOSp($N|2K$)-symmetry. 
In fact, there is an important class of VBS states\cite{Hasebe-T-unpub-12} 
which is a SUSY generalization of 
a series of SO($2n+1$)-invariant and USp($2K$)-invariant states considered respectively 
in Refs.~\onlinecite{Scalapino-Z-H-98,Tu-Z-X-08} and in Ref.~\onlinecite{Schuricht-R-08}.  
However, most of the conclusions obtained here hold for those models as well. 

The UOSp($N|2K$) SVBS state \eqref{typeNSVBS} may be rewritten as 
\begin{align}
|\text{SVBS}(N|2K)\rangle^{(M)} &= \prod_i (\psi_i^{\text{t}} \mathcal{R}_{N|2K}
\psi_{i+1})^M|\text{vac}\rangle\nonumber\\
& \equiv\prod_i (\Psi_i^{\text{t}}  R^{(M)}_{N|2K} \Psi_{i+1})|\text{vac}\rangle,
\end{align}
where $\Psi_i$ is a graded fully symmetric representation of UOSp($N|2K$) 
of the order $M$ and $R^{(M)}_{N|2K}$ is the metric for 
this representation.\cite{hasebe-2011}  
Another equivalent form (a matrix-product form)\cite{Hasebe-T-11} may be useful for practical purposes:
\begin{equation}
|\text{SVBS}(N|2K)\rangle^{(M)} =\mathcal{A}_1\mathcal{A}_2\cdots \mathcal{A}_L, 
\end{equation}
where the matrix $\mathcal{A}_i $ is defined as:
\begin{equation}
\mathcal{A}_i \equiv R^{(M)}_{N|2K}\Psi_i\Psi_i^{\text{t}} |\text{vac}\rangle_i \; .  
\label{eqn:A-matrix-general-SUSY}
\end{equation}

\subsection{UOSp(1$|$2) SVBS states}

Let us begin with the simplest case\cite{Arovas-H-Q-Z-09,Hasebe-T-11} $(N,K)=(1,1)$.  
The graded Schwinger operator is given by 
\begin{equation}
\psi_{i}
=({b_{i}^1}^{\dagger},{b_{i}^2}^{\dagger},{f_{i}}^{\dagger})^{\text{t}}\equiv 
(a_{i}^{\dagger},b_{i}^{\dagger},f_{i}^{\dagger})^{\text{t}}\; ,
\end{equation}
and the corresponding  SVBS state, which we call  the UOSp(1$|$2) SVBS state 
(precisely, this is the one dubbed {\em type-I} in Ref.~\onlinecite{Hasebe-T-11}),  
is given by: 
\begin{equation}
|\text{SVBS}(1|2)\rangle^{(M)} 
=\prod_{i} (a_i^{\dagger}b_{i+1}^{\dagger}-b_i^{\dagger}a_{i+1}^{\dagger}-rf_i^{\dagger}f_{i+1}^{\dagger})^M|\text{vac}\rangle \; , 
\label{eqn:SUSY-I-def}
\end{equation}
where  we have added the fermion  doping parameters 
$r$ by hand. However, such a parameter may be absorbed in the redefinition 
of the normalization of fermions ($f^{\dagger}\mapsto f^{\dagger}/\sqrt{r}$, $f\mapsto \sqrt{r} f$) and the SVBS states possess the SUSY 
even for finite values of the parameter $r$. 

\subsubsection{$\mathcal{S}=1$}

Let us consider the superspin $\mathcal{S}=1$ case. 
Since $\mathcal{S}$ is related to the number $M$ of SUSY valence bonds through  
\eqref{eqn:superspin-to-M}, the case $M=1$ of eq.\eqref{eqn:SUSY-I-def} 
corresponds to $\mathcal{S}=1$.  

The SVBS state on a finite open chain is specified its edge states, 
$\alpha$  and $\beta$, 
respectively on the site $1$ and $L$:   
\begin{equation}
|\text{SVBS}(1|2)\rangle^{(1)}_{\alpha\beta}=(\mathcal{R}_{1|2} \psi_1)^{\alpha}
~\prod_{i=1}^{L-1}
(\psi_i^{\text{t}}\mathcal{R}_{1|2}\psi_{i+1})~\psi_L^{\beta}|\text{vac}\rangle, 
\label{eqn:type-I-M-1-SVBS}
\end{equation}
where $\psi_j^{\text{t}}=(a_j^{\dagger},~b_j^{\dagger},~\sqrt{r}f_j^{\dagger})$ 
and the UOSp(1$|$2) metric $\mathcal{R}_{1|2}$ is defined in (\ref{osp2Nmetric}). 
The state  $|\text{SVBS-I}\rangle^{(M=1)}_{\alpha\beta}$ can be expressed as a 
product of the matrices $\mathcal{A}^{(1)}_{i}$ defined on a each site: 
\begin{equation}
|\text{SVBS}(1|2)\rangle_{\alpha\beta}^{(1)}=(\mathcal{A}_1^{(1)}\mathcal{A}_2^{(1)}\cdots\mathcal{A}_L^{(1)})_{\alpha\beta}, 
\end{equation}
where $\mathcal{A}^{(1)}_j$ is given by 
\begin{align}
\mathcal{A}_j^{(1)}&=\mathcal{R}^{(2)}_I\psi_j\psi_j^{\text{t}}|\text{vac}\rangle_j\nonumber\\
&=
\begin{pmatrix}
|0\rangle_j & \sqrt{2}|-1\rangle_j & \sqrt{r}|{-1/2}\rangle_j\\
-\sqrt{2}|1\rangle_j & -|0\rangle_j & -\sqrt{r}|{1/2}\rangle_j \\
-\sqrt{r}|{1/2}\rangle_j & -\sqrt{r}|{-1/2}\rangle_j & 0 
\end{pmatrix}\nonumber\\
&=\sum_{a=-1,0,1} A(a)|a\rangle + \sum_{\sigma=-1/2,1/2} A(\sigma)|\sigma\rangle,   
\end{align}
with 
\begin{align}
& A(1)=\begin{pmatrix}
0 & 0 & 0 \\
-\sqrt{2} & 0 & 0 \\
0 & 0 & 0 
\end{pmatrix},~~ A(0)=
\begin{pmatrix}
1 & 0 & 0 \\
0 & -1 & 0 \\
0 & 0 & 0 
\end{pmatrix},\nonumber\\
& A(-1)=
\begin{pmatrix}
0 & \sqrt{2} & 0 \\
0 & 0 & 0 \\
0 & 0 & 0 
\end{pmatrix},\nonumber\\
& A(1/2)=\begin{pmatrix}
0 & 0 & 0 \\
0 & 0 & -\sqrt{r} \\
-\sqrt{r}& 0 & 0 
\end{pmatrix},~~ A(-1/2)=
\begin{pmatrix}
0 & 0 & \sqrt{r} \\
0 & 0 & 0 \\
0 & -\sqrt{r} & 0 
\end{pmatrix} \; . 
\label{canonicalosp12matr}
\end{align}
The five basis states corresponding to 
the $\mathcal{S}=1$ irreducible representation (denoted by $\bold{5}$) are given by 
\begin{align}
&|1\rangle =\frac{1}{\sqrt{2}}{a^{\dagger}}^2|\text{vac}\rangle, 
~~|0\rangle ={a^{\dagger}}b^{\dagger}|\text{vac}\rangle,
~~~|-1\rangle =\frac{1}{\sqrt{2}}{b^{\dagger}}^2|\text{vac}\rangle,\nonumber\\
&|1/2\rangle =a^{\dagger}f^{\dagger}|\text{vac}\rangle, 
~~ |-1/2\rangle =b^{\dagger}f^{\dagger}|\text{vac}\rangle \; ,
\end{align}
where $|\text{vac}\rangle$ is the vacuum of both the boson and the fermion:  
$a|\text{vac}\rangle=b|\text{vac}\rangle=f|\text{vac}\rangle=0$. 
The first three states corresponds to the spin-1 ($\bold{3}$) representation of SU(2), 
and the second two states constitute $\bold{2}$ with spin-$1/2$. 

The parent Hamiltonian of the state \eqref{eqn:type-I-M-1-SVBS} is 
constructed\cite{Arovas-H-Q-Z-09,Hasebe-T-11} 
in such a way that the local Hamiltonian $h_{j,j+1}$ acting on the bond $(j,j+1)$ 
annihilates all the nine states 
appearing in the product $\mathcal{A}_{j}\mathcal{A}_{j+1}$.  
Therefore, the ground state on a finite open chain 
is nine-fold degenerate with respect to the matrix indices.
Since the $\psi_j$ and $\psi_j^{\text{t}}$ represent the two auxiliary  
degrees of freedom at the site $j$, 
the above nine-fold degeneracy 
reflects the existence of the three edge degrees of freedom on both edges of an open chain: 
\begin{equation}
|\!\uparrow\rangle\!\rangle=a^{\dagger}|\text{vac}\rangle,
~~~|\!\downarrow\rangle\!\rangle=b^{\dagger}|\text{vac}\rangle,~~~
|0\rangle\!\rangle=f^{\dagger}|\text{vac}\rangle \; .
\end{equation} 

As the {\em doping parameter} $r$ is changed, the state \eqref{eqn:type-I-M-1-SVBS} 
interpolates between the two well-known states: 
at $r\rightarrow 0$,  $|\text{SVBS}(1|2)\rangle^{(1)}$  is reduced 
to the original VBS state\cite{Affleck-K-L-T-87,Affleck-K-L-T-88a}  $|\text{VBS}\rangle$
\begin{equation}
|\text{SVBS}(1|2)\rangle^{(1)} \rightarrow |\text{VBS}\rangle^{(1)}
=\prod_{i}(a^{\dagger}_i b_{i+1}^{\dagger}-b_i^{\dagger}a_{i+1})|\text{vac}\rangle, 
\end{equation}
while, at $r\rightarrow \infty$,   $|\text{SVBS-I}\rangle$  is reduced 
to the Majumdar-Ghosh (MG) dimer state\cite{Majumdar-G-69,*Majumdar-70}  
$|\text{MG}\rangle$ 
\begin{equation}
|\text{SVBS}(1|2)\rangle^{(1)}  \rightarrow \prod_i  f^{\dagger}_i |\text{MG}\rangle, 
\end{equation}
where 
\begin{equation}
|\text{MG}\rangle=(\prod_{i:\text{even}}-\prod_{i:\text{odd}})
(a^{\dagger}_i b^{\dagger}_{i+1}-b^{\dagger}_i a^{\dagger}_{i+1})|\text{vac}\rangle.
\end{equation}
In the discussion of the entanglement spectra (section \ref{sect:entanglementSVBS}), 
we will see in the two limits, the entanglement entropy nicely interpolates 
between that of the VBS state and the MG state. 

\subsubsection{Higher-$\mathcal{S}$}

It is easy to generalize the above strategy to the cases with general superspin-$\mathcal{S}$. 
In Ref.~\onlinecite{Hasebe-T-11}, the expression of the $\mathcal{A}$-matrix for 
superspin-$\mathcal{S}$ type-I SVBS state is given as:
\begin{equation}
\mathcal{A}^{(\mathcal{S})}_{ab}(j)={\cal F}_{a}^{\text{L}}(a^{\dagger}_{j},b^{\dagger}_{j},f^{\dagger}_{j})
{\cal F}_{b}^{\text{R}}(a^{\dagger}_{j},b^{\dagger}_{j},f^{\dagger}_{j})
|\text{vac}\rangle_{j}  \; ,
\label{eqn:A-OSp12-higher-S}
\end{equation}
where the $\mathcal{S}$-th order polynomials ${\cal F}_{a}^{\text{L}}$ 
and ${\cal F}_{b}^{\text{R}}$ 
are defined in eqs.(C3a) and (C3b) of Ref.~\onlinecite{Hasebe-T-11}.  
The above expression may be readily rewritten into the standard 
form \eqref{eqn:A-matrix-general-SUSY}: 
\begin{subequations}
\begin{equation}
\mathcal{A}^{(\mathcal{S})}_{ab}(j) = R_{1|2}^{(\mathcal{S})} \Psi_j  \Psi_j^{\text{t}} |\text{vac}\rangle_{j} \; ,
\end{equation}
where
\begin{equation}
\begin{split}
& (\Psi_{j})_{a} \equiv {\cal F}_{a}^{\text{R}}(a^{\dagger}_{j} \; , \; b^{\dagger}_{j}  \; , \;
f^{\dagger}_{j}) \quad (1\leq a \leq 2\mathcal{S}+1) \; , \\
& (R^{(\mathcal{S})}_{1|2})_{ab} \\
& \equiv 
\begin{cases}
(-1)^{a-1}\delta_{b,(\mathcal{S}+2)-a} & (1\leq a,b \leq \mathcal{S}+1) \\
(-1)^{\mathcal{S}-(a-1)}\delta_{b,(3\mathcal{S}+3)-a} & 
(\mathcal{S}+2 \leq a,b \leq 2\mathcal{S}+1) \; ,
\end{cases} 
\end{split}
\end{equation}
\end{subequations}  

\subsection{UOSp(1$|$4) SVBS states}
\label{sec:SO5-SVBS}

Now we proceed to the case $(N,K)=(1,2)$ (one fermion species and four  
bosonic).  
For UOSp(1$|$4), the graded Schwinger operator is given as:
\begin{equation}
\psi=({b^1}^{\dagger}, {b^2}^{\dagger},{b^3}^{\dagger},{b^4}^{\dagger}, 
\sqrt{r} f^{\dagger})^{\text{t}} \; . 
\label{eqn:UOSp14-spinor}
\end{equation}
These five operators correspond to the five-dimensional representation ($\mathbf{5}$) 
of UOSp(1$|$4); the first four (${b^{1}}^{\dagger}, {b^{2}}^{\dagger}, {b^{3}}^{\dagger}, {b^{4}}^{\dagger}$) 
respectively create the four bosonic states 
\begin{equation}
\begin{split}
& |1\rangle = \left|\frac{1}{2},\frac{1}{2}\right\rangle \, , \; 
|2\rangle = \left|-\frac{1}{2},-\frac{1}{2}\right\rangle \, , \\
& |3\rangle = \left|\frac{1}{2},-\frac{1}{2}\right\rangle \, , \; 
|4\rangle = \left|-\frac{1}{2},\frac{1}{2}\right\rangle 
\end{split}
\label{eqn:weight-spinor}
\end{equation} 
which are already contained in the spinor representation of SO(5) 
and the last one $f^{\dagger}$ creates the fermionic state $|5 \rangle = |f \rangle$.  
We prepare $z$ copies of {\bf 5}s to construct the physical Hilbert space 
at each site of the lattice with the coordination number $z$ and,  
according to which representation is chosen from the tensor product of 
$z$ $\mathbf{5}$s, we can obtain several different types of MPSs. 
For instance, since a pair of {\bf 5}s is decomposed as 
\begin{equation}
\mathbf{5} \otimes \mathbf{5} \sim 
\mathbf{1} \oplus \mathbf{10} \oplus \mathbf{14} \; ,
\label{eqn:CG-decomp-OSp14}
\end{equation}
two different SVBS states ({\bf 10} and {\bf 14}) are obtained in one dimension ($z=2$).  

Following the general method described in section \ref{sec:general-idea}, 
one can construct the following UOSp(1$|$4) SVBS state:
\begin{equation} 
\begin{split}
& |\text{SVBS}(1|4)\rangle^{(M)}
=\prod_{\langle i,j\rangle} (\psi_{i}^{\text{t}}\mathcal{R}_{1|4}\psi_{j})^M 
|\text{vac}\rangle \\
& = \prod_{\langle i,j\rangle}(
{b^{1}_i}^{\dagger}{b^{2}_j}^{\dagger}-{b^{2}_i}^{\dagger}{b^1_j}^{\dagger}
+{b^3_i}^{\dagger}{b^4_j}^{\dagger}-{b^4_i}^{\dagger}{b^3_j}^{\dagger} 
-r f^{\dagger}_{i}f^{\dagger}_{j} 
)^{M} |\text{vac}\rangle 
\end{split}
\label{s05svbswavefun}
\end{equation}
where the summation is taken over the nearest-neighbor pairs 
$\langle i,j\rangle$ 
and $r$ denotes a real parameter varying from 0 to $\infty$.   
The state has the same structure as the UOSp(1$|$2) SVBS state except for 
the metric $\mathcal{R}_{1|4}$ defined in \eqref{eqn:def-general-metric} 
or \eqref{osp4Nmetric}.   
The superspin $\mathcal{S}$ in this state is given as
\begin{equation}
2\mathcal{S} = \sum_{\sigma=1}^{4}{b_i^{\sigma}}^{\dagger}{b_i^{\sigma}}+f^{\dagger}_{i}f_{i} =zM \; .
\end{equation}
The dimension of the local physical Hilbert space (i.e. the size of the representation $\mathcal{S}$) 
\eqref{eqn:dim-physical-space} reads for $(N,K)=(1,2)$:
 \begin{equation}
 \begin{split}
 d_{\mathcal{S}}(1|4) &=\begin{pmatrix} 2\mathcal{S}+3 \\3 \end{pmatrix}+ 
 \begin{pmatrix} 2\mathcal{S}+2 \\ 3 \end{pmatrix}  \\
 & =\frac{(4\mathcal{S}+3)(2\mathcal{S}+1)(\mathcal{S}+1)}{3} \; .
 \end{split} 
 \end{equation}
In the following, we consider the one-dimensional case ($z=2$) 
with $M=1(=\mathcal{S}$) 
where the SO(5) spin magnitude takes the following two values:
 \begin{equation}
 S_i =\frac{1}{2}\sum_{\sigma=1}^{4}{b_i^{\sigma}}^{\dagger}{b_i^{\sigma}}=M,~~M-\frac{1}{2} 
 \end{equation}
and $d_{1}(1|4)=14$.  

 On a finite one-dimensional chain, the UOSp(1$|$4) SVBS state \eqref{s05svbswavefun} may be written as
\begin{equation}
\begin{split}
& |\text{SVBS(T)}\rangle_{\alpha_{\text{L}},\alpha_{\text{R}}}  \\
&=\left\{\mathcal{R}_{1|4}\psi_{1}\right\}_{\alpha_{\text{L}}}
\prod_{j=1}^{L-1} (\psi^{\text{t}}_j 
\mathcal{R}_{1|4} \psi_{j+1}) \left\{\psi_{L}^{\text{t}} \right\}_{\alpha_{\text{R}}}   
|\text{vac}\rangle  \\
&=(\mathcal{A}^{\text{(T)}}_1\mathcal{A}^{\text{(T)}}_2\cdots
\mathcal{A}^{\text{(T)}}_L)_{\alpha_{\text{L}},\alpha_{\text{R}}}
\; ,
\end{split}
\label{eqn:OSP14-MPS-adj}
\end{equation}
where $\mathcal{R}_{1|4}$ is given by (\ref{osp4Nmetric}) with $N=1$. 
The matrix $\mathcal{A}$ is defined by 
\begin{equation}
\begin{split}
&\mathcal{A}^{\text{(T)}} = \mathcal{R}_{1|4}\psi\psi^{\text{t}} \nonumber\\
&=\begin{pmatrix}
|1,2\rangle & \sqrt{2}|2,2\rangle  & |2,3\rangle
& |2,4\rangle & \sqrt{r} |2,f\rangle \\
-\sqrt{2}|1,1\rangle & -|1,2\rangle & -|1,3\rangle 
& -|1,4\rangle & -\sqrt{r}  |1,f\rangle  \\
|1,4\rangle &|2,4\rangle & |3,4\rangle & \sqrt{2}|4,4\rangle 
& \sqrt{r} |4,f\rangle \\
-|1,3\rangle & -|2,3\rangle & -\sqrt{2}|3,3\rangle 
& -|3,4\rangle & -\sqrt{r} |3,f\rangle  \\
-\sqrt{r}|1,f\rangle  & -\sqrt{r}|2,f\rangle  
& -\sqrt{r}|3,f\rangle  & - \sqrt{r}|4,f\rangle  & 0
\end{pmatrix}  \\
& \equiv \sum_{\sigma\le \tau =1}^4 A^{\text{(B)}}_{\text{T}}(\sigma,\tau)|\sigma,\tau\rangle 
+\sum_{\sigma=1}^{4}A^{\text{(F)}}_{\text{T}}(\sigma)|\sigma,f \rangle \; , 
\end{split}
\label{eqn:def-OSp14-VBS-adj}
\end{equation}
where the $D=14$ basis states are given in terms of the graded Schwinger operators 
in \eqref{eqn:UOSp14-spinor} as ($\sigma,\tau=1,2,3,4$):
\begin{equation}
\begin{split}
&|\sigma,\sigma\rangle \equiv \frac{1}{\sqrt{2}}({b^{\sigma}}^{\dagger})^2|\text{vac}\rangle \; ,\\
&|\sigma,\tau\rangle 
\equiv {b^{\sigma}}^{\dagger}{b^{\tau}}^{\dagger}|\text{vac}\rangle \;\; (\sigma < \tau) \; , \\
&|\sigma,f \rangle \equiv {b^{\sigma}}^{\dagger}f^{\dagger}|\text{vac}\rangle \; . 
\end{split}
\end{equation}
The expressions of the 14 matrices $A(\sigma,\tau)$ and $A(\sigma)$ are given 
in appendix \ref{sec:UOSp14-A-matrices-adj}.  

Since the Schwinger operators are used, it is obvious that 
the physical Hilbert space thus constructed is  
the $\mathcal{S}=1$ (i.e. $\mathbf{14}$) fully symmetric representation 
in the tensor-product decomposition \eqref{eqn:CG-decomp-OSp14}: 
\begin{equation}
(\bold{5}\otimes \bold{5})_{\text{fully-sym.}}= \bold{14}~\xrightarrow{\text{SO(5)}} 
~\bold{10}\oplus \bold{4},
\end{equation}
where `$\rightarrow$' denotes the decomposition into the SO(5) irreducible representations.  
As in the case of UOSp(1$|$2) ($(N,K)=(1,1)$), the physical Hilbert space contains 
two irreducible representations of SO(5): the spinor- ($\mathbf{4}$) and the adjoint 
($\mathbf{10}$) representations.   
Since all the 14 basis correspond to the components of the rank-2 symmetric tensor 
made of the two constituent spinors ({\bf 5}), 
we call the MPS thus constructed {\em tensor-type} and use the suffix ``T''.   

A remark is in order here about other possible MPSs.  
In fact, as has been mentioned before, another important MPS is 
obtained\cite{Hasebe-T-unpub-12} if we use the 10-dimensional anti-symmetric representation (vector representation; hence the MPS may be called 
`vector-type'), in stead of the 14-dimensional one 
\begin{equation}
(\mathbf{5}\otimes \mathbf{5})_{\text{anti-sym.}}= \mathbf{10}~\xrightarrow{\text{SO(5)}} 
~ \mathbf{5}\oplus \mathbf{4} \oplus \mathbf{1} \; . 
\end{equation}
The MPS obtained in this way is a direct generalization of the SO(5)-invariant MPS considered 
in Refs.~\onlinecite{Scalapino-Z-H-98,Tu-Z-X-08}.  The details of this class of MPS 
will be reported elsewhere\cite{Hasebe-T-unpub-12}.   

\subsubsection{Limiting Cases}

Now let us consider the two important limiting cases $r\rightarrow 0$ and 
$r \rightarrow \infty$. 
In the limit $r=0$, the UOSp(1$|$4) SVBS states \eqref{s05svbswavefun} or 
\eqref{eqn:OSP14-MPS-adj} reduce to the following VBS states 
\begin{equation}
|\text{VBS}\rangle =\prod_{\langle i,j\rangle}({b^1_i}^{\dagger}{b^2_j}^{\dagger}-{b^2_i}^{\dagger}{b^1_j}^{\dagger}
+{b^3_i}^{\dagger}{b^4_j}^{\dagger}-{b^4_i}^{\dagger}{b^3_j}^{\dagger} )^M|\text{vac}\rangle, 
\end{equation}
dubbed {\em bosonic SO(5) VBS state} in Ref.~\onlinecite{Tu-Z-X-X-N-09}.  

\begin{figure}[floatfix]
\begin{center}
\includegraphics[scale=0.4]{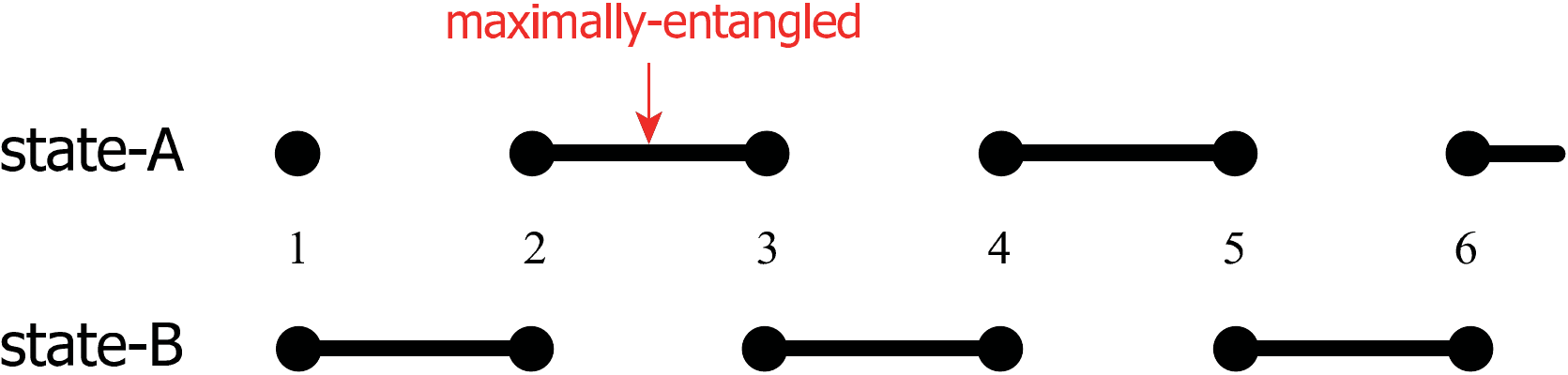}
\caption{(Color online) The $r\rightarrow \infty$ limit of $\mathcal{S}=1$ 
SVBS state. Filled circles denote the bosonic qubits ($S=1/2$ spins for UOSp(1$|$2) 
and 4-dimensional SO(5) spinors for UOSp(1$|$4)).  
On a chain with even number of sites, 
the MPS is block diagonal with the (1,1)-block $\mathcal{B}_{1,1}$ and the (2,2)-block 
$\mathcal{B}_{2,2}$ corresponding to state-A and B, respectively. 
\label{fig:MG-state-general}}
\end{center}
\end{figure}

 In the other limit $r\rightarrow \infty$, the dominant part of $\mathcal{A}^{\text{(T)}}$ 
 reads (after dropping factors proportional to $\sqrt{r}$) 
 \begin{equation}
 \mathcal{A}^{\text{(T)}}_{\infty}(j) = 
 \begin{pmatrix}
 0 & 0 & 0 & 0 & |2\rangle_{j} \\
 0 & 0 & 0 & 0 & -|1\rangle_{j} \\
 0 & 0 & 0 & 0 &  |4\rangle_{j} \\
 0 & 0 & 0 & 0 & -|3\rangle_{j} \\
- |1\rangle_{j} &  - |2 \rangle_{j} & - |3\rangle_{j} & - |4\rangle_{j} & 0
\end{pmatrix} \; .
\end{equation}  
Then, the two-site MPS $\mathcal{A}^{\text{(T)}}_{\infty}(j) \mathcal{A}^{\text{(T)}}_{\infty}(j+1)$ 
takes the following block-diagonal form 
\begin{equation}
\mathcal{A}^{\text{(T)}}_{\infty}(j) \mathcal{A}^{\text{(T)}}_{\infty}(j+1)
=  \pm 
\begin{pmatrix}
\mathcal{B}_{1,1}(j,j+1) & 0 \\
0 & \mathcal{B}_{2,2}(j,j+1)
\end{pmatrix}
\; ,
\end{equation} 
where $|1\rangle,\ldots,|4\rangle$ are defined in eq.\eqref{eqn:weight-spinor} and 
the $(2,2)$-block is the SO(5)-singlet made up of two spinors: 
\begin{equation}
\mathcal{B}_{2,2}(j,j+1) =
 |1\rangle_{j}|2\rangle_{j{+}1}-|2\rangle_{j}|1\rangle_{j{+}1}
 +|3\rangle_{j}|4\rangle_{j{+}1}-|4\rangle_{j}|3\rangle_{j{+}1} \; .
 \end{equation}
When the $4{\times}4$ matrix $\mathcal{B}_{1,1}(j,j+1)$ is multiplied by 
$\mathcal{B}_{1,1}(j+2,j+3)$ from the right, a new SO(5)-singlet is inserted at the bond 
$(j{+}1,j{+}2)$. 
Therefore, one sees that 
the string of $\mathcal{A}^{\text{(T)}}_{\infty}$ represents an SO(5)-generalization 
of the Majumdar-Ghosh valence-bond crystal\cite{Majumdar-G-69,Majumdar-70} 
[see Fig.~\ref{fig:MG-state-general}].   
The vector-type UOSp(1$|$4) SVBS state mentioned above shares the same 
property.\cite{Hasebe-T-unpub-12} 

\section{String Order}
\label{sec:stringorder}

One of the striking features of these VBS states is the existence of non-local 
order called {\em string order}.  
In the usual spin systems, it is known\cite{Kennedy-T-92-PRB,Kennedy-T-92-CMP} 
that the string order is a manifestation of the spontaneous 
$\mathbb{Z}_2{\times} \mathbb{Z}_2$ symmetry 
breaking in the ground state.  

\subsection{UOSp(1$|$2) SVBS states}

In the case of the usual (pure) spin systems, 
the string order parameters are defined by the infinite-distance limit of 
the string correlation functions\cite{denNijs-R-89}:
\begin{subequations}
\begin{align}
&O^{z}_{\text{string}} \equiv  \lim_{n\nearrow \infty}
\Biggl\langle S_{j}^{z}\, \exp\left[
i\pi \sum_{k=j}^{j+n-1}S^{z}_{k}
\right]S^{z}_{j+n}\Biggr\rangle \;, 
\label{eqn:Ostring-z} \\
& O^{x}_{\text{string}} \equiv  \lim_{n\nearrow \infty}
\Biggl\langle S_{j}^{x}\, \exp\left[
i\pi \sum_{k=j+1}^{j+n}S^{x}_{k}
\right]S^{x}_{j+n}\Biggr\rangle   \; .
\label{eqn:Ostring-x}
\end{align}
\end{subequations}

It is straightforward to generalize the string order parameters to the case 
with SUSY by replacing the spin operators $S^{a}$ 
to their $(4\mathcal{S}{+}1)$-dimensional expressions. 
For superspin $\mathcal{S}=1$, it is given by\cite{Hasebe-T-11} 
($O^{x}_{\text{string}}=O^{z}_{\text{string}}$ by SU(2)-symmetry):
\begin{equation}
O_{\text{string}}^{x,z}(r)=
\frac{4 \left\{ r^4+14 r^2+18 +2 \left(r^2+3\right) \sqrt{8 r^2+9}
\right\}}{\left(8 r^2+9\right)
   \left(\sqrt{8 r^2+9}+3\right)^2}  \; .
\end{equation}
In the limit $r\rightarrow 0$, the above string expression reproduces 
the well-known value\cite{Kennedy-T-92-PRB,Kennedy-T-92-CMP} $4/9$ 
(perfect string correlation).  
In the opposite limit $r\nearrow \infty$, 
the string order parameter ${\cal O}_{\text{string}}^{\infty}$ 
approaches to a finite value $1/16$, which implies that the string 
order survives in the $r\nearrow \infty$ limit.  
This agrees with the fact that the spin-1 Haldane state is 
adiabatically connected to the spin-1/2 dimer state.\cite{Hida-92} 

\begin{figure}[floatfix]
\begin{center}
\includegraphics[scale=0.7]{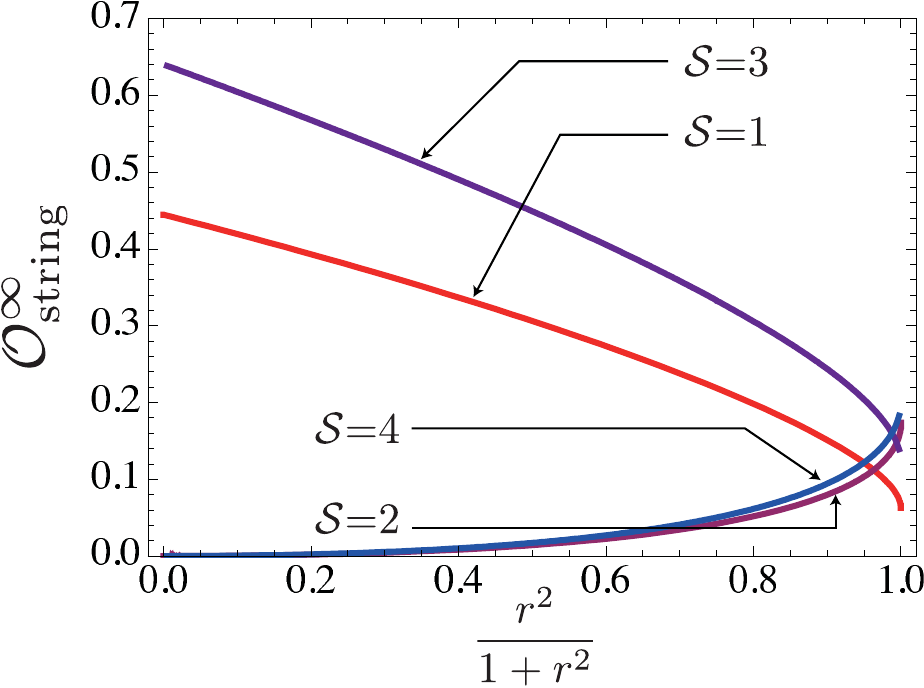}
\caption{(Color online) The string order parameter  
${\cal O}_{\text{string}}^{\infty}$
for several values of superspin $\mathcal{S}$ plotted as a function of $r$ [Ref.~\onlinecite{Hasebe-T-11}].  
Note that ${\cal O}^{\infty}_{\text{string}}(r{=}0)=0$ 
for even-$\mathcal{S}$ corresponding to the vanishing of string order parameter 
for even-$S$. 
\label{fig:Ostring-model1}}
\end{center}
\end{figure}

One can readily generalize the above results to the higher-$\mathcal{S}$ 
cases,\cite{Arovas-H-Q-Z-09} 
which are SUSY-analogues of the higher-spin (bosonic) VBS state introduced in Ref.~\onlinecite{Arovas-A-H-88}.  
In the original spin-$S$ VBS states ($r=0$), the string order parameters have been 
investigated\cite{Oshikawa-92,Totsuka-S-mpg-95} and it has been concluded that 
they vanish for even integer $S$.  
In contrast, for finite values of the doping parameter $r$, the string order parameters 
revive\cite{Hasebe-T-11} due to the existence of SUSY (see Fig.~\ref{fig:Ostring-model1}).  
This interesting behavior will be discussed in section \ref{sec:relation-string-top} in the light of 
symmetry-protected topological order. 

\subsection{UOSp(1$|$4) SVBS states}

In Ref.~\onlinecite{Tu-Z-X-08}, it has been pointed out that the idea of 
hidden-symmetry breaking\cite{Kennedy-T-92-PRB,Kennedy-T-92-CMP} 
and the associated string order parameters\cite{denNijs-R-89} 
can be generalized to a class of models with higher symmetry SO($2n{+}1$) 
by using the $2^{n}$-dimensional spinor representation as the auxiliary Hilbert 
space.  

The four string order parameters for the SO(5) ($n=2$) VBS state are 
defined \cite{Tu-Z-X-08} by analogy with their SU(2) cousin:
\begin{equation}
O_{\text{string}}^{ab} \equiv  \lim_{n\nearrow \infty}
\Biggl\langle L_{j}^{ab}\, \exp\left[
i\pi \sum_{k=j}^{j+n-1}L^{ab}_{k}
\right]L^{ab}_{j+n}\Biggr\rangle  
\label{eqn:def-string-OP-SO5}
\end{equation}
($L^{ab}=-L^{ba}$ are the SO(5)-generators).  
The set of integers $(a,b)$ (with $a,b=1,2,3,4,5$) labels the ten generators 
and we may choose e.g. $(a,b)=(1,2)$, $(2,5)$, $(3,4)$ and $(4,5)$. 

Since, by the SO(5) symmetry, the string order parameters are independent of the SO(5) indices $a,b$, 
we can assume $(a,b)=(1,2)$ without a loss of generality.  
In Ref.~\onlinecite{Tu-Z-X-08}, it has been argued that 
the string order of the SO(5) VBS state is a consequence of 
the hidden $(\mathbb{Z}_2{\times} \mathbb{Z}_2)^2$ symmetry breaking.  
In the original SU(2) case, we pick up a pair $\{S^{z},S^{x}\}$ 
and the two commuting $\mathbb{Z}_2$s are generated by $\be^{i\pi S^{x}}$ and $\be^{i\pi S^{z}}$, 
the former of which plays the role of the flipping operator of $S^{z}$.  
In the case SO(5), we have two(=rank of SO(5)) such pairs (e.g. $\{L^{12},L^{25}\}$ and $\{L^{34},L^{45}\}$) 
and this is why the square of $\mathbb{Z}_2{\times} \mathbb{Z}_2$ appears.   
Similarly, as we already know that the generalized string order exists\cite{Hasebe-T-11} in 
the UOSp(1$|$2) SVBS state, we can expect finite string order in the case of UOSp(1$|$4) as well 
by considering two pairs of string order parameters. 

First we set $r=0$ and consider the SO(5) limit. 
By plotting the eigenvalues of local $(L^{12},L^{34})$ appearing 
in the string \eqref{eqn:OSP14-MPS-adj} 
of $\mathcal{A}^{\text{(T)}}$, one can easily see\cite{Tu-Z-X-X-N-09} that both $L_{12}$ and $L_{34}$ exhibit 
a kind of hidden antiferromagnetic order which is essentially the same as that observed\cite{denNijs-R-89} 
in the $S=1$ VBS state.  In fact, the string order parameter \eqref{eqn:def-string-OP-SO5} 
for $(a,b)=(1,2)$ ($(a,b)=(3,4)$) removes the effects of the randomly inserted 
zeros in the $L^{12}$ ($L^{34}$) configuration to pick up the hidden antiferromagnetic order. 

The generalization of eq.\eqref{eqn:def-string-OP-SO5} to  
the UOSp(1$|$4) SVBS state with arbitrary superspin-$\mathcal{S}$ 
is straightforward; for $\mathcal{S}=1$, the bosonic generators $L^{ab}$ 
are replaced by the 14-dimensional matrices (the explicit forms of them are not very important). 
The MPS formalism enables us to obtain the following result:
\begin{equation}
\begin{split}
O_{\text{string}}^{ab} &=
\frac{\left\{4 r^2+3 \left(\sqrt{16 r^2+25}+5\right)\right\}^2}{\left(16 r^2+25\right) 
\left(\sqrt{16 r^2+25}+5\right)^2} \\
&\rightarrow 
\begin{cases}
\frac{9}{25}  \quad & (r\rightarrow 0) \\
\frac{1}{16}  \quad & (r\rightarrow \infty) \; .
\end{cases}
\end{split}
\end{equation}
In order to highlight qualitatively different behaviors with respect to 
the superspin $\mathcal{S}$, 
we plot the result in Fig.~\ref{fig:String-UOSP(1,4)-(S=1)} together with 
that of the superspin-2 case
\begin{equation}
O_{\text{string}}^{ab}=
\frac{49 \left(7- \sqrt{40 r^2+49}\right)^2}{400 \left(40 r^2+49\right)}
\; .
\end{equation} 
From this plot, one can clearly see that, 
for finite doping, {\em both} the $\mathcal{S}=1$ and $2$ states are topological, 
while the latter is non-topological (i.e. non-Haldane) at $r=0$ (see also Fig.~\ref{fig:Ostring-model1}).   
The limiting value $1/16$ is equal to the string order of the $\mathcal{S}=1$ UOSp(1$|$2) SVBS 
at $r\rightarrow \infty$.  
Similar results have been obtained\cite{Hasebe-T-unpub-12} for the vector-type MPS mentioned 
in section \ref{sec:SO5-SVBS}.   
\begin{figure}[floatfix]
\begin{center}
\includegraphics[scale=0.7]{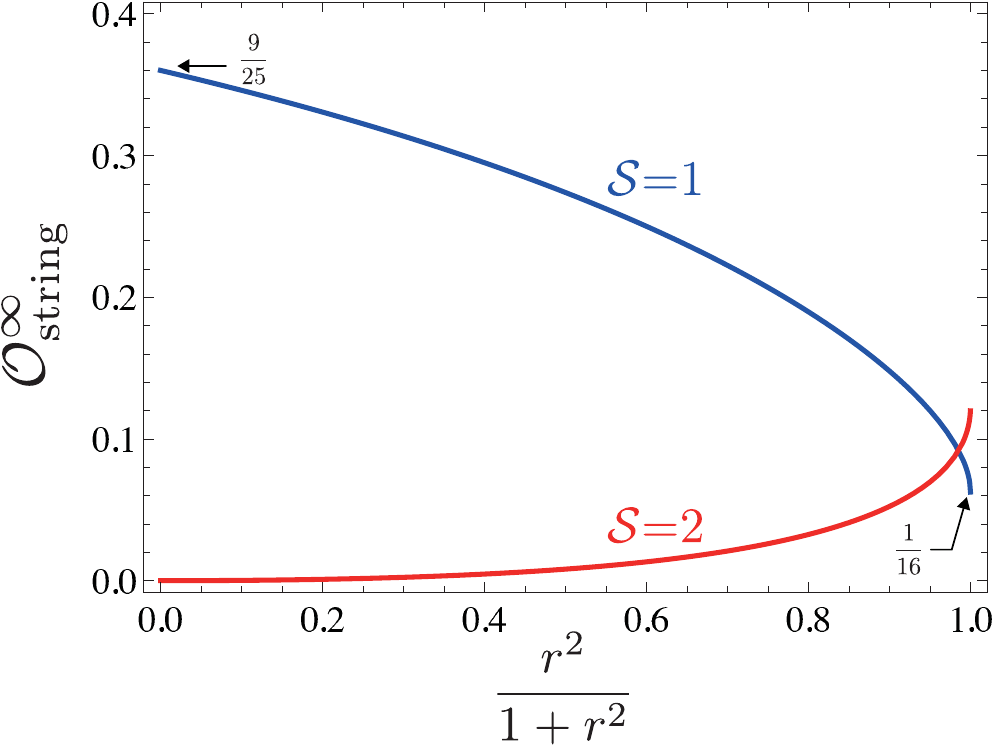}
\caption{(Color online) The infinite-distance limit of 
the string correlation function for UOSp(1$|$4) states. At $r=0$, the string order reproduces 
the known result\cite{Tu-Z-X-X-N-09} $9/25=0.36$.  
Also plotted is the string order of the $\mathcal{S}=2$ ($M=2$) state.  
As in the UOSp(1$|$2) case, the string order vanishes at $r=0$ and revives after doping.  
\label{fig:String-UOSP(1,4)-(S=1)}}
\end{center}
\end{figure}

\section{Entanglement Spectra of SVBS States}
\label{sect:entanglementSVBS}

In the pioneering paper, Li and Haldane\cite{Li-H-08} argued 
that the entanglement spectrum, which is obtained 
by taking logarithm of the Schmidt eigenvalues (or, the eigenvalues of the reduced density matrix) 
of the ground-state wave function, 
might be the fingerprint of the physical edge states that reflect the topological order in the bulk. 
Specifically, the entanglement levels below the entanglement gap reflect the structure of 
the physical edge excitations.\cite{Li-H-08,Thomale-S-R-B-10} 
Later, the entanglement spectrum has been proven useful 
in uncovering the bulk topological properties in a variety of systems 
(e.g. quantum-Hall systems \cite{Thomale-S-R-B-10,Regnault-B-H-09,Lauchli-B-S-H-10}, 
topological insulators \cite{Fidkowski-10,Prodan-H-B-10,Turner-Z-V-10} 
and spin chains \cite{Thomale-A-B-10,Pollmann-T-B-O-10}) 
only by looking at their ground-state wave functions.  
Since entanglement cut creates point boundaries in one dimension, we may expect that 
the discrete level structure of the entanglement spectrum reflects the bulk topological order.  

In order to carry out the explicit calculation of the Schmidt coefficients 
(or, entanglement spectrum), 
we adopt the SMPS formalism introduced in our previous paper.\cite{Hasebe-T-11}
One of the biggest merits of using the SMPS formalism is that the Schmidt decomposition, 
which is the essential step of the calculation, is {\em almost} done already 
when we write down the SMPS expression.  
Therefore, all we have to do is to rewrite the SMPS into the form of the Schmidt decomposition 
by using the singular-value decomposition.\cite{Vidal-03,Orus-V-08}  
However, when the (S)MPSs with different edge states 
are asymptotically orthogonal to each other in the infinite-size limit 
(this is the case in all (S)MPSs discussed below), the entanglement spectrum is most easily 
obtained from the (infinite-size) norms for different edge states:  
\begin{equation}
\lambda_{\alpha}= \lim_{j,L-j,L \nearrow \infty}
\sqrt{\frac{{\mathcal{N}_{j}(\alpha_{\text{L}},\alpha)\mathcal{N}_{L-j}
(\alpha,\alpha_{\text{R}})}}{{\mathcal{N}_{L}(\alpha_{\text{L}},\alpha_{\text{R}})}}}
\; ,
\end{equation}
where $\mathcal{N}_{j}$ is the squared norm of the MPS on a length-$j$ system 
\begin{equation}
\mathcal{N}_{j}(\alpha,\beta) 
\equiv  |(\mathcal{A}_1\mathcal{A}_2 \cdots   \mathcal{A}_j)_{\alpha,\beta}|^2
\; .
\end{equation}

\subsection{UOSp(1$|$2) SVBS states}
\label{subsec:typeISVBS}

\subsubsection{$\mathcal{S}=1$}

\begin{figure}[floatfix]
\centering
\includegraphics[width=7cm]{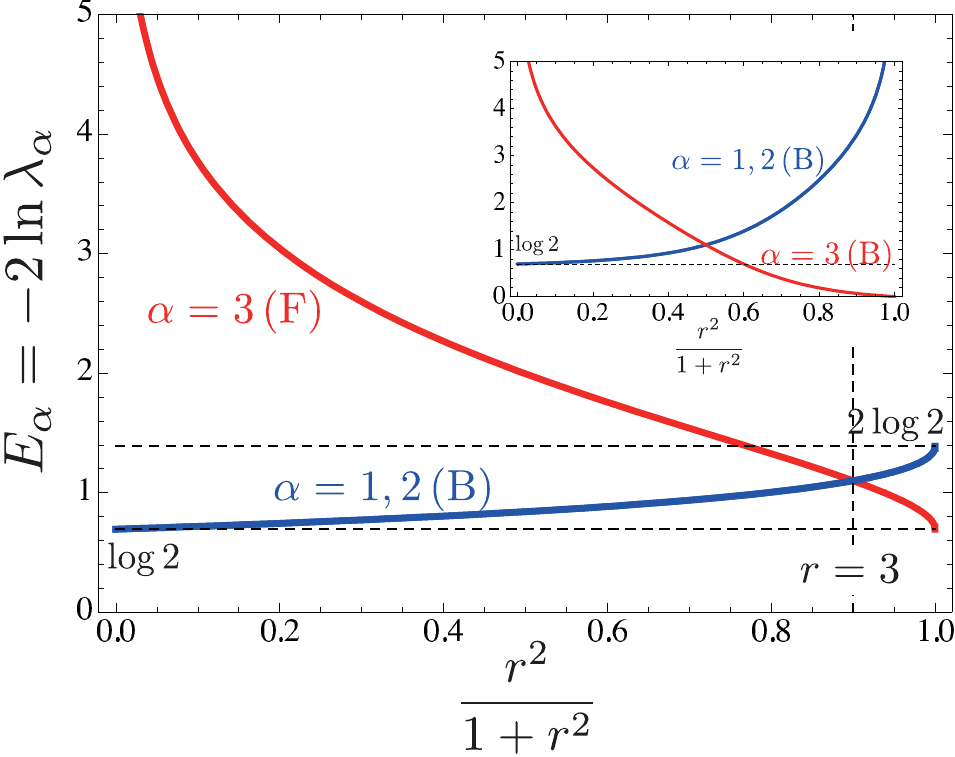}
\caption{(Color online) The behavior of entanglement spectrum 
of the $\mathcal{S}=1$ UOSp(1$|$2) SVBS state (the inset is for the bosonic-pair VBS state).  
`B' and `F' denote bosonic- and fermionic part 
of the spectrum, respectively. \label{coefficientsSL11.fig}}
\end{figure}

\begin{figure}[floatfix]
\centering
\includegraphics[width=7cm]{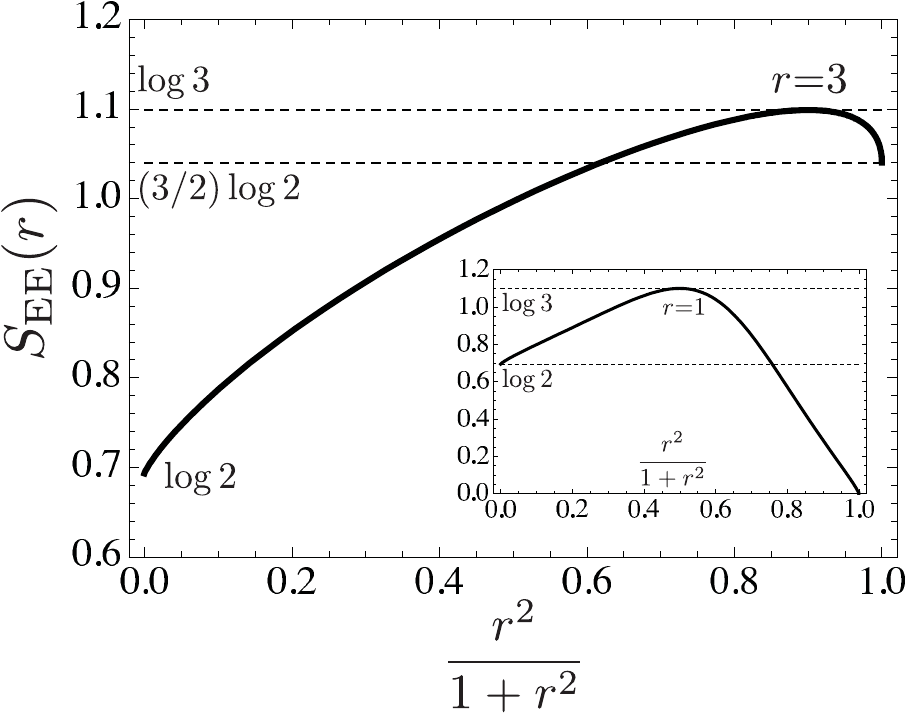}
\caption{The behavior of the entanglement entropy of the $\mathcal{S}=1$ UOSp(1$|$2) SVBS state. 
(The inset is for the bosonic-pair VBS state.) 
\label{entropySL11.fig} }
\end{figure}

By utilizing the SMPS,  the Schmidt coefficients of the SVBS infinite chain, are readily derived as 
\begin{subequations}
\begin{align}
&{\lambda_{\text{B}}}^2\equiv {\lambda_1}^2={\lambda_2}^2=\frac{1}{4}+\frac{3}{4\sqrt{9+8r^2}} \; ,\\
&{\lambda_{\text{F}}}^2\equiv {\lambda_3}^2=\frac{1}{2}-\frac{3}{2\sqrt{9+8r^2}} \; ,
\end{align}
\label{s=1osp(12)Schmidt}
\end{subequations}
which are shown in Fig.~\ref{coefficientsSL11.fig}, and 
the corresponding entanglement entropy
\begin{equation}
S_{\text{EE}}=-\sum_{\alpha}{\lambda_{\alpha}}^2 \text{log}_2 {\lambda_{\alpha}}^2 
\end{equation}
is also depicted in Fig.\ref{entropySL11.fig}.  
From the entanglement spectra, we find that the bosonic and the fermionic sectors exhibit distinct behaviors.  
As mentioned in section \ref{subsec:typeISVBS}, 
the SVBS chain interpolates the original VBS ($r=0$) and the MG dimer chains ($r\rightarrow \infty$). 
Then, we expect the entanglement entropy of SVBS chain also reduces that of VBS at $r=0$, and that of MG at $r\rightarrow\infty$.  
Indeed, in such two limits, the entanglement entropy gives those of the VBS and MG dimer chains:     
\begin{equation}
\begin{split}
\lim_{r\rightarrow 0}S_{\text{EE}}(r) &=\log 2, \\
\lim_{r\rightarrow \infty} S_{\text{EE}}(r) &= \frac{3}{2}\log 2. 
\end{split}
\label{asymptoticentanglemententropy1}
\end{equation}
The states are maximally entangled when   
\begin{equation}
 {\lambda_1}^2={\lambda_2}^2={\lambda_3}^2={1}/{3}    ~~~~~(\text{at} ~~r=3),   
\end{equation}
where the entanglement entropy takes the maximal value 
$S_{\text{EE}}^{\text{(max)}}=\log 3$.  
In contrast to the usual bosonic VBS states,\cite{
Fan-K-R-04, 
katsura-2007-76, katsura-2008-41} the entanglement entropy 
$S_{\text{EE}}$ of the SVBS states differs from what is expected from 
the dimension of the MPS matrices (i.e. bond dimension); 
they attain the maximal entanglement {\em only} at a particular value  
of the doping parameter $r$, which is different from the position of the maximal 
entanglement of the corresponding maximally-entangled pairs 
[for more details, see the Supplementary Material Ref.~\onlinecite{suppl}].  

The `level crossing point' ($r=3$) between the bosonic and the fermionic spectra  
generally does not imply a quantum phase transition, 
in the sense  that divergence of physical quantities, e.g.  
spin-spin correlation length, does not occur at the point.  
The (open) $\mathcal{S}=1$ SVBS chain accommodates  $\mathcal{S}=1/2$ superspins 
at the edges, $i.e.$ the number of the edge degrees of freedom is 3 corresponding to $a^{\dagger}|\text{vac}\rangle$, $b^{\dagger}|\text{vac}\rangle$ and $f^{\dagger}|\text{vac}\rangle$. 
Therefore, as has been found \cite{katsura-2007-76, katsura-2008-41} 
in the usual bosonic VBS states, one sees that the entanglement entropy 
is bounded by the logarithm of the number of the edge degrees of freedom. 
However, here is one remarkable point; 
since the parameter $r$ controls the contributions of the bosonic- and the fermionic degrees of freedom, 
one might expect that the entanglement is maximal at $r=1$ where they appear with equal amplitudes 
(indeed, this is the case for a system of two $\mathcal{S}=1/2$ superqubits 
[see Ref.~\onlinecite{suppl}]).    
Contrary to this naive expectation, the explicit calculation indicates that 
the maximally entangled point is located at $r=3$ due to many-body effect of SUSY.  
Note still in the bosonic many-body case, the entanglement is maximal at $r=1$ 
(see the inset in Fig.\ref{entropySL11.fig}). 

To see a property peculiar to the SUSY states, let us introduce a ``boson-pair VBS state'': 
\begin{align}
|\text{b-p.-VBS}\rangle 
=
\prod_{j}(a_j^{\dagger}b_{j+1}^{\dagger}-b_{j}^{\dagger}a_{j+1}^{\dagger}-rc_j^{\dagger}c_{j+1}^{\dagger})|\text{vac}\rangle, 
\label{bosonpairsvbswav}
\end{align}
where $c_i^{\dagger}$ denotes the creation operator for a bosonic holes that satisfies $[c_i,c^{\dagger}_j]=\delta_{ij}$ 
and $a^{\dagger}_{j}a_{j}+b^{\dagger}_{j}b_{j}+c^{\dagger}_{j}c_{j}=2$.  
The new state $|\text{b-p.-VBS}\rangle$ derived simply 
by replacing the fermionic operator $f^{\dagger}$ in the SVBS state 
\eqref{eqn:SUSY-I-def} with bosonic one $c^{\dagger}$ 
neither has the inversion symmetry with respect to the center of 
a link ({\em link-inversion}) 
nor has the UOSp(1$|$2) symmetry.  
More importantly,  
The entanglement spectrum is plotted  in the inset of Fig.~\ref{coefficientsSL11.fig}.  
As in the $\mathcal{S}=1$ SVBS state, the boson-pair VBS chain has three Schmidt eigenvalues,  
two of which are doubly degenerate and the other is non-degenerate. 
On the other hand, the entanglement entropy (see the inset of Fig.~\ref{entropySL11.fig}) exhibits 
a different asymptotic behavior for $r\rightarrow \infty$ since 
$|\text{b-p.-VBS}\rangle$ reduces, in the limit $r\rightarrow \infty$, to the product state   
$~\prod_j c_j^{\dagger} |\text{vac}\rangle$, while the SUSY version $|\text{SVBS}(1|2)\rangle$ 
still retains finite entanglement due to SUSY.  

\subsubsection{$\mathcal{S}=2$}

\begin{figure}[floatfix]
\centering 
\includegraphics[scale=0.7]{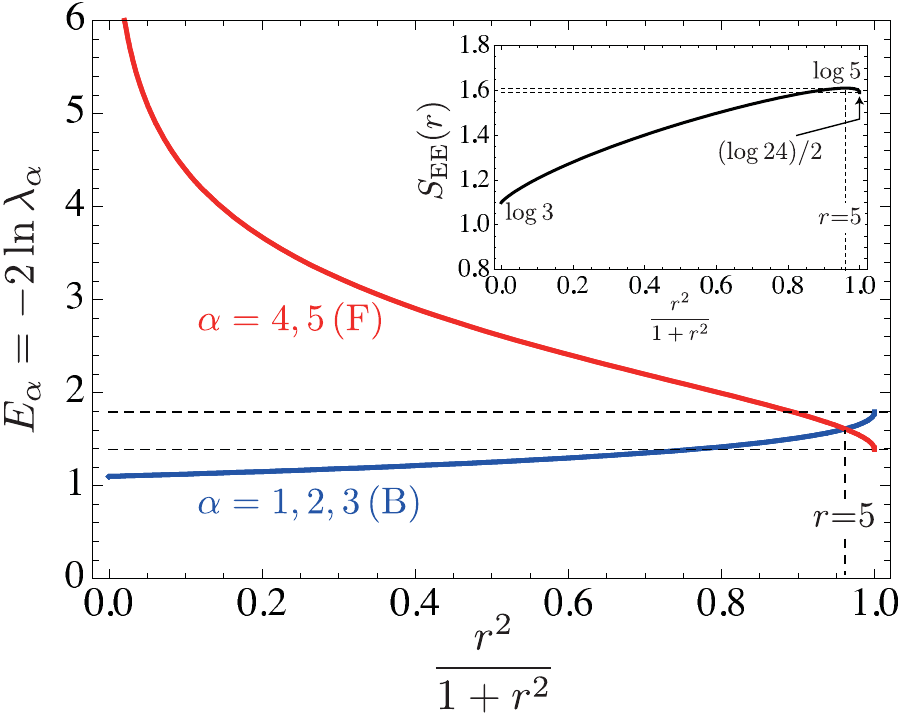}
\caption{(Color online) 
The entanglement spectrum and the entanglement entropy (inset) of the $\mathcal{S}=2$  
UOSp(1$|$2) SVBS chain. `B' and `F' denote bosonic- and fermionic part 
of the spectrum, respectively. 
\label{entropySL12.fig} }
\end{figure}

Next, we proceed to the $\mathcal{S}=2$ SVBS chain.  The bulk superspin is $\mathcal{S}=2$ which consists of SU(2) $S=2$ and $S=3/2$ spins.   
Therefore, we have five Schmidt coefficients, three of which (bosonic part) come from SU(2) $S=1$ and the remaining two (fermionic part) come from SU(2) $S=1/2$. 
The Schmidt coefficients are calculated as 
\begin{subequations}
\begin{align}
&{\lambda_{\text{B}}}^2\equiv {\lambda_1}^2={\lambda_2}^2={\lambda_3}^2=\frac{1}{6}+ \frac{5(4+\sqrt{25+24r^2})}{6(25+24r^2+4\sqrt{25+24r^2})},\\
&{\lambda_{\text{F}}}^2\equiv {\lambda_4}^2={\lambda_5}^2= \frac{1}{4}- \frac{5(4+\sqrt{25+24r^2})}{4(25+24r^2+4\sqrt{25+24r^2})} .
\end{align}\label{s=2osp(12)Schmidt}
\end{subequations}
The bosonic part is triply degenerate as in the case of original $S=2$ VBS chain, while the fermionic part,  which  newly appeared in SUSY case, is doubly degenerate. 
Such double degeneracy is a fingerprint of a symmetry-protected topological (Haldane) phase 
in 1D.\cite{Pollmann-T-B-O-10,*Pollmann-B-T-O-12}   
In the absence of fermionic holes ($r=0$), the fermionic part of the spectrum is 
infinitely higher-lying (see Fig.~\ref{entropySL12.fig}) 
and the entanglement of the system is completely determined only by 
the bosonic part which does not show the signature of the Haldane phase. 

In the SUSY case, on the other hand, the fermionic levels appear 
above the finite entanglement gap and there always exists 
doubly degeneracy in the Schmidt coefficients 
which accounts for the topological stability 
of the SVBS state regardless of the parity of the bulk superspin $\mathcal{S}$. 
We will revisit this in section \ref{sect:supersymmetryprotectedtopologicalorder}. 
As shown in Fig.\ref{entropySL12.fig},  the five Schmidt coefficients take the same value $1/5$ at $r=5$, and  
the asymptotic behaviors of the entanglement entropy are 
\begin{equation}
\begin{split}
\lim_{r\rightarrow 0}S_{\text{EE}}(r) &=\log 3, \\
\lim_{r\rightarrow \infty} S_{\text{EE}}(r) &= \log 2 + \frac{1}{2} \log 6 \; .
\end{split}
\label{asymptoticentanglemententropy2}
\end{equation}
Thus, at $r\rightarrow \infty$, the $\mathcal{S}=2$ SVBS state 
supports the finite entanglement entropy and does not reduce 
to a simple product state as in the $\mathcal{S}=1$ SVBS chain.  

\subsection{UOSp(1$|$4) SVBS states}

\begin{figure}[floatfix]
\centering
\includegraphics[width=6.5cm]{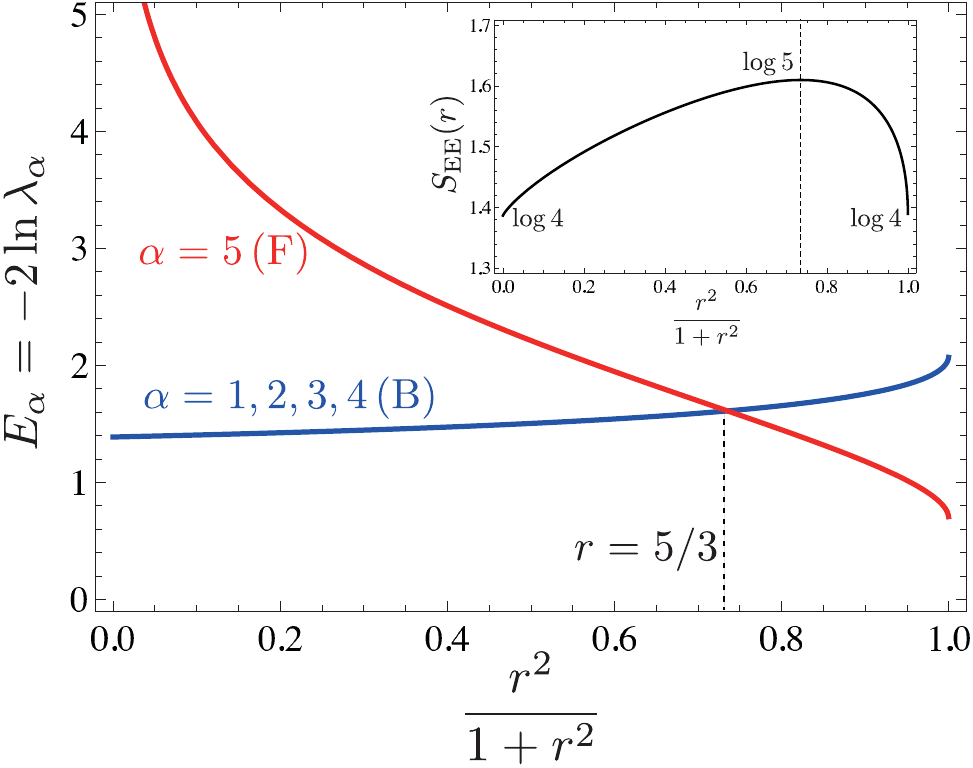}
\caption{(Color online) The entanglement spectrum of the  
UOSp(1$|$4) SVBS state \eqref{eqn:OSP14-MPS-adj}.  
The entanglement entropy of the same state is shown in the inset.  
\label{fig:ES-Osp14-adj} }
\end{figure}

In the case of UOSp(1$|$4) ($(N,K)=(1,2)$), we obtain the entanglement spectrum 
of the MPS \eqref{s05svbswavefun} as:
\begin{equation}
\begin{split}
& (\lambda_{\sigma}(r))^{2}= \frac{1}{8} + \frac{5}{8 \sqrt{16 r^2+25}} \;\; (\sigma=1,2,3,4) \\
& (\lambda_{5}(r))^{2} = \frac{1}{2}-\frac{5}{2 \sqrt{16 r^2+25}} \; , 
\end{split} 
\end{equation}
which are plotted in Fig.~\ref{fig:ES-Osp14-adj} together with the corresponding entanglement 
entropy.  
The bosonic part of the spectrum is quadratically degenerate while the fermionic part is non-degenerate. 
In both cases, the entanglement entropy $S_{\text{EE}}(r)$ 
takes its maximal value $\log 5$ at intermediate value of $r=5/3$ 
where all the five Schmidt coefficients coincide.  
The entanglement entropy $S_{\text{EE}}(r)$ exhibits the following asymptotic behaviors: 
\begin{equation}
\lim_{r\rightarrow 0}S_{\text{EE}}(r)
=\lim_{r\rightarrow \infty} S_{\text{EE}}(r) = \log 4   \; .
\label{asymptoticosp14entanglemententropy}
\end{equation}
If we had a boson ${b^{5}}^{\dagger}$ instead of the fermion $f^{\dagger}$ in \eqref{s05svbswavefun} 
as in the boson-pair VBS state eq.\eqref{bosonpairsvbswav}, 
entanglement would vanish in the limit $r \rightarrow \infty$.  
Therefore, the existence of finite entanglement even in the $r\rightarrow \infty$ limit 
may be attributed to the fermionic property of the holes. 

Here it should be emphasized that all the limiting behaviors 
\eqref{asymptoticentanglemententropy1}, 
\eqref{asymptoticentanglemententropy2} and \eqref{asymptoticosp14entanglemententropy} 
can be understood from the viewpoint of the edge states; 
basically, the limiting value of $S_{\text{EE}}(r)$ is determined solely by information of 
the irreducible representation which describes the emergent edge states.  
In fact, the general formulas \eqref{eqn:SEE-r-0-limit} and \eqref{eqn:SEE-r-infty-limit} given 
in appendix \ref{sec:edge-ES-relation} reproduce the above results.  

\section{Supersymmetry-protected Topological Order}
\label{sect:supersymmetryprotectedtopologicalorder}

In this section, we show that a family of SVBS states $|\text{SVBS}(1|2K)\rangle$  
 exhibits the generalized topological order which will be characterized below. 
Our argument is a SUSY generalization of the one presented in 
Ref.~\onlinecite{Pollmann-T-B-O-10,*Pollmann-B-T-O-12}.   
In the following arguments, we utilize the SMPS formalism. 
The SMPS formalism itself is defined independent of the super Lie group symmetries, 
and is a general formalism to treat a system of boson-fermion mixture 
whose ground state is represented by a supermatrix.  
Therefore, though we mainly discuss the SVBS states which have specific underlying 
particular super Lie group symmetries, 
the following arguments apply to any boson-fermion mixture systems.

Before going into the detail, we first characterize the symmetry operation 
(both unitary and anti-unitary) within the framework of MPS.\cite{Garcia-W-S-V-C-08} 
The MPS $\mathcal{A}_{1}\mathcal{A}_{2}\cdots$ is said to be invariant under 
the (anti-)unitary operation if the transformed 
state $\mathcal{A}^{\prime}_{1}\mathcal{A}^{\prime}_{2}\cdots$ coincides 
with the original one up to an overall phase.  
Then, it can be shown\cite{Garcia-W-S-V-C-08} that the invariance of a pure MPS 
is equivalent to the existence of a $D$-dimensional ($D$ being the size of the MPS 
matrix $A$) unitary matrix $U$ which satisfies
\begin{equation}
A^{\prime}(m) = \be^{i\theta} U^{\dagger}A(m) U \; .
\label{eqn:A-to-UAU}
\end{equation}
The phase $\theta$ is not universal and depends, in general, on the symmetry operation in question.  

The ($c$-number) unitary matrix $U$ in \eqref{eqn:A-to-UAU} may be postulated as: 
\begin{equation}
U_{I} =
\begin{pmatrix}
U_{\text{B}} & 0 \\
0 & U_{\text{F}}
\end{pmatrix}, 
\end{equation}
where $U_{\text{B}}$ and $U_{\text{F}}$ are unitary matrices that act on 
the two bosonic subspaces having different fermion numbers. 
The reason for choosing the above form may be seen as follows. 
First we note that eq.\eqref{eqn:A-to-UAU} implies that the MPS transforms like
\begin{equation}
|\Psi\rangle \mapsto  
\str ( U^{\dagger}\mathcal{A}_1 \mathcal{A}_2\cdots \mathcal{A}_{2n+1}U) \; ,
\end{equation}
where {\em supertrace} is defined as
\begin{equation}
\text{str} 
\begin{pmatrix}
A_{\text{B}}^{(1)} & A_{\text{F}}^{(1)} \\
A_{\text{F}}^{(2)} & A_{\text{B}}^{(2)} 
\end{pmatrix}
= \text{tr}A_{\text{B}}^{(1)} - \text{tr}A_{\text{B}}^{(2)} \; .
\end{equation}
While in the case of bosonic MPS, this, combined with $\text{tr}(AB)=\text{tr}(BA)$, 
immediately implies $\mathcal{I}|\Psi\rangle \propto |\Psi\rangle$, 
the relation $\str(AB)=\str(BA)$ 
holds only when $A$ and $B$ are super-matrices 
(that contain the Grassmann-odd blocks in their off-diagonal parts). 
In fact, if $A$ and $B$ were merely the $c$-number matrices, 
$A$ and $B$, in general, would not commute inside $\str({\cdot})$: $\str(AB)\neq \str(BA)$.  
To satisfy $\str(AB)=\str(BA)$ only with $c$-number matrices, 
either $A$ or $B$ is forbidden to have $c$-number components 
in the off-diagonal blocks.  

Physically, the above relation states that the original symmetry operation 
(acting on the physical Hilbert space on each site) 
`fractionalizes' into the ones ($U$ and $U^{\dagger}$) which act on the edge 
states on both ends of the system.   

In what follows, we parametrize the $A(m)$-matrices in terms of the $D{\times}D$-matrices 
$(\Lambda,\{\Gamma(m)\})$ as $A(m)= \Gamma(m)\Lambda$.   
The diagonal matrix $\Lambda$ contains the Schmidt eigenvalues in its diagonal elements 
($\tr(\Lambda^{2})=1$) and commutes with the unitary matrix: $[\Lambda,U]=0$.  
In what follows, we use the symbol $\Gamma$ for the MPS $A$-matrices in the {\em canonical} 
form.\cite{Vidal-03} 

Then, the $\Gamma$-matrices satisfy the condition for the canonical MPS 
on infinite-size systems\cite{Orus-V-08}
\begin{equation}
\sum_{m}\Gamma^{\dagger}(m)\Lambda^{2}\Gamma(m) = \mathbf{1}_{D} \; .
\label{eqn:canonical-condition}
\end{equation}
(For more details about the properties of $U$, see appendix \ref{sect:proofsymmetryprotencted}.)  
In terms of these $\Gamma$ matrices, eq.\eqref{eqn:A-to-UAU} reads as 
\begin{equation}
\Gamma^{\prime}(m) = \be^{i\theta} U^{\dagger}\Gamma(m) U \; .
\label{eqn:U-Gamma-U}
\end{equation}
Now let us determine the properties of $U$ satisfying the above equation for 
specific symmetry operations. 

\subsection{Inversion symmetry}\label{subsec;Inversion}

A matrix product state on a circle is given by 
\begin{equation}
|\Psi\rangle =\str(\mathcal{A}_1\mathcal{A}_2\cdots \mathcal{A}_{2n+1}) \; ,
\end{equation}
where `str' denotes the super-trace. 
By the inversion with respect to a given link, the state is transformed as 
\begin{equation}
\mathcal{I}|\Psi\rangle = \str(\mathcal{A}_{2n+1}\cdots \mathcal{A}_2 \mathcal{A}_1). 
\end{equation}
Here, we use the property of the supertrace: 
$\str(M_1M_2)= \str((M_1M_2)^{\st})= \str(M_2^{\st}M_1^{\st})$ to rewrite the above as 
 \begin{equation}
 \mathcal{I}|\Psi\rangle = \str ( \mathcal{A}_1^{\st} \mathcal{A}_2^{\st} 
 \cdots \mathcal{A}_{2n+1}^{\st}) \; ,
 \label{eqn:inverted-MPS}
 \end{equation}
where {\em supertransposition} `$\st$' is defined as 
\begin{equation}
\begin{pmatrix}
M_1 & N_1 \\
N_2 & M_2
\end{pmatrix}^{\st}\equiv 
\begin{pmatrix}
M_1^{\text{t}} & N_2^{\text{t}} \\
-N_1^{\text{t}} & M_2^{\text{t}}
\end{pmatrix}.
\label{eqn:super-tanspos-def}
\end{equation}
Therefore, the link-inversion $\mathcal{I}$ amounts, in terms of $\mathcal{A}$, 
to
\begin{equation}
\mathcal{A}_{i} \xrightarrow{\mathcal{I}} {\mathcal{A}_{i}}^{\st}  \; .
\end{equation}
If we write 
\begin{equation} 
\mathcal{A}_{i} = \sum_{m} \Lambda \Gamma(m)|m\rangle_{i} \; ,
\end{equation} 
we see that $\mathcal{I}$ acts on $\Gamma(m)$ as 
\begin{equation}
\Gamma(m) \xrightarrow{\mathcal{I}}
\Gamma'(m)=\Gamma(m)^{\st} \; .
\end{equation}
Here, $m$ labels both bosonic and fermionic components 
and $\Gamma(m)$ are given by 
\begin{equation}
\begin{split}
&\Gamma(m)=\begin{pmatrix}
M_1(m) & 0 \\
0 & M_2(m)
\end{pmatrix} \;\; (\text{$m$: bosonic})\\
&\Gamma(m)=\begin{pmatrix}
0 & N_1(m)\\
N_2(m) & 0
\end{pmatrix} \;\; (\text{$m$: fermionic}) \; .
\end{split}
\end{equation}
Originally, $M_1,M_2,N_1$ and $N_2$ are all {\em c-number} coefficient matrices.  
However, for practical reasons, it is often convenient to assume that the basis states 
are commuting and take into account the anti-commuting properties of the fermionic 
states by supermatrices. 

If $\mathcal{I}$ leaves the MPS invariant up to a phase, 
the general relation\cite{Garcia-W-S-V-C-08} \eqref{eqn:U-Gamma-U} implies that 
there exists a unitary matrix $U_{I}$ satisfying
\begin{equation}
 \Gamma(m)^{\st}=e^{i\theta_{I}} U_{I}^{\dagger}\Gamma(m)U_{I} .  
\label{sinversionsymm2}
\end{equation}
In fact, we can prove that $\theta$ can take the only two values, $0$ and $\pi$, namely 
\begin{equation}
{U_{I}}^{\dagger}\Gamma(m)U_{I} =\pm \Gamma(m)^{\st} \; .
\end{equation}
For later convenience, we introduce the following diagonal matrix 
having the same block diagonal structure as $U_{I}$:
\begin{equation}
P \equiv 
\begin{pmatrix}
\mathbf{1}_{\text{B}} & 0 \\
0 & -\mathbf{1}_{\text{F}} 
\end{pmatrix} \;\; 
(U_{I}P=PU_{I} ) \; .
\label{definversionP}
\end{equation} 

Then, the fact that the link-inversion squares to unity leads to 
an important conclusion that $U_I$ is a `symmetric' or `antisymmetric' unitary matrix: 
\begin{equation}
U^{\text{t}}_{I}=\pm PU_{I} 
\label{eqn:UI-transpose}
\end{equation}
The appearance of $P$ is closely related to the property of 
supertransposition:
\begin{equation}
(A^{\st})^{\st}= PAP \; .
\end{equation}
We give the outline of the proof in the appendix \ref{sect:proofsymmetryprotencted}. 

By computing the determinant of the above, one can show that 
either fermionic (when the sign + occurs) or bosonic ($-$) sector has even-fold degeneracy 
in each entanglement level, which we will use as the fingerprint of 
the SUSY-protected topological order.    

\subsection{Time-Reversal Symmetry}

Before discussing the properties of SMPS under time-reversal, let us define 
the time-reversal operation in the SUSY case. 
Under the time reversal transformation $\mathcal{T}$, the spin is transformed as 
\begin{equation}
S_a \overset{\mathcal{T}}\rightarrow -S_a. 
\end{equation}
In the usual matrix representation, the above relation can be expressed as 
\begin{equation}
S_a \rightarrow -S_a =(e^{i\pi S_y} K) S_a (K e^{-i\pi S_y})=R^y_{ab}(\pi)S_b^{*},  
\end{equation}
where  $K$ is the complex conjugation operator and $R^y(\pi)$ represents 
the $\pi$-rotation around the $y$-axis: 
\begin{equation}
R^y(\pi)=
\begin{pmatrix}
-1 & 0 & 0 \\
0 & 1 & 0 \\
0 & 0 & -1 
\end{pmatrix}. 
\end{equation}

As in the usual case, time reversal operation is defined as 
\begin{equation}
\begin{split}
&S_a \overset{\mathcal{T}}\rightarrow (e^{i\pi S_y} K)S_a (K e^{-i\pi S_y}) =-S_a, \\
&S_{\sigma} \overset{\mathcal{T}}\rightarrow 
(e^{i\pi S_y} K)S_{\sigma} (K e^{-i\pi S_y})=\epsilon_{\sigma\tau}S_{\tau}, 
\end{split}
\label{trevfermionicS}
\end{equation}
where $\text{UOSp}(1|2)$ superspin matrices $S_a$ $(a=x,y,z)$ 
and $S_{\sigma}$ $(\sigma=\theta_1,\theta_2)$ are defined as  
\begin{equation}
S_a=\frac{1}{2}\begin{pmatrix} 
\sigma_a & 0 \\
0 & 0 
\end{pmatrix},~~~S_{\sigma}=\frac{1}{2}
\begin{pmatrix}
0 & \tau_{\sigma} \\
-(i\sigma_2 \tau_{\sigma})^{\text{t}} & 0 
\end{pmatrix},\label{TRsuperspinmat}
\end{equation}
with the Pauli matrices $\sigma_a$ and $\tau_{1}=(1,0)^{\text{t}}$ and  $\tau_{2}=(0,1)^{\text{t}}$. 
The fermionic generators $S_{\sigma}$ have the off-diagonal blocks 
which transform as different irreducible representations of $\text{SU}(2)$ 
and act as spin-1/2 raising- 
and lowering matrices. 
 In the Schwinger operator representation, 
$S_{\sigma}$ are explicitly given by  
$S_{\theta_1}=\frac{1}{2}(a^{\dagger}f+f^{\dagger}b)$, 
$S_{\theta_2}=\frac{1}{2}(b^{\dagger}f-f^{\dagger}a)$.  
Under the time-reversal transformation, the SU(2) spinor states are interchanged: 
$|\!\!\uparrow \rangle=a^{\dagger}|0\rangle \rightarrow |\!\!\downarrow\rangle=b^{\dagger}|0\rangle $, 
$|\!\!\downarrow \rangle =b^{\dagger}|0\rangle\rightarrow -|\!\!\uparrow\rangle=-a^{\dagger}|0\rangle$, 
and the spin-less fermion state remains the same: 
$f^{\dagger}|0\rangle \rightarrow f^{\dagger}|0\rangle$. 
This implies that the time reversal transformation of $S_{\sigma}$ 
is given by (\ref{trevfermionicS}). 
Then we have $\mathcal{T}^2 S_{\sigma}=-S_{\sigma}$, 
so the relation $\mathcal{T}^2=-1$ for half-integer spins appear for the
 ``fermionic spins''.  
 
In fact, for integer superspins, $\mathcal{T}$ satisfies\footnote{%
When $\mathcal{S}$ is half-odd-integer, $\mathcal{T}^{2} =-P$ which generalizes 
$\mathcal{T}^{2}=-\mathbf{1}$ for the SU(2) case.}
\begin{equation}
\mathcal{T}^{2} = \mathcal{P} \; , \;\; (\mathcal{P})_{mn} = \delta_{mn}(-1)^{F(n)} \; , 
\label{squaretrel}
\end{equation}
where $\mathcal{P}$ acting on the physical Hilbert space is analogous to $P$ 
in eq.\eqref{definversionP} acting 
on the auxiliary space and, due to the fermion number operator $F(n)$ 
($F(n)=0$ or $F(n)=1$ when $n$ labels the bosonic or fermionic variables),  
$(-1)^{F(n)}$ gives a minus sign 
for the fermionic sector of the (physical) Hilbert space.  
 
Using the above properties, one can readily see that 
the time reversal operation transforms $\Gamma(m)$ as: 
\begin{equation}
\Gamma(m) \xrightarrow{\mathcal{T}} 
\Gamma(m)'=\sum_{n}R^{y}_{mn}(\pi)\Gamma(n)^*. 
\end{equation}
Then,  time reversal invariance of the SMPS means that there exists 
a unitary $U_{T}$ such that\cite{Garcia-W-S-V-C-08} 
\begin{equation}
\sum_n R_{mn}^y(\pi)\Gamma^*(n)=e^{i\theta_T}U_{T}^{\dagger}\Gamma(m)U_{T}. 
\label{susytimereversal}
\end{equation}
The property $\mathcal{T}^{2}=\mathcal{P}$ (for integer superspin) 
requires that the unitary matrix $U_{T}$ should satisfy   
\begin{equation}
U_{T}^{\text{t}}=\pm P U_{T} \; . 
\label{eqn:UT-transpose}
\end{equation}
Since this is exactly the same as eq.(\ref{eqn:UI-transpose}) for 
the link-inversion, a similar conclusion is drawn 
about the entanglement spectrum. 

\subsection{$\mathbb{Z}_2 \times \mathbb{Z}_2$ symmetry}
\label{sec:UOSp12-Z2-Z2}

The $\mathbb{Z}_2 {\times} \mathbb{Z}_2$ symmetry\cite{Kennedy-T-92-PRB,Kennedy-T-92-CMP} 
in the original bosonic case is generated by the two commuting $\pi$ rotations 
around $x$- and $z$ axes. 
However, the symmetry around each axis alone does not directly imply 
the double degeneracy of the entanglement spectrum. 
Rather, it has been shown\cite{Pollmann-T-B-O-10} that 
their combination leads to a non-trivial conclusion concerning the entanglement spectrum.  
In the following, we show that an analogous symmetry leads to a similar conclusion even in the presence 
of SUSY. 

The $\pi$ rotation around the $x$ ($z$) axis $\hat{u}_{x}(\pi)$ 
($\hat{u}_{z}(\pi)$) acts on SMPS as: 
\begin{equation}
\Gamma(m) \xrightarrow{\hat{u}_{a}(\pi)}
\Gamma(m)'= \sum_n R_{mn}^{a}(\pi)\Gamma(n) \;\; (a=x,z) \; , 
\end{equation}
where $R_{mn}^{a}(\pi)$ is the $(4\mathcal{S}+1)$-dimensional 
rotation matrix of UOSp(1$|$2)  (see, e.g., eq.\eqref{eqn:OSp12-rotation-y}). 
The right hand side is equivalent to the action of a unitary matrix 
$U_{a}$\cite{Garcia-W-S-V-C-08} 
\begin{equation}
\sum_n R_{mn}^{a}(\pi)\Gamma(n)
=\be^{i\theta_{a}}U_{a}^{\dagger}\Gamma(m)U_{a} \;\; (a=x,z) 
\; .
\end{equation}
Then, the property $(R^{a})^{2}=\mathcal{P}$ implies the following 
\begin{align}
&e^{2i\theta_x}=1 \Rightarrow e^{i\theta_x}=\pm 1, \nonumber\\
&U_a P U_a =e^{i\phi_a}\mathbf{1} \; . 
\end{align}
The phase factor $e^{i\phi_{a}}$ can be absorbed in the definition of $U_{a}$ 
and we may assume $U^{\dagger}_{a} = PU_{a}$ ($a=x,z$) hereafter. 

On the other hand, for the combination of the rotations $\hat{u}_{x}(\pi)$ and 
$\hat{u}_{z}(\pi)$, we obtain (see appendix \ref{sec:proof-Z2-Z2-sym} for detail)
\begin{equation}
(U_z P U_x)(U^{\dagger}_{z}U^{\dagger}_{x})=e^{i\phi_{xz}}\mathbf{1}. 
\end{equation}
By using $U_{a}^{\dagger} = PU_{a}$ obtained above, one can show 
$\be^{i\phi_{xz}}=\pm 1$ and the following exchange property:
\begin{equation}
U_x U_z  =\pm P U_{z}U_{x} \; . 
\label{eqn:UxUz-exchange-SUSY}
\end{equation}
In terms of the block components $U_{a,\text{B}}$ and $U_{a,\text{F}}$,   
this reads:
\begin{equation}
U_{x,\text{B}}U_{z,\text{B}} = \pm U_{z,\text{B}}U_{x,\text{B}} \; , \;\;
U_{x,\text{F}}U_{z,\text{F}} = \mp U_{z,\text{F}}U_{x,\text{F}} \; ,
\end{equation}
which immediately implies the same degenerate structure of the entanglement 
spectrum as in the two previous cases. 

\subsection{$(\mathbb{Z}_2\times \mathbb{Z}_2)^2$ symmetry in UOSp(1$|$4) SVBS }
\label{subsec:z2z2UOSp(1|4)}

Now let us discuss the entanglement spectrum in the systems with SO(5)-symmetry 
and its SUSY generalization UOSp(1$|$4).  
Inversion symmetry acts independently of the internal symmetry and leads to exactly the same 
conclusion as above. 
The crucial difference from the SU(2) case is the existence 
of $(\mathbb{Z}_2 \times \mathbb{Z}_2 )^2$-symmetry\cite{Tu-Z-X-08}  
in a class of the SO(5) VBS states.\footnote{
Specifically, $(\mathbb{Z}_2 \times \mathbb{Z}_2 )^2$-symmetry can be defined 
for the SO(5) states where all the allowed weights at each site are integers 
(e.g. the vector- and the adjoint representations). 
} 
Specifically, the group $(\mathbb{Z}_2{\times} \mathbb{Z}_2)^2$ consists of 
the following 16 elements: 
\begin{align}
\overbrace{(1,R^{12}(\pi))\times (1,R^{25}(\pi))}^{\mathbb{Z}_2 
\times \mathbb{Z}_2 }\times \overbrace{(1,R^{34}(\pi))
\times (1,R^{45}(\pi))}^{\mathbb{Z}_2 \times \mathbb{Z}_2 } \; ,
\end{align}
with $R^{ab}(\pi) \equiv \exp(i\pi \sigma_{ab})$ ($\sigma_{ab}$: SO(5) generators).  
The four-fold degeneracy of the entanglement spectra of the SO(5) VBS states has been 
discussed\cite{TuandOrusPRB2011} from the viewpoint of  
$(\mathbb{Z}_2 \times \mathbb{Z}_2 )^2$-symmetry. 

It is straightforward to generalize the above symmetry to the UOSp(1$|$4) case; 
now the matrices $R^{ab}(\pi)$ satisfying ${R^{ab}(\pi)}^{2}=\mathbf{1}$ are replaced by 
the block-diagonal matrices of the form\footnote{%
This is the case for the class of UOSp(1$|$4) states discussed here. 
For the vector representation, for instance, we have a slightly different form of 
$R^{ab}$. 
}
\begin{equation}
R^{ab}(\pi) = 
\begin{pmatrix}
R_{ab}^{(\text{B})} & 0 \\
0 & R_{ab}^{(\text{F})}
\end{pmatrix}
\; .
\label{eqn:Rab-UOSp14}
\end{equation}
For instance, in the superspin-1 UOSp(1$|$4) SVBS state discussed in 
section \ref{sec:SO5-SVBS}, $R_{ab}^{(\text{B})}$ and $R_{ab}^{(\text{F})}$ are given 
by $R^{ab}(\pi)$ in the adjoint- ({\bf 10}) and the spinor ({\bf 4}) representation of SO(5), 
respectively.  
It is easy to show that the above matrices satisfy 
\begin{subequations}
\begin{align}
& R^{ab}(\pi)R^{ab}(\pi) = \mathcal{P}_{4|10}~~ \text{(no sum for $a$ and $b$)}
\label{eqn:RR-rel-1} \\
\begin{split}
& R^{12}(\pi)R^{25}(\pi) =
\mathcal{P}_{4|10} R^{25}(\pi)R^{12}(\pi) \\
& R^{34}(\pi)R^{45}(\pi) =
\mathcal{P}_{4|10} R^{45}(\pi)R^{34}(\pi) 
\end{split}
\label{eqn:RR-rel-2}
\\
& R^{25}(\pi)R^{45}(\pi) = \mathcal{P}_{4|10} R^{45}(\pi)R^{25}(\pi) \\
\begin{split}
& R^{12}(\pi)R^{34}(\pi) = R^{34}(\pi)R^{12}(\pi) \; , \\
& R^{12}(\pi)R^{45}(\pi) = R^{45}(\pi)R^{12}(\pi) \; ,\\
& R^{25}(\pi)R^{34}(\pi) = R^{34}(\pi)R^{25}(\pi) 
  \; ,
\end{split}
\end{align}
\end{subequations}
with 
\begin{equation}
\mathcal{P}_{4|10} \equiv 
\begin{pmatrix}
1_{10} & 0 \\
0 & -1_{4}
\end{pmatrix} \; .
\end{equation} 

Now we can apply the argument in section \ref{sec:UOSp12-Z2-Z2} 
since we have the same exchange relations \eqref{eqn:RR-rel-1}, 
\eqref{eqn:RR-rel-2} as before.  Then, we immediately conclude that 
there exist two sets of the corresponding unitary matrices $\{U_{12},U_{25}\}$ 
and $\{U_{34},U_{45}\}$ satisfying 
\begin{equation}
\begin{split}
& \sum_{n}[R^{ab}(\pi)]_{mn}\Gamma(n) 
= \be^{i\theta_{ab}} U^{\dagger}_{ab}\Gamma(m)U_{ab} \, , \; 
U^{\dagger}_{ab} =  P U_{ab} \\
& U_{12} U_{25}  =\pm P U_{25}U_{12} \; , \;\; 
U_{34}U_{45} = \pm P U_{45}U_{34}  \; ,
\end{split} 
\label{eqn:UabUab-exchange-SUSY}
\end{equation}
where the matrix $P$ is defined in eq.\eqref{definversionP}.  
Note that the same sign should be chosen for the two exchange relations above by 
the SO(5) symmetry. 

The role of the unitary transformation $U_{ab}$ is clear. 
First we note that, as in the SO(5) case, the following two 
are mutually commuting generators of the same block-diagonal 
form as $R^{ab}(\pi)$ [Eq.\eqref{eqn:Rab-UOSp14}] 
\begin{equation}
L^{ab} = 
\begin{pmatrix}
\sigma_{ab}^{(\text{B})} & 0 \\
0 & \sigma_{ab}^{(\text{F})} \; 
\end{pmatrix} 
\end{equation}
and can be used as the weight of UOSp(1$|$4).  
Since $R^{25}$ and $R^{45}$ act on the weight $(L^{12}, L^{34})$ as
\begin{equation}
\begin{split}
& {R^{25}}^{\dagger}L^{12}R^{25}=-L^{12}\, , \; 
{R^{45}}^{\dagger}L^{12}R^{45}= L^{12}  \\
& {R^{25}}^{\dagger}L^{34}R^{25}= L^{34}\, , \; 
{R^{45}}^{\dagger}L^{34}R^{45}= - L^{34}  \; ,
\end{split}
\end{equation}
it is legitimate to assume that the algebra is represented in the product space 
$V_{1}{\otimes}V_{2}$ where $V_{1}$ and $V_{2}$ respectively correspond to 
$\{U_{12},U_{25}\}$ and $\{U_{34},U_{45}\}$.  
For instance, the two unitary operations $U_{25}$ and $U_{45}$ actually mean 
\begin{equation}
\begin{split}
& U_{25} \otimes \mathbf{1} \, , \; 
\mathbf{1} \otimes U_{45} \; . \\
& (U_{25} \otimes \mathbf{1})(\mathbf{1} \otimes U_{45} )
= U_{25} \otimes U_{45} \\
& (\mathbf{1} \otimes U_{45} )(U_{25} \otimes \mathbf{1})
= (P U_{25}) \otimes U_{45} \; .
\end{split}
\end{equation}
Now we use the fact that $V_{1}$ and $V_{2}$ should always have even-dimensional 
sectors $V^{(\text{e})}_{1}$ and $V^{(\text{e})}_{2}$ 
(they have the same dimensions by the SO(5)-symmetry) to show that 
the dimension of $V^{(\text{e})}_{1}{\otimes} V^{(\text{e})}_{2}$ should be integer-multiple of four.  
This explains the existence of the four-fold-degenerate entanglement level in 
the UOSp(1$|$4) SVBS states (see also the argument in appendix \ref{append:z2zesquare}).

\section{Relations between String order parameter and topological order}
\label{sec:relation-string-top}

Later, the use of the string order parameters in detecting the Haldane phase 
was criticized \cite{Gu-W-09} since they are well-defined only in a restricted 
class of models and fail to capture the robustness of the Haldane 
phase as a symmetry-protected topological phase 
(see Refs.~\onlinecite{Haegeman-G-C-S-12,Pollmann-T-12} for the attempts at 
alternative order parameters). 
Now a natural question arises; under what conditions the string order parameters 
(\ref{eqn:Ostring-z}) and (\ref{eqn:Ostring-x}) correctly capture the topological 
nature of the Haldane phase?  
Below we will uncover the explicit relationship between the string order 
and the topological order to answer to this question. 

\subsection{String Order Parameters in MPS Framework}

Let us first consider the structure of the string order parameters 
(\ref{eqn:Ostring-z}) and (\ref{eqn:Ostring-x}) from the MPS point of view.  
\cite{Totsuka-S-mpg-95,Garcia-W-S-V-C-08}   
In evaluating them using MPS, the following matrices are necessary 
\begin{equation}
\begin{split}
& [T^{a}]_{\bar{\alpha},\alpha;\bar{\beta},\beta} \equiv \sum_{m,n=1}^{d}
\left[A^{\ast}(m)\right]_{\bar{\alpha},\bar{\beta}}
\left[A(n)\right]_{\alpha,\beta} 
\langle m|S^{a}|n\rangle \\
& [T_{\text{string}}]_{\bar{\alpha},\alpha;\bar{\beta},\beta} \equiv \sum_{m,n=1}^{d}
\left[A^{\ast}(m)\right]_{\bar{\alpha},\bar{\beta}}
\left[A(n)\right]_{\alpha,\beta} 
\langle m|\be^{i \pi S^{a}}|n\rangle   \\
& [T_{\text{string}}^{a}]_{\bar{\alpha},\alpha;\bar{\beta},\beta} \equiv \sum_{m,n=1}^{d}
\left[A^{\ast}(m)\right]_{\bar{\alpha},\bar{\beta}}
\left[A(n)\right]_{\alpha,\beta} 
\langle m|S^{a}\be^{i \pi S^{a}}|n\rangle \\
& \quad (a=x,z)
\end{split}
\end{equation}
as well as the usual transfer matrix.  
For instance, the MPS expression of the string order parameter $\mathcal{O}^{z}_{\text{string}}$ 
(for an open chain) reads:
\begin{equation}
\begin{split}
O^{z}_{\text{string}} & \equiv  
\Biggl\langle S_{j}^{z}\, \exp\left[
i\pi \sum_{k=j}^{j+n-1}S^{z}_{k}
\right]S^{z}_{j+n}\Biggr\rangle \\
& = 
T^{N_{\text{L}}}T_{\text{string}}^{z} (T_{\text{string}})^{n-1}T^{z}\,T^{N_{\text{R}}}  \; ,
\end{split}
\end{equation}
where we have omitted the denominator necessary to normalize the MPS.  
The two parts $T^{N_{\text{L}}}$ ($N_{\text{L}}=j-1$) and $T^{N_{\text{R}}}$  ($N_{\text{R}}=L-n-j$) 
are straightforward; for the canonical MPS, they reduce, in the infinite-size limit, to:
\begin{equation}
\begin{split}
& [T^{N_{\text{L}}}]_{\bar{\alpha}_{\text{L}},\alpha_{\text{L}};\bar{\beta},\beta}
 \xrightarrow{N_{\text{L}}\nearrow \infty} 
 \delta_{\bar{\alpha}_{\text{L}},\alpha_{\text{L}}}
\delta_{\bar{\beta},\beta} \; ,   \\
& [T^{N_{\text{R}}}]_{\bar{\alpha},\alpha;\bar{\beta}_{\text{R}},\beta_{\text{R}}}
 \xrightarrow{N_{\text{R}}\nearrow \infty} 
 \delta_{\bar{\alpha},\alpha} \delta_{\bar{\beta}_{\text{R}},\beta_{\text{R}}} \; .
 \end{split}
\end{equation}
The boundary dependent factors $\delta_{\bar{\alpha}_{\text{L}},\alpha_{\text{L}}}$ and 
$\delta_{\bar{\beta}_{\text{R}},\beta_{\text{R}}}$ are canceled by those coming from 
the denominator.  Therefore, all we have to compute is the infinite-distance limit 
($n \nearrow \infty$) of the following quantity:
\begin{equation}
\sum_{\alpha,\beta}[T_{\text{string}}^{z} 
(T_{\text{string}})^{n-1}T^{z}]_{\alpha,\alpha;\beta,\beta}  \; .
\label{eqn:Ostring-by-Ts}
\end{equation}

\subsection{String Order Parameters and Entanglement Spectrum}

Now we show that the existence of non-vanishing string order parameters 
serves as the {\em sufficient condition} for the symmetry-protected topological 
order discussed in the previous section.  Let us begin with the simpler case of 
the usual VBS states.  

Since we are interested in the long-distance limit $|i-j|\nearrow \infty$, 
we need to know the asymptotic behavior of the string $(T_{\text{string}})^{|i-j|}$.  
To this end, we can borrow the results of Ref.~\onlinecite{Garcia-W-S-V-C-08} 
(Theorem 2); according to the theorem, the MPS should be invariant under both 
of the $\pi$-rotations
\begin{equation} 
\hat{u}_{x}=\otimes_{j}\be^{-i\pi S^{x}_{j}} \; , \;\; 
\hat{u}_{z}=\otimes_{j} \be^{-i\pi S^{z}_{j}} 
\end{equation} 
in order for the string $(T_{\text{string}})^{|i-j|}$ not to vanish in the long-distance limit. 
Then, Lemma 1 of Ref.~\onlinecite{Garcia-W-S-V-C-08} guarantees 
that there exists a pair of unitary matrices  
$U_{x}$ and $U_{z}$ which are unique and satisfy:
\begin{equation}
\begin{split}
& \sum_{n=1}^{d}R^{(S)}_{a}(\pi)_{mn} A(n) 
= \be^{i\theta_{a}} \, U^{\dagger}_{a} A(m) U_{a}  \\
& \quad 
(a=x,z; \; \be^{i\theta_{a}} = \pm 1) \\
&  (U_{a})^{2} = \mathbf{1} \;\; , \;\; 
U_{x}U_{z} = \pm U_{z}U_{x} \; ,
\end{split}
\label{eqn:RA-UAU}
\end{equation}
where the two sign choices are {\em independent}. 
The above exchange property between $U_{x}$ and $U_{z}$ has a very important 
implication to the structure of the entanglement spectrum\cite{Pollmann-T-B-O-10}:
\begin{equation}
\begin{split}
\det\left\{(U_{x}U_{z})_{\lambda}\right\} &= \det\left\{(U_{x})_{\lambda}\right\}
\det\left\{(U_{z})_{\lambda}\right\} \\
&= (\pm 1)^{d_{\lambda}}\det\left\{(U_{z}U_{x})_{\lambda}\right\} \\
&= (\pm 1)^{d_{\lambda}}\det\left\{(U_{x})_{\lambda}\right\}
\det\left\{(U_{z})_{\lambda}\right\}
 \; (\neq 0) \; .
 \end{split}
 \label{eqn:UxUz-degeneracy}
\end{equation}
Therefore, the degree of degeneracy $d_{\lambda}$ of each entanglement level $\lambda$ 
should be even when $U_{x}$ and $U_{z}$ are anti-commuting. 
Typically, this happens in the VBS states with {\em odd}-integer-$S$.  

\begin{figure}[floatfix]
\begin{center}
\includegraphics[scale=0.4]{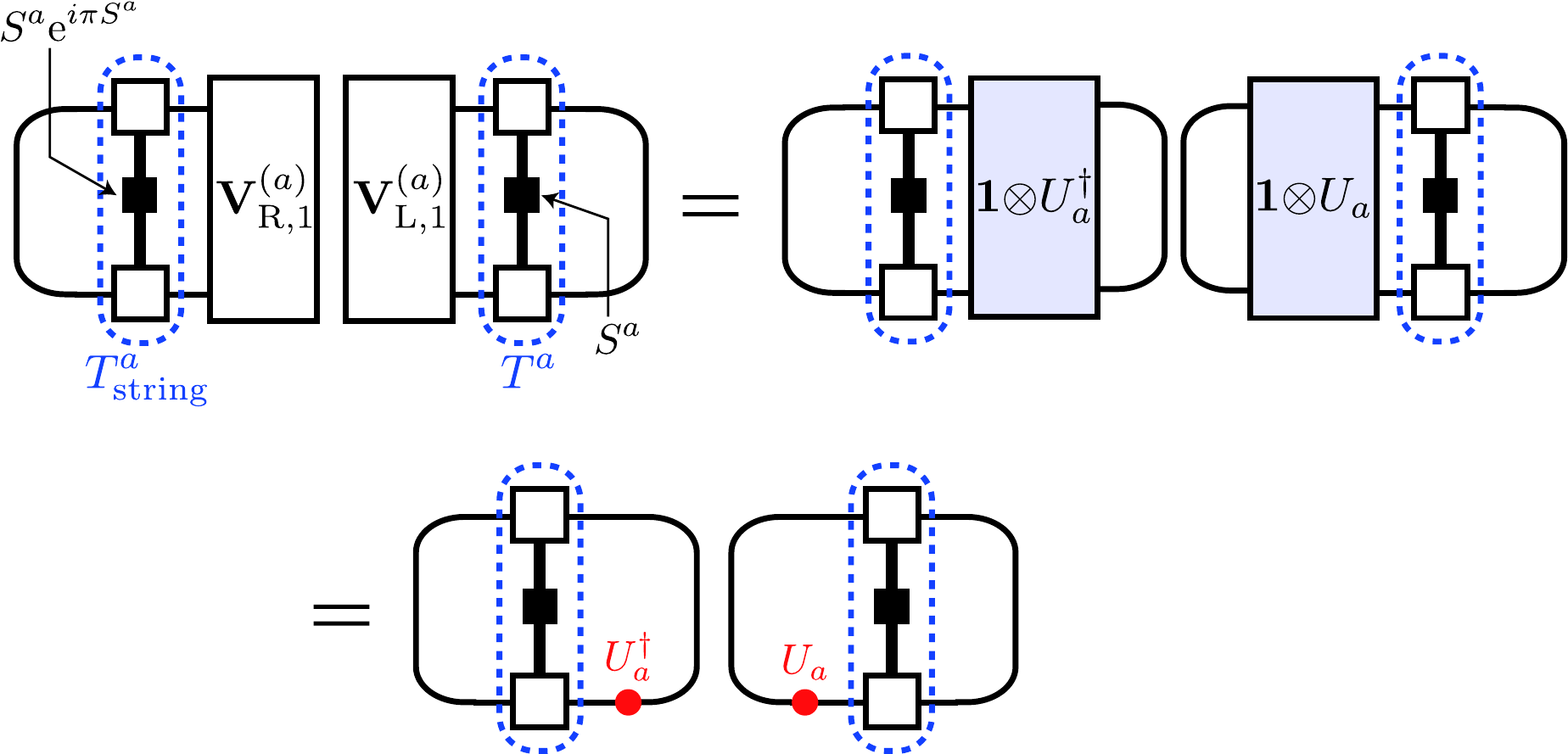}
\end{center}
\caption{(Color online) Diagrammatic representation of the main part of 
string correlation function $\left\{
(T_{\text{string}}^{z} \mathbf{V}^{(u)}_{\text{R},1})
(\mathbf{V}^{(u)}_{\text{L},1}T^{z})
\right\}$. 
\label{fig:string-in-MPS}}
\end{figure}

Now we show that when the string order parameters 
are non-vanishing $\mathcal{O}_{\text{string}}^{z,x}\neq 0$, 
the minus sign realizes (i.e. $U_{x}$ and $U_{z}$ anti-commute) 
in eq.(\ref{eqn:UxUz-degeneracy}) and the entanglement spectrum has 
the degenerate structure. 
To this end, we investigate eq.(\ref{eqn:Ostring-by-Ts}). 
First of all, the invariance of the MPS under $\hat{u}_{x,z}$ implies that 
the string part $(T_{\text{string}})^{n-1}$ reduces essentially to 
a phase $(\be^{i\theta_{a}})^{n-1}=(\pm 1)^{n-1}$.  
This is a direct consequence of Theorem 2 of Ref.~\onlinecite{Garcia-W-S-V-C-08} 
and is easily understood since the overlap 
$\langle \Psi|\hat{u}_{a}|\Psi\rangle = (T_{\text{string}})^{L}$ vanishes otherwise. 
The price to pay is the boundary factors appearing at the two end points 
of the string correlation functions (see Fig.\ref{fig:string-in-MPS}):
\begin{equation}
\begin{split}
& \sum_{\alpha,\beta}\left\{T_{\text{string}}^{z} 
\left(
\sum_{n=1}^{D^{2}}\mathbf{V}^{(u)}_{\text{R},n}\mathbf{V}^{(u)}_{\text{L},n}
\right)
(T_{\text{string}})^{n-1}T^{z}\right\}_{\alpha,\alpha;\beta,\beta} \\
& \quad 
\xrightarrow{|i-j|\nearrow \infty}
\sum_{\alpha,\beta}\left\{
(T_{\text{string}}^{z} \mathbf{V}^{(u)}_{\text{R},1})
(\mathbf{V}^{(u)}_{\text{L},1}T^{z})
\right\}_{\alpha,\alpha;\beta,\beta}  \\
&= \sum_{\alpha,\beta}\left\{
(T_{\text{string}}^{z} \left\{ \mathbf{1}{\otimes} U^{\dagger}_{a}\right\}\mathbf{1})
(\mathbf{1}\left\{ \mathbf{1}{\otimes} U_{a}\right\}T^{z})
\right\}_{\alpha,\alpha;\beta,\beta}  \; ,
\end{split}
\end{equation}
where $\mathbf{V}^{(u)}_{\text{L/R},n}$ denotes the left (L) and the right (R) eigenvectors 
of $T_{\text{string}}$.  

\begin{figure}[floatfix]
\begin{center}
\includegraphics[scale=0.4]{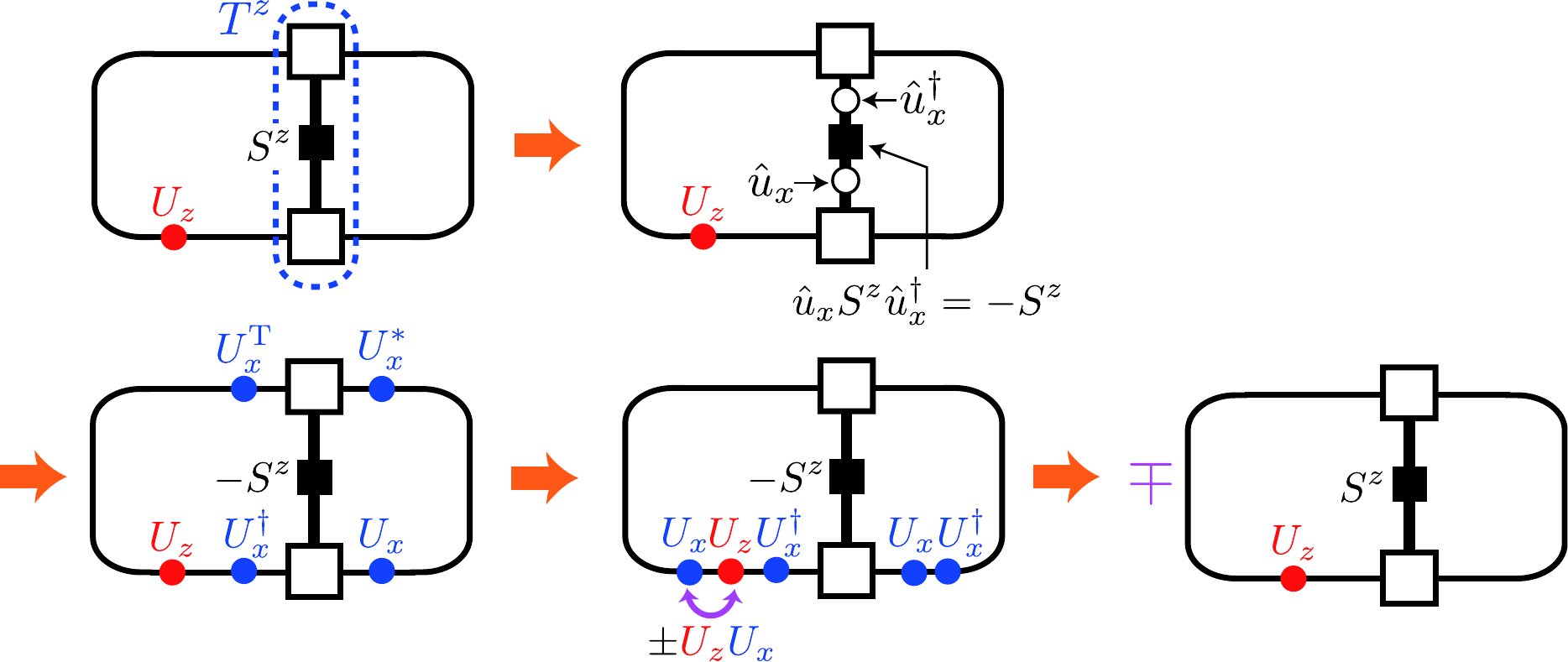}
\end{center}
\caption{(Color online) Rewriting the boundary factor (for $a=z$) using $\hat{u}_{x}$. 
When $U_{x}$ and $U_{z}$ are anti-commuting, the minus sign coming from 
$\hat{u}_{x}S^{z}\hat{u}^{\dagger}_{x}=-S^{z}$ is canceled and an overall 
plus sign is recovered. 
\label{fig:boundary-factor}}
\end{figure}

To see whether the boundary factors are non-vanishing or not, 
we consider the right-boundary factor 
$(\mathbf{1}\left\{ \mathbf{1}{\otimes} U_{z}\right\}T^{z})$ 
of $\mathcal{O}^{z}_{\text{string}}$ (i.e. $a=z$).  
First we rewrite it by using (see the second figure of Fig.\ref{fig:boundary-factor}):
\begin{equation}
S^{z} 
= \hat{u}^{\dagger}_{x}\hat{u}_{x}S^{z}\hat{u}_{x}^{\dagger}\hat{u}_{x}
= \hat{u}^{\dagger}_{x}(-S^{z})\hat{u}_{x}  \quad 
(\hat{u}_{x}=\otimes_{k} \be^{-i\pi S^{x}})  \; .
\end{equation}
The unitary operators $\hat{u}^{\dagger}_{x}$ and $\hat{u}_{x}$ 
appearing on both sides of $-S^{z}$ can be absorbed into the MPS matrices 
by using eq.(\ref{eqn:RA-UAU}) (the third figure of Fig.\ref{fig:boundary-factor}). 
By re-arranging the unitary matrices $U_{x}U_{z}$ (the fourth figure of 
Fig.\ref{fig:boundary-factor}), we arrive at the expression:
\begin{equation}
\begin{split}
\mathbf{1}\left\{ \mathbf{1}{\otimes} U_{z}\right\}T^{z}
& = \mathbf{1}\left\{ \mathbf{1}{\otimes} (U_{x}U_{z}U_{x}^{\dagger})\right\}(-T^{z}) \\
&= \mathbf{1}\left\{ \mathbf{1}{\otimes} (
\pm U_{z}U_{x}U_{x}^{\dagger})\right\}(-T^{z}) \\
&= \mp \mathbf{1}\left\{ \mathbf{1}{\otimes} U_{z}\right\}T^{z} \; .
\end{split}
\label{eqn:boundary-factor}
\end{equation}
Therefore, we see that the boundary factors, and hence the string order parameter 
itself, vanish when 
$U_{x}$ and $U_{z}$ are commuting (as, e.g., in the even-$S$ VBS states). 
On the other hand, if both of the string order parameters are finite, this immediately 
implies that the ground state MPS is not only invariant 
under the two $\pi$-rotations\cite{Garcia-W-S-V-C-08} 
$\hat{u}_{x}$ and $\hat{u}_{z}$, but also has the adjoint $U_{x,z}$ matrices satisfying
\begin{equation}
U_{x}U_{z}= - U_{z}U_{x} \; . 
\label{eqn:UxUz-exchange}
\end{equation}
By the argument in Ref.~\onlinecite{Pollmann-T-B-O-10,*Pollmann-B-T-O-12}, the ground state is topologically 
non-trivial in the sense that each entanglement level is even-fold degenerate. 
Therefore, the finiteness of the pair of string order parameters $\mathcal{O}^{x,z}_{\mathrm{string}}$ 
is the {\em sufficient} condition for the topological phase.  
It is crucial that {\em both} $O^{x}_{\text{string}}$ and $O^{z}_{\text{string}}$ are non-zero 
for the existence of the topological order. For instance, one can construct a solvable spin-1 
model\cite{Klumper-S-Z-92} which 
exhibits a kind of ``hidden order'' similar to the one in the VBS model and has\cite{Totsuka-S-94} 
$O^{x}_{\text{string}}=0$ and $O^{z}_{\text{string}}\neq 0$.  In fact, in this case, 
the two entanglement eigenvalues are no longer degenerate and the state is not topological. 

\subsection{Case of SMPS}

\begin{figure}[floatfix]
\begin{center}
\includegraphics[scale=0.4]{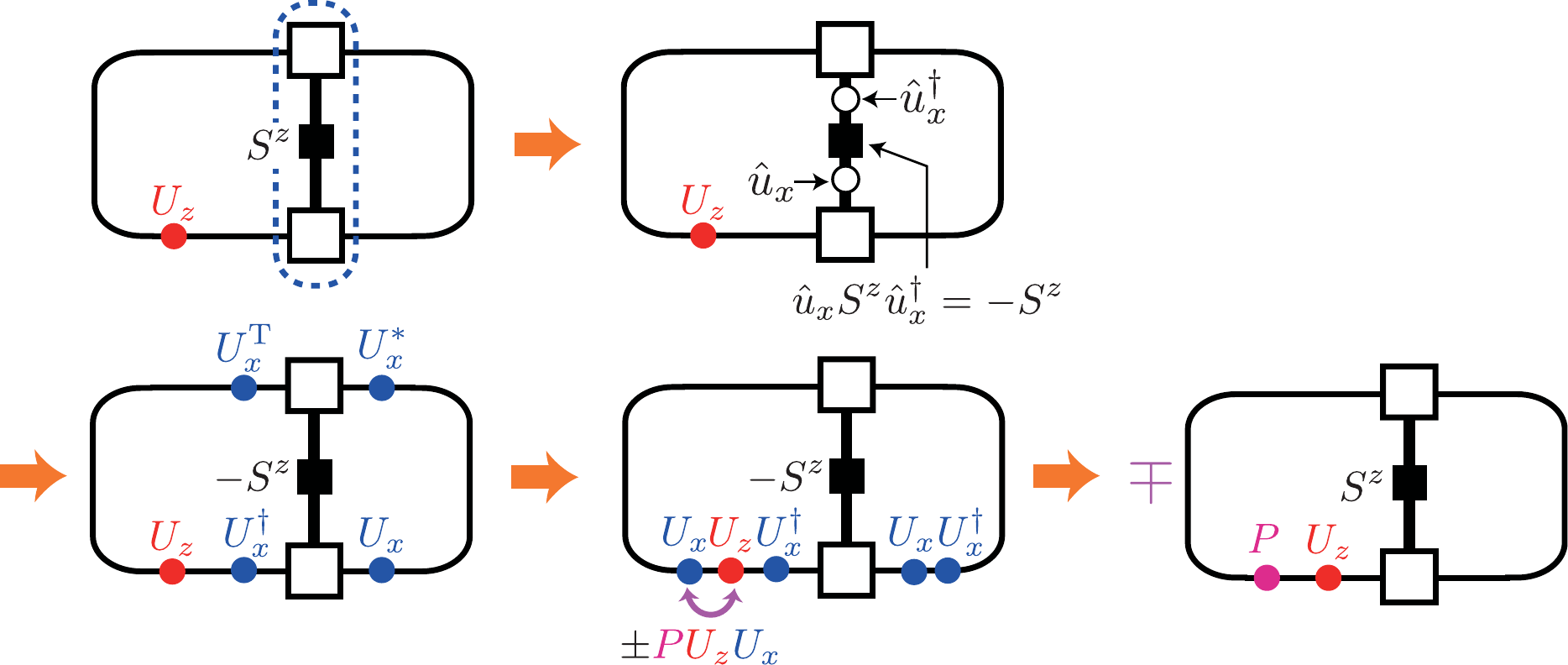}
\end{center}
\caption{(Color online) Rewriting the boundary factor (for $a=z$) using $\hat{u}_{x}$. 
When $U_{x}$ and $U_{z}$ are anti-commuting, the minus sign coming from 
$\hat{u}_{x}S^{z}\hat{u}^{\dagger}_{x}=-S^{z}$ is canceled and an overall 
plus sign is recovered. Note that an extra $P$-matrix appears in the SUSY case. 
\label{fig:boundary-factor-SUSY}}
\end{figure}

Basically, we follow the same line of arguments to show that finite string correlation 
implies the topological phase.  
The only difference is that now we have the $P$ matrix \eqref{definversionP} in the key equation 
(\ref{eqn:UxUz-exchange}):
\begin{equation}
U_xU_z =\pm P U_xU_z \; . 
\end{equation}   
Correspondingly, the last step 
(see Fig.~\ref{fig:boundary-factor}) in evaluating 
the boundary factor is modified. 
Specifically, in stead of eq.(\ref{eqn:boundary-factor}), we have 
(see Fig.~\ref{fig:boundary-factor-SUSY}):
\begin{equation}
\begin{split}
\mathbf{1}\left\{ \mathbf{1}{\otimes} U_{z}\right\}T^{z}
& = \mathbf{1}\left\{ \mathbf{1}{\otimes} (U_{x}U_{z}U_{x}^{\dagger})\right\}(-T^{z}) \\
&= \mp \mathbf{1}\left\{ \mathbf{1}{\otimes} PU_{z}\right\}T^{z} \; .
\end{split}
\label{eqn:boundary-factor-SUSY}
\end{equation}
Therefore, one of the two components (bosonic and fermionic) vanishes 
just by symmetry:
\begin{equation}
\begin{split}
& 
\raisebox{-4.0ex}{\includegraphics[scale=0.5]{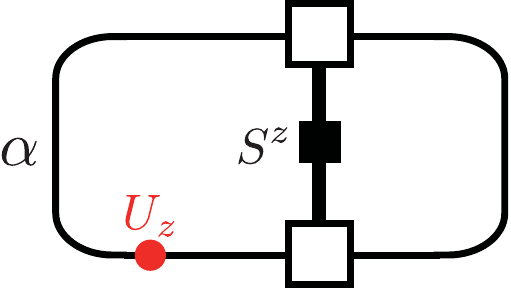}} \\
& = 
\begin{cases}
\sum_{\alpha \in \text{F}} \;\; 
\raisebox{-3.0ex}{\includegraphics[scale=0.4]{boundary-SUSY-B}} & 
\text{when } \be^{i\phi_{xz}}=+1 \\
\sum_{\alpha \in \text{B}} \;\; 
\raisebox{-3.0ex}{\includegraphics[scale=0.4]{boundary-SUSY-B}} &
\text{when } \be^{i\phi_{xz}}=-1   \; .
\end{cases}
\end{split}
\end{equation}
Therefore, if the two string order parameters are both non-vanishing, 
either the bosonic- or the fermionic sector exhibits the degenerate structure 
mentioned in section \ref{sect:supersymmetryprotectedtopologicalorder} 
and the ground state is topologically non-trivial. 

Now it is straightforward to generalize the above argument to the case of 
UOSp(1$|$4) to show that when {\em all} the four string order parameters 
\begin{equation}
\mathcal{O}_{\text{string}}^{ab} \equiv  \lim_{|i{-}j|\nearrow \infty}
\Bigl \langle L_{i}^{ab} \exp\left[ i\pi \sum_{k=i}^{j-1} L^{ab}_{k}\right]L_{j}^{ab}
\Bigr\rangle 
\end{equation}
(where $(a,b)=(1,2)$, $(2,5)$, $(3,4)$ and $(4,5)$, and $L_{ab}$ are 
the SO(5) generators) are non-zero, $2^{2}{\times}$(integer)-fold degeneracy 
occurs in some (bosonic or fermionic) sectors of the entanglement spectrum. 

\section{Summary and Discussions}
\label{sect:summary}

We investigated the effects of doped fermionic holes on the topological phases 
in quantum antiferromagnets.   
To this end, we first introduced a family of SVBS  states 
which may be thought of as the hole-doped version of the usual (bosonic) VBS states 
e.g. spin-$S$ SU(2)- states, the SO(5)- and the Sp($N$) VBS states. 
One of the standard ways of looking at the topological properties in these states is 
to investigate the string order parameters. 
We explicitly evaluated the behaviors of the string order 
parameters of the UOSp(1$|$2)- and the UOSp(1$|$4) SVBS states 
for various values of superspin-$\mathcal{S}$, 
and found that even when the string order parameters vanish identically in the absence of 
doping, they revive immediately after holes are introduced in the system.  
This might suggest that the doped holes changes the property of the ground state 
and thereby stabilizes the topological phase.  

To better understand the nature of the states, we calculated the entanglement spectrum. 
Basically, the spectrum consists of the bosonic and the fermionic sectors; 
at zero doping $r=0$, the fermionic sector is separated from the bosonic sector, 
which constitutes the low-``energy'' part of the spectrum, by an infinitely large entanglement gap. 
Upon doping, the fermionic sector starts participating in the entanglement.  
The point is that the existence of supersymmetry allows the coexistence of the two 
sectors having different entanglement structures.  
In addition to that, the entanglement spectra in the SUSY systems 
exhibit the following salient features:   
(i) In contrast to naive expectation, the SUSY entanglement spectra 
for the bosonic- and the fermionic sectors 
do not coincide with each other at $r=1$,  as a consequence of SUSY 
many-body effect.    
(ii) In the two extreme limits of the doping parameter, $r\rightarrow 0$ and $\infty$, 
the entanglement spectra of the SVBS states indeed reproduce those of the original bosonic VBS state 
and the Majumdar-Ghosh-type states, respectively.  

On the basis of the observations made for the particular states (UOSp(1$|$2) SVBS and 
UOSp(1$|$4) SVBS), we characterized, 
with the help of the SMPS formalism,  
the symmetry-protected topological orders in the SUSY systems in terms of 
the entanglement spectrum.  According to the results, there always exists a topologically-protected 
sector (whose degenerate structure depends on the symmetry of the SMPS in question) 
in the spectrum of the SUSY systems. 
Also, by using the SMPS formalism, we clarified an intimate connection between 
the finiteness of the string order parameters and the degenerate structure 
of the entanglement spectra;  
the finite string order is the {\em sufficient} condition for the degeneracy 
in the entanglement spectrum, which is the fingerprint of the (topological) Haldane state in the bulk.  
These explain the revival of the string order upon doping. 

The above remarkable features can be understood in the light of SUSY edge state picture.   
Intuitively, the degenerate structure can be understood 
by the existence of  fictitious `edge' superspins that appear at the entanglement cut of the chain. 
When the bulk system has superspin $\mathcal{S}$, 
two superspins $\mathcal{S}/2$s, 
which consist of the SU(2) spin $\mathcal{S}/2$ and 
its super-partner $\mathcal{S}/2-1/2$, emerge at the edges:
\begin{equation}
\mathcal{S}/2~~\overset{\text{SUSY}}\longleftrightarrow ~~\mathcal{S}/2-1/2 \; . 
\end{equation}
Then, there always exist half-odd-integer spins at the edges {\em regardless of the parity of 
the bulk superspin}, 
since SUSY, being the symmetry that relates the state with integer spin and 
that with half-odd-integer spin, guarantees the coexistence of both.   
Such half-odd-integer `edge' spins bring the even-fold degeneracy 
to the entanglement spectrum of the UOSp(1$|$2)-symmetric systems.  
Therefore, if we have a topological phase (e.g. Haldane phase) characterized by 
the above type of degenerate structures in the entanglement spectrum, 
it exists for {\em all} values of superspin $\mathcal{S}$.  
A similar argument applies, with due modification, to cases with other types of SUSY.  
In this sense, one may say that SUSY plays a unique role in stabilizing the topological phases 
of matter in 1D.  

Since our study presented here is restricted to a particular class of VBS states with SUSY, 
one obvious future direction would be to extend it to more generic models. 
The argument for symmetry-protected topological orders presented in this paper 
can be generally applied to {\em any} system whose ground-state wavefunction is given 
by the (S)MPS states.  
Thus, it would be interesting to see, for instance, the robustness of the Haldane phase 
in the SUSY Heisenberg model with respect to the parity of the bulk superspin $\mathcal{S}$.  
This might highlight the unique behavior of SUSY topological phases in comparison to the bosonic 
counterparts studied in Ref.~\onlinecite{Pollmann-T-B-O-10}.  

Another future direction is the generalizations to higher dimensions.    
In higher dimensions, the SVBS states generally interpolate between the bosonic VBS states 
and the resonating-valence-bond (RVB) type of states 
\cite{Anderson-73,*Fazekas-A-74,Rokhsar-K-88}, 
where the wave function is given by the summation over all possible dimer coverings of 
singlet (i.e. $(a_{i}^{\dagger}b_{j}^{\dagger}-b_{i}^{\dagger}a_{j}^{\dagger})$) bonds 
(in 1D, we have the Majumdar-Ghosh valence-bond crystals). 
The latter is well-known to have non-trivial topological properties \cite{Rokhsar-K-88} 
and it would be interesting 
to study the change in the entanglement properties and the edge-state structure 
as the doping is varied by using the techniques of projected entangled pair states (PEPS) 
\cite{Schuch-P-C-G-12}. 

Application to other topologically non-trivial states of matter, 
such as quantum Hall states or various topological states in cold atom systems, 
is even more interesting.  
For instance, the SUSY-extended Laughlin wave function, which has a close analogy 
with the SVBS states studied here, interpolate between 
different quantum-Hall ground states, such as the Laughlin states and  the Moore-Read 
Pfaffian states.   
In this respect, 
as the SVBS states in 1D provided a unifying way of deriving the entanglement spectra of 
the (bosonic) VBS state and the MG dimer state, 
the study of the entanglement spectra of the SUSY Laughlin wavefunction will naturally give 
a unifying understanding of the entanglement structure of various quantum Hall ground states.   

Finally, we would like to comment on the recent work on the non-local
order parameters for the symmetry-protected topological order.  
When completing this paper, we became aware of a recent preprint  
by Pollmann and Turner (Ref.~\onlinecite{Pollmann-T-12})  
which also discusses the string order parameter from the entanglement point
of view.  Although some of the conclusions obtained there overlap 
with ours, the main goal there is to go beyond the string order parameter 
and is different from that of this paper.

\section*{Acknowledgement}

We are very grateful to Hosho Katsura and Frank Pollmann for useful discussions and email correspondences.  
K.H. would like to thank the condensed matter group in YITP for warm hospitality during his stay.    
This work was supported in part by 
Grants-in-Aid for Scientific Research 
(B) 23740212 (K.H.), (C) 20540375, 
(C) 24540402 (K.T.) and 
by the global COE (GCOE) program `The next generation of 
physics, spun from universality and emergence' of Kyoto University.

\appendix

\section{$A$-matrices for UOSp(1$|$4) SVBS states}

\subsection{Superspin-1 SVBS}
\label{sec:UOSp14-A-matrices-adj}

The fourteen 5$\times$5 $A$-matrices for the $\mathcal{S}=1$ SVBS state discussed in 
section \ref{sec:SO5-SVBS} are explicitly given as:
\begin{subequations}
\begin{equation}
\begin{split}
\allowdisplaybreaks
& A(1,1)= -A(2,2)^{\text{t}}=-\sqrt{2}
\begin{pmatrix} 
\sigma_- & 0 & 0 \\
0 & 0_2 & 0 \\
0 & 0 & 0 
\end{pmatrix}, \\
& A(3,3)= -A(4,4)^{\text{t}}=-\sqrt{2}
\begin{pmatrix}
0_2 & 0 & 0 \\
0 & \sigma_- & 0 \\
0 & 0 & 0 
\end{pmatrix}, \\
& A(1,2)=
\begin{pmatrix}
\sigma_3 & 0 & 0 \\
0 & 0_2 & 0 \\
0& 0 & 0 
\end{pmatrix}, \\
& A(1,3)= -A(2,4)^{\text{t}}=-
\begin{pmatrix}
0_2 & \sigma_- & 0 \\
\sigma_- & 0_2 & 0 \\
0& 0 & 0 
\end{pmatrix}, \\
& A(1,4)=A(2,3)^{\text{t}}=\frac{1}{2}
\begin{pmatrix}
0_2 & -1_2+\sigma_3 & 0 \\
1_2+\sigma_3 & 0_2 & 0 \\
0 & 0 & 0 
\end{pmatrix},   \\
& A(3,4)=
\begin{pmatrix}
0_2 & 0 & 0 \\
0 & \sigma_3 & 0 \\
0 & 0 & 0 
\end{pmatrix},
\end{split}
\end{equation}
\begin{equation}
\begin{split}
& A(1)=A(2)^{\st}=
\begin{pmatrix}
0 & 0 & 0 & 0 & 0 \\
0 & 0 & 0 & 0 & -\sqrt{r} \\
0 & 0 & 0 & 0 & 0 \\
0 & 0 & 0 & 0 & 0 \\
-\sqrt{r} & 0 & 0 & 0 & 0 
\end{pmatrix}, \\
& A(3)=A(4)^{\st}=\begin{pmatrix}
0 & 0 & 0 & 0 & 0 \\
0 & 0 & 0 & 0 & 0 \\
0 & 0 & 0 & 0 & 0 \\
0 & 0 & 0 & 0 &  -\sqrt{r}\\
0 & 0 & -\sqrt{r} & 0 & 0 
\end{pmatrix},
\end{split}
\label{so5dashgammaalpha}
\end{equation}
\end{subequations}
where the symbols `$\text{t}$' and `$\st$' denote the transposition and supertransposition 
(\ref{eqn:super-tanspos-def}), respectively.   
They can be represented by linear combinations of the UOSp(1$|$4) generators. 
 
\subsection{Properties}
\label{sect:propertiesofSO(5)gammas}

As has been discussed in section \ref{subsec;Inversion}, the link-inversion symmetry is implemented 
in the SMPS as
\begin{equation}
\mathcal{I}: \; A(m) \mapsto A(m)^{\st} \; , 
\end{equation}
or to write the bosonic- and the fermionic component separately
\begin{equation}
\mathcal{I}: \; A(\sigma,\tau) \mapsto A(\sigma,\tau)^{\text{t}} \; , \; 
A(\sigma) \mapsto A(\sigma)^{\st} \; .
\end{equation}
Then, it can be shown 
\begin{equation}
A(m)^{\st}  = \mathcal{W}^{\dagger} A(m)\mathcal{W} \; ,
\end{equation}
where 
\begin{equation}
\mathcal{W}=\begin{pmatrix}
W & 0 \\
0 & 1 
\end{pmatrix},
\label{eqn:def-W-matrix}
\end{equation}
with 
\begin{equation}
W=
\begin{pmatrix}
0 & i\sigma_2 \\
 i\sigma_2 & 0 
\end{pmatrix}. 
\end{equation}

\section{Edge States and General Asymptotic Behavior of Entanglement}
\label{sec:edge-ES-relation}

The asymptotic behaviors eqs.\eqref{asymptoticentanglemententropy1}, 
\eqref{asymptoticentanglemententropy2} and \eqref{asymptoticosp14entanglemententropy} 
can be understood from a more general point of view. 
Let us consider the UOSp(1$|2K$) SVBS state with bulk-superspin $\mathcal{S}$. 
The UOSp(1$|2K$) SVBS has $N=1$ supersymmetry, and consists of one bosonic sector and one  fermionic sector. 
For the bulk-superspin $\mathcal{S}$, the emergent superspin-$\mathcal{S}/2$ objects appear 
at the edges and the UOSp(1$|2K$)  SVBS state accommodates the graded fully symmetric 
representation\cite{hasebe-2011} at each edge:
\begin{subequations}
\begin{align}
&|m_1,m_2,\cdots,m_{2K}\rangle\nonumber\\&=\frac{1}{\sqrt{m_1!m_2!\cdots m_{2K}!}}(b_1^{\dagger})^{m_1}(b_2^{\dagger})^{m_2}\cdots (b_{2K}^{\dagger})^{m_{2K}}|\text{vac}\rangle,\\
&|n_1,n_2,\cdots,n_{2K}\rangle\nonumber\\&=\frac{1}{\sqrt{n_1!n_2!\cdots n_{2K}!}}(b_1^{\dagger})^{n_1}(b_2^{\dagger})^{n_2}\cdots (b_{2K}^{\dagger})^{n_{2K}}f^{\dagger}|\text{vac}\rangle, 
\end{align}
\end{subequations}
with $m_1+m_2+\cdots+m_{2K}=n_1+n_2+\cdots+n_{2K}+1=\mathcal{S}$. 
Then,  the number of the bosonic- and fermionic states on each edge are respectively given by 
\begin{subequations}
\begin{align}
&D_{\text{B}}= \begin{pmatrix}  
\mathcal{S}+2K-1 \\
\mathcal{S} 
\end{pmatrix}  
=\frac{(\mathcal{S}+2K-1)!}{(2K-1)!\mathcal{S}!},\\
&D_{\text{F}}=\begin{pmatrix}  
\mathcal{S}+2K-2 \\
\mathcal{S}-1
\end{pmatrix} 
=\frac{(\mathcal{S}+2K-2)!}{(2K-1)!(\mathcal{S}-1)!}. 
\end{align}
\end{subequations}
(The bosonic degrees of freedom coincide with the fully symmetric representation of USp($2K$)\cite{Schuricht-R-08}.) 
For instance, for the UOSp(1$|$2) $(K=1)$ SVBS state, we have  
\begin{equation}
D_{\text{B}} ={\mathcal{S}+1} \; , \;\;
D_{\text{F}} ={\mathcal{S}} \; ,
\end{equation}
while for the UOSp(1$|$4) $(K=2)$ SVBS state, 
\begin{equation}
\begin{split}
&D_{\text{B}} = 
\frac{1}{6}(\mathcal{S}+1)(\mathcal{S}+2)(\mathcal{S}+3),\\
&D_{\text{F}} = 
\frac{1}{6}\mathcal{S}(\mathcal{S}+1)(\mathcal{S}+2).  
\end{split}
\label{degbosonfermion}
\end{equation}
In the infinite chain limit, the spin degrees of freedom are equivalent 
\begin{equation}
\begin{split}
&{\lambda_1}^2={\lambda_2}^2=\cdots = {\lambda_{D_{\text{B}}}}^2 \equiv {\lambda_{\text{B}}}^2,\\
&{\lambda_{D_{\text{B}}+1}}^2={\lambda_{D_{\text{B}}+2}}^2=\cdots 
= {\lambda_{D_{\text{B}}+D_{\text{F}}}}^2 \equiv{\lambda_{\text{F}}}^2,  
\end{split}
\end{equation}
and the normalization condition of the Schmidt coefficients, 
$\sum_{\alpha=1}^{D_{\text{B}}+D_{\text{F}}}|\lambda_{\alpha}|^2=1$, is rewritten as  
\begin{equation}
D_{\text{B}} {\cdot} {\lambda_{\text{B}}}^2 +D_{\text{F}} {\cdot} {\lambda_{\text{F}}}^2=1. 
\label{eqn:Schmidt-normalization}
\end{equation}
Then, the entanglement entropy is expressed as    
\begin{equation}
\begin{split}
S_{\text{EE}}(r) 
&=-\sum_{\alpha=1}^{D_{\text{B}}}|\lambda_{\alpha}|^2\log |\lambda_{\alpha}|^2
- \sum_{\alpha=1}^{D_{\text{F}}}|\lambda_{D_{\text{B}}
+\alpha}|^2\log |\lambda_{D_{\text{B}}+\alpha}|^2   \\
&=-D_{\text{B}} |\lambda_{\text{B}}|^2 \log  |\lambda_{\text{B}}|^2 
-D_{\text{F}} |\lambda_{\text{F}}|^2 \log  |\lambda_{\text{F}}|^2 .
\end{split}
\end{equation}
At $r=0$, only the Schmidt coefficients of boson sector survive and eq.\eqref{eqn:Schmidt-normalization} 
implies
\begin{equation}
{\lambda_{\text{B}}}^2=\frac{1}{D_{\text{B}}},~~~~{\lambda_{\text{F}}}^2=0,   
\end{equation}
and hence
\begin{equation}
\lim_{r \rightarrow 0} S_{\text{EE}}(r) =\log D_{\text{B}} \; . 
\label{eqn:SEE-r-0-limit}
\end{equation}
Thus, the entanglement entropy of the spin $S$ original VBS states is reproduced. 

\begin{figure}[floatfix]
\begin{center}
\includegraphics[scale=0.38]{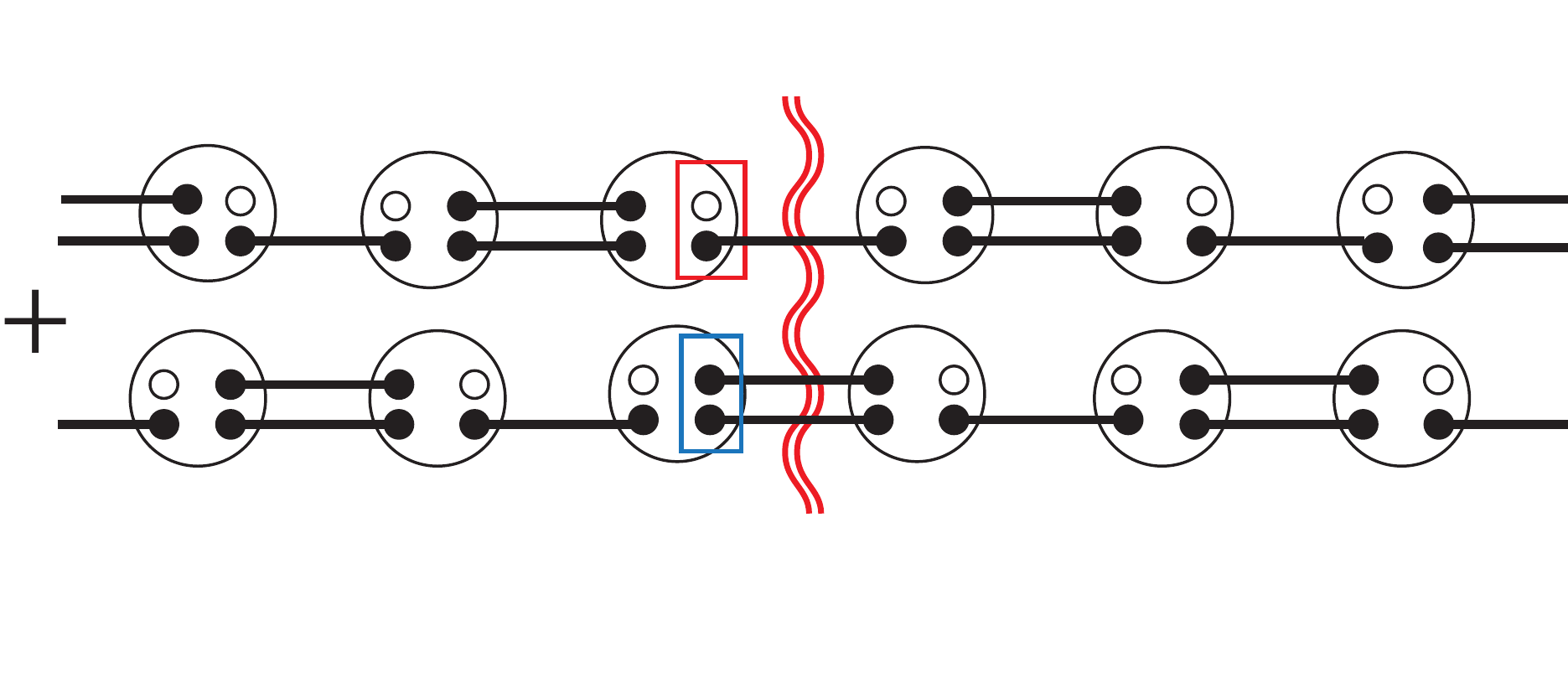}
\caption{(Color online) 
The $r\rightarrow \infty$ limit of the bulk superspin $\mathcal{S}=2$ UOSp(1$|$4) SVBS state 
is given by the superposition of the two partially dimerized states related to each other by 
one-site translation. 
When we make an entanglement cut at an arbitrary bond 
(shown by wavy lines), we always have two different kinds of sections: 
one with four `fermionic' edge states (upper) and the one with ten `bosonic' edge states (lower).  
These two different sections respectively yield four-fold- and ten-fold degenerate entanglement 
levels.  
\label{fig:EE-cut-MG}}
\end{center}
\end{figure}

On the other hand, in the limit $r\rightarrow\infty $, the SVBS states reduce 
to the (partially) dimerized  states [see Fig.\ref{fig:EE-cut-MG}].   
In the upper state in Fig.~\ref{fig:EE-cut-MG}, the fermionic edge states appear, while 
in the lower the edge states are bosonic.   Since both cases appear with equal weights, 
the sum of the Schmidt coefficients for the bosonic sector and that for the fermionic sector should be equal:   
\begin{equation}
 \sum_{\alpha=1}^{D_{\text{B}}}{\lambda_{\alpha}}^2
= \sum_{\alpha=1}^{D_{\text{F}}} {\lambda_{D_{\text{B}}+\alpha}}^2=1/2.
\end{equation}
Therefore, we have 
\begin{subequations}
\begin{align}
{\lambda_{\text{B}}}^2=\frac{1}{2D_{\text{B}}},~~~~{\lambda_{\text{F}}}^2=\frac{1}{2D_{\text{F}}} 
\end{align}
\end{subequations}
for $r\rightarrow \infty$, and the corresponding entanglement entropy is derived as 
\begin{equation}
\lim_{r\rightarrow \infty} S_{\text{E.E.}}(r) =\log \left( 2\sqrt{D_{\text{B}} D_{\text{F}}} \right) \; , 
\label{eqn:SEE-r-infty-limit}
\end{equation}
with $D_{\text{B}}$ and $D_{\text{F}}$ given by eq.(\ref{degbosonfermion}). 
Thus, from the entanglement point of view, the role of SUSY is two-fold. 
First, it necessitates two different Schmidt eigenvalues corresponding to the $N=1$ SUSY.   
Second, it enables the system to support finite entanglement even in the limit $r\rightarrow \infty$. 
 
 For the superspin-$\mathcal{S}$ UOSp(1$|$2) SVBS states,\cite{Hasebe-T-11}  
 the entanglement entropy behaves as 
\begin{subequations}
\begin{align}
\lim_{r=0} S_{\text{EE}}(r) &=\log (\mathcal{S}+1), \\
\lim_{r\rightarrow \infty} S_{\text{EE}} (r) 
&= \log 2+\frac{1}{2}\log (\mathcal{S}(\mathcal{S}+1)) \; ,
\end{align}
\end{subequations} 
which, for $\mathcal{S}=1$ and $\mathcal{S}=2$, reproduces 
the results (\ref{asymptoticentanglemententropy1}) 
and (\ref{asymptoticentanglemententropy2}).  
For the superspin-$\mathcal{S}$ UOSp(1$|$4) SVBS states, on the other hand, 
\begin{subequations}
\begin{align}
& \lim_{r=0} S_{\text{EE}}(r) =\log (\mathcal{S}+1)(\mathcal{S}+2)(\mathcal{S}+3)-\log 6, \\
\begin{split}
& \lim_{r\rightarrow \infty} S_{\text{EE}} (r)  \\
& \quad = -\log 3+  \log (\mathcal{S}+1)(\mathcal{S}+2) +\frac{1}{2}\log \mathcal{S}(\mathcal{S}+3).  
\end{split}
\end{align}
\end{subequations}
Setting $\mathcal{S}=1$, we reproduce the previous result (\ref{asymptoticosp14entanglemententropy}).

\section{Proofs}
\label{sect:proofsymmetryprotencted}

In this appendix, we outline the proof of the important relations 
(\ref{eqn:UI-transpose}), (\ref{eqn:UT-transpose}) and (\ref{eqn:UxUz-exchange-SUSY}).  
For later convenience, we derive a useful property of pure canonical MPSs. 

Suppose that we have a pure MPS whose canonical form is characterized by the MPS 
data\cite{Garcia-V-W-C-07,Orus-V-08} 
$(\Lambda,\Gamma)$ and that it satisfies the following relation for some unitary 
matrix $U$:
\begin{equation}
\Gamma(m) = \be^{i\theta_{U}} \, U^{\dagger} \Gamma(m) U \; .  
\label{eqn:VeqVGammaV}
\end{equation}
Since the MPS is canonical, the following holds:
\begin{equation}
\sum_{m}\Gamma^{\dagger}(m)\Lambda^{2}\Gamma(m) = \mathbf{1}_{D} \; .
\label{eqn:canonical-condition-2}
\end{equation}
Physically, it states that the $D^{2}$-dimensional vector $\mathbf{V}_{\mathrm{L}}^{(0)}$ 
\begin{equation}
(\mathbf{V}_{\mathrm{L}}^{(0)})_{a;b} \equiv \delta_{ab} \quad 
(1 \leq a,b \leq D) 
\end{equation}
is the dominant left-eigenvector of the left transfer matrix
\begin{equation}
(T_{\mathrm{L}})_{\bar{a},a;\bar{b},b} 
\equiv \sum_{m}(\Lambda\Gamma^{\ast}(m))_{\bar{a}\bar{b}}
(\Lambda\Gamma(m))_{ab} \; .
\end{equation}
Plugging   
$\Gamma^{\dagger}(m) = \be^{-i\theta_{U}} \, U^{\dagger} \Gamma^{\dagger}(m) U$ into 
(\ref{eqn:canonical-condition-2}), we obtain: 
\begin{equation}
\be^{-i\theta_{U}} \,\sum_{m} U^{\dagger} \Gamma^{\dagger}(m) U
\Lambda^{2}\Gamma(m)
=  \mathbf{1}_{D} \; , 
\end{equation}
or equivalently
\begin{equation}
\sum_{m} \Gamma^{\dagger}(m) \Lambda U
\Lambda\Gamma(m)
=\be^{i\theta_{U}} \,  U \; . 
\end{equation}
This implies that the unitary matrix 
\begin{subequations}
\begin{equation}
U_{\bar{b}b}=\sum_{a}
\left\{ \mathbf{1}{\otimes} U \right\}_{aa;\bar{b}b}
\equiv \sum_{a}\delta_{a\bar{b}}U_{ab} \; ,
\end{equation}
when viewed as a $D^{2}$-dimensional 
vector, is the left-eigenvector of $T_{\mathrm{L}}$ with the eigenvalue $\be^{i\theta_{U}}$: 
\begin{equation}
UT_{\mathrm{L}} 
= \be^{i\theta_{U}}U  \; .
\end{equation}
\end{subequations}
Since, by assumption of canonical MPS, $\mathbf{1}_{D}$ is the unique left-eigenvector 
with the eigenvalue $|\lambda|=1$, we conclude 
\begin{equation}
\be^{i\theta_{U}} =1 \; , \;\; U = \be^{i \phi} \mathbf{1}_{D} \; . 
\label{eqn:V-as-dominanteigenv}
\end{equation}
Since in deriving the above, we have only assumed that the (infinite-system) MPS in question  
is pure and takes the canonical form, (\ref{eqn:V-as-dominanteigenv}) holds 
for any MPS (including SMPS) satisfying the assumption. 

\subsection{Inversion-symmetry}

We use the property $\mathcal{I}^{2}=1$ to derive the important property 
(\ref{eqn:UI-transpose}) of the adjoint $U_{I}$ matrix.  Applying supertransposition st on 
(\ref{sinversionsymm2}) and using $(A^{\st})^{\st}=PAP$, we obtain 
\begin{equation}
\Gamma(m) = \be^{2i\theta_{I}}\, (U_{I}P U_{I}^{\ast})^{\dagger}
\Gamma(m)\, (U_{I} P U_{I}^{\ast})  \; .
\end{equation}
Postulate $U$ is the block diagonal matrix  
\begin{equation}
U=\begin{pmatrix}
U_{\text{B}} & 0 \\
0 & U_{\text{F}}
\end{pmatrix}.  
\end{equation}

By eqs.(\ref{eqn:VeqVGammaV}) and (\ref{eqn:V-as-dominanteigenv}), 
this implies that the $D{\times}D$ matrix 
$(U_{I} P U_{I}^{\ast})$ should be equal (up to an overall phase) to the unit matrix: 
\begin{equation}
(U_{I} P U_{I}^{\ast}) = \be^{i\Phi_{I}}\mathbf{1}_{D}  \; .
\end{equation}
After multiplying $U^{\text{t}}_{I}$ from the right and making transposition, we deduce 
\begin{equation}
U_{I} = \be^{-2i\Phi_{I}}P^{2} U_{I}
= \be^{-2i\Phi_{I}}U_{I}  \;  \Leftrightarrow \; \be^{-i\Phi_{I}} = \pm 1
\end{equation}
Therefore, we obtain eq.(\ref{eqn:UI-transpose}):
\begin{equation}
U^{\text{t}}_{I} = \pm P U_{I}  \; .
\end{equation}

It is interesting to calculate $U_{I}$ for superspin-$S$ UOSp(1$|$2) SVBS states. 
For the $\mathcal{S}=1$ SVBS state, $U$ is identified as 
\begin{equation}
U_{I} =\mathcal{R}_{1|2}=
\begin{pmatrix}
0 & 1 & 0 \\
-1 & 0 & 0 \\
0 & 0 & -1 
\end{pmatrix},
\end{equation}
which satisfies 
\begin{align}
& U_{I}^{\dagger} \Gamma(m)U_{I} 
=-\Gamma(m)^{\st},\nonumber\\
& U_{I}^{\text{t}}=-PU_{I} \; .
\end{align}

For the $\mathcal{S}=2$ SVBS state, $U$ is identified as 
\begin{equation}
U_{I}=\begin{pmatrix}
0 & 0 & 1  & 0 & 0 \\
0 & -1 & 0 & 0 & 0 \\
1 & 0 & 0 & 0 & 0 \\
0 & 0 & 0 & 0 & -1 \\
0 & 0 & 0 & 1 & 0 
\end{pmatrix},
\end{equation}
and $\Gamma(m)$ satisfy the relation 
\begin{subequations}
\begin{align}
&U_{I}^{\dagger}\Gamma(m)U_{I}=+\Gamma(m)^{\st},\\
&U_{I}^{\text{t}}=PU_{I}. 
\end{align}
\end{subequations}

For the $\mathcal{S}=1$ UOSp(1$|$4) SVBS state, we use the relations given in 
appendix \ref{sec:UOSp14-A-matrices-adj} to show that 
\begin{equation}
\begin{split}
& \Gamma(m)^{\st} = U^{\dagger}_{I}\Gamma(m) U_{I} \\
& U_{I}^{\text{t}} = - PU_{I} 
\end{split}
\end{equation}
with $U_{I} = \mathcal{W}$ defined in eq.\eqref{eqn:def-W-matrix}.  
This is consistent with the existence of the four-fold degenerate entanglement 
level in this state (see Fig.~\ref{fig:ES-Osp14-adj}).

\subsection{Time-reversal symmetry }

If the MPS is invariant under time-reversal, the $\Gamma$-matrices satisfy%
\cite{Garcia-W-S-V-C-08}
\begin{equation}
\sum_n R_{mn}^y(\pi)\Gamma^*(n)=e^{i\theta_{T}}U_{T}^{\dagger}\Gamma(m)U_{T} \; , 
\end{equation}
where the rotation matrix $R_{mn}^y(\pi)$ takes the block-diagonal form
\begin{equation}
R^y(\pi) = 
\begin{pmatrix}
R^{y}_{\mathcal{S}}(\pi) &  \mathbf{0} \\
\mathbf{0} & R^{y}_{\mathcal{S}{-}1/2} (\pi)
\end{pmatrix}  
\label{eqn:OSp12-rotation-y}
\end{equation}
with $R^{y}_{\mathcal{S}}(\pi)$ and $R^{y}_{\mathcal{S}{-}1/2} (\pi)$ being 
the ordinary rotation matrices for spin-$\mathcal{S}$ and $(\mathcal{S}-1/2)$, respectively. 
Since $\mathcal{T}^{2}=\mathcal{P}$ [see Eq. \eqref{squaretrel}],  
\begin{equation}
\begin{split}
(-1)^{F(l)} \,  \Gamma(l) &=  
\sum_{m=1}^{d}R^{y}_{lm}\left\{
\sum_{n=1}^{d}R^{y}_{mn}\Gamma^{\ast}(n) \right\}^{\ast} \\
&= \sum_{m=1}^{d}R^{y}_{lm}\left\{
\be^{-i\theta_{T}} U^{\text{t}}_{T} 
\Gamma^{\ast}(m) U^{\ast}_{T}
\right\}   \\
&= \left\{ U_{T}U^{\ast}_{T} \right\}^{\dagger} 
\Gamma(l) 
\left\{U_{T}U^{\ast}_{T}\right\}  \; ,
\end{split}
\end{equation}
or equivalently
\begin{equation}
\Gamma(l) = 
\left\{ U_{T}U^{\ast}_{T} \right\}^{\dagger} 
(-1)^{F(l)} \, \Gamma(l) 
\left\{U_{T}U^{\ast}_{T}\right\}\; .
\label{eqn:TR-gamma-adjoint-SUSY}
\end{equation}
By using the property
\begin{equation}
(-1)^{F(l)}\Gamma(l) = P \Gamma(l) P \; ,
\end{equation}
eq.(\ref{eqn:TR-gamma-adjoint-SUSY}) may be rewritten as:
\begin{equation}
\Gamma(l)=
\left\{ U_{T}PU^{\ast}_{T} \right\}^{\dagger} 
 \Gamma(l) 
\left\{U_{T}PU^{\ast}_{T}\right\} \; .
\end{equation}
Now we can apply eqs.(\ref{eqn:VeqVGammaV}) and (\ref{eqn:V-as-dominanteigenv}) 
to conclude
\begin{equation}
U^{\text{t}}_{T} = \pm P U_{T} \; .
\end{equation}

For $\mathcal{S}=1$ UOSp(1$|$2) SVBS state, 
with $\Gamma(1)=\mathcal{A}(1)$, $\Gamma(2)=\mathcal{A}(0)$, 
$\Gamma(3)=\mathcal{A}(-1)$, $\Gamma(4)=\mathcal{A}(1/2)$, 
$\Gamma(5)=\mathcal{A}(-1/2)$ (\ref{canonicalosp12matr}), and 
\begin{align}
&U_{T}=\begin{pmatrix}
0 & 1 & 0 \\
-1 & 0 & 0 \\
0 & 0 & 1
\end{pmatrix},\nonumber\\
&R^y(\pi)=
\begin{pmatrix}
0 & 0 & 1 & 0 & 0 \\
0 & -1 & 0 & 0 & 0 \\
1 & 0 & 0 & 0 & 0 \\
0 & 0 & 0 & 0 &  1 \\
0 & 0 & 0 & -1 & 0
\end{pmatrix}, 
\end{align}
we have 
\begin{equation}
\sum_n R_{mn}^y(\pi)\Gamma^*(n)
=U_{T}^{\dagger}\Gamma(m)U_{T} \; , 
\end{equation}
and 
\begin{equation}
U_{T}^{\text{t}}= - PU_{T} \; .
\end{equation}

For $\mathcal{S}=2$ UOSp(1$|$2) SVBS state, with 
\begin{align}
&U_{T} =\begin{pmatrix}
0 & 0 &  1 & 0 & 0 \\
0 & -1 & 0 & 0 & 0 \\
1 & 0 & 0 & 0 & 0 \\
0 & 0 & 0 & 0 & -1\\
0 & 0 & 0 & 1 & 0 
\end{pmatrix},\nonumber\\
&R^y(\pi)=
\begin{pmatrix}
0 & 0 & 0 & 0 & 1 & 0 & 0 & 0 & 0 \\
0 & 0 & 0 & -1 & 0 & 0 & 0 & 0 & 0\\
0 & 0 & 1 & 0 & 0 & 0 & 0 & 0 & 0\\
0 & -1 & 0 & 0 & 0 & 0 & 0 & 0 & 0\\
1 & 0 & 0 & 0 & 0 & 0 & 0 & 0 & 0 \\
0 & 0 & 0 & 0 & 0 & 0 & 0 & 0 & -1 \\
0 & 0 & 0 & 0 & 0 & 0 & 0 & 1 & 0 \\
0 & 0 & 0 & 0 & 0 & 0 & -1 & 0 & 0 \\
0 & 0 & 0 & 0 & 0 & 1 & 0 & 0 & 0
\end{pmatrix}, 
\end{align}
we have 
\begin{equation}
\sum_n R_{mn}^y(\pi)\Gamma^*(n)
=+U_{T}^{\dagger}\Gamma(m)U_{T} \; , 
\end{equation}
and 
\begin{equation}
U_{T}^{\text{t}}= + PU_{T} \; .
\end{equation}

\subsection{$\mathbb{Z}_2{\times}\mathbb{Z}_2$-symmetry}
\label{sec:proof-Z2-Z2-sym}

Finally consider the $\pi$ rotation around the $x$- and the $z$-axis, 
\begin{equation}
\Gamma(m)\rightarrow \sum_n R_{mn}^{a}(\pi)\Gamma(n) \;\; (a=x,z) \; . 
\end{equation}
Instead of $(R^{a})^{2}=\mathbf{1}$ in the bosonic case, $R^{a}$ in the SUSY case 
satisfies $(R^{a})^{2}=\mathcal{P}$ $((\mathcal{P})_{mn} = \delta_{mn}(-1)^{F(n)})$.  
Therefore, the use of the terminology `$\mathbb{Z}_2{\times}\mathbb{Z}_2$-symmetry' is not 
precise. However, to underline the connection to its bosonic counterpart we use 
the terminology in the SUSY cases as well. 

When the MPS has such a symmetry, we have\cite{Garcia-W-S-V-C-08} 
\begin{equation}
\sum_n R_{mn}^{a}(\pi)\Gamma(n)=e^{i\theta_{a}}U_{a}^{\dagger}\Gamma(m)U_{a} \;\; 
(a=x,z) 
\label{eqn:UGammaU-rotation}
\end{equation}
for some block diagonal unitary matrix:
\begin{equation}
U_{a} =\begin{pmatrix}
U_{a,\text{B}} & 0 \\
0 & U_{a,\text{F}}
\end{pmatrix}
\; .  
\end{equation}

Now let us consider what (\ref{eqn:UGammaU-rotation}) implies. 
We begin by $(R^{a})^{2}=\mathcal{P}$ (valid for integer superspin-$S$): 
\begin{equation}
\begin{split}
(\mathcal{P})_{nn}\Gamma(n) & = P \Gamma(n) P \\
& =e^{i\theta_{a}}\sum_m R_{mn}^{a}(\pi)
U_{a}^{\dagger}\Gamma(m)U_{a} \\
&=e^{2i\theta_{a}}(U_{a}^{\dagger})^2 \Gamma(n)U_{a}^2 \; , 
\end{split}
\label{z2superkey}
\end{equation}
which, after $P$s are rearranged, reads
\begin{equation}
\Gamma(n) = e^{2i\theta_{a}}(U_{a}PU_{a})^{\dagger} 
\Gamma(n)(U_{a}PU_{a})  
\end{equation}
implying 
\begin{equation}
(U_{a}PU_{a}) = \be^{i\phi_{a}} \mathbf{1}_{D} \; .
\end{equation}
The phase $\be^{i\phi_{a}}$ can be absorbed in the definition of $U_{a}$ 
and we have:
\begin{equation}
(U_{a}PU_{a}) =  \mathbf{1}_{D} \; \Leftrightarrow \; 
U^{\dagger}_{a} = PU_{a} \; .
\end{equation}

Next, we consider the product of the  two rotations $R^{x}$ and $R^{z}$. 
In the case of SUSY, they obey the following exchange relation:
\begin{equation}
R^{x}R^{z} = \mathcal{P} R^{z}R^{x} \;\; .
\end{equation}
When combined with eq.(\ref{eqn:UGammaU-rotation}), 
this translates into the following relation for $\Gamma$:
\begin{equation}
(U_{x}U_{z})^{\dagger}\Gamma(m)(U_{x}U_{z})
= (U_{z}PU_{x})^{\dagger}\Gamma(m)(U_{z}PU_{x}) \; .
\end{equation}
After rearranging the $U$s, we arrive at the form to which 
eqs.(\ref{eqn:VeqVGammaV}) and (\ref{eqn:V-as-dominanteigenv}) are applicable:
\begin{equation}
\Gamma(m) = (U_{z}PU_{x}U_{z}^{\dagger}U_{x}^{\dagger})^{\dagger}
\Gamma(m) 
(U_{z}PU_{x}U_{z}^{\dagger}U_{x}^{\dagger}) \; .
\end{equation}
Therefore we have 
\begin{equation}
U_{z}PU_{x}U_{z}^{\dagger}U_{x}^{\dagger}
= \be^{i\phi_{xz}} \mathbf{1}_{D} 
\end{equation}
with $\be^{i\phi_{xz}}=\pm 1$.  
The resulting equation
\begin{equation}
U_{x}U_{z} = \pm P U_{z}U_{x} 
\label{eqn:UxUz-to-UzUx-SUSY}
\end{equation}
or
\begin{equation}
U_{x,1}U_{z,1} = \pm U_{z,1}U_{x,1} \; , \;\;
U_{x,2}U_{z,2} = \mp U_{z,2}U_{x,2} 
\end{equation}
implies the degenerate structure of the entanglement spectrum. 

Let us calculate $U$-matrices for superspin-$\mathcal{S}$ UOSp(1$|$2) SVBS states.   
For odd-$\mathcal{S}$, they assume the following form:
\begin{subequations}
\begin{equation}
U_{a}^{(\mathcal{S})}  = -i 
\begin{pmatrix}
R^{a}_{\mathcal{S}/2}(\pi) &  \mathbf{0} \\
\mathbf{0} & R^{a}_{(\mathcal{S}{-}1)/2} (\pi)
\end{pmatrix}
\end{equation}
which satisfy
\begin{equation}
U_{x}U_{z} = - PU_{z}U_{x} \; .
\end{equation} 
\end{subequations}
Therefore, the degenerate spectrum appears in the bosonic sector. 

For even-$\mathcal{S}$, on the other hand, they are given by:
\begin{subequations}
\begin{equation}
U_{a}^{(\mathcal{S})}  = 
\begin{pmatrix}
R^{a}_{\mathcal{S}/2}(\pi) &  \mathbf{0} \\
\mathbf{0} & R^{a}_{(\mathcal{S}{-}1)/2} (\pi)
\end{pmatrix}  
\end{equation}
satisfying
\begin{equation}
U_{x}U_{z} = + PU_{z}U_{x} \; 
\label{eqn:UxUz-to-UzUx-SUSYeven} ,
\end{equation} 
\end{subequations}
which implies that the fermionic spectrum exhibits the degenerate structure. 

\subsection{ $(\mathbb{Z}_2{\times}\mathbb{Z}_2)^2$ symmetry}
\label{append:z2zesquare}
In this appendix, we summarize some useful relations concerning 
the $A$-matrices of the UOSp(1$|$4) $\mathcal{S}=1$ SVBS states given in 
appendix \ref{sec:UOSp14-A-matrices-adj}.  

The invariance of the MPS under $R^{ab}(\pi)$ defined in eq.\eqref{eqn:Rab-UOSp14} 
implies\cite{Garcia-W-S-V-C-08} the existence of the $5{\times}5$ unitary matrices $U_{ab}$ 
satisfying   
\begin{equation}
\sum_{n=1}^{14}[R^{ab}(\pi)]_{mn}A({n})
= + U_{ab}^{\dagger} A({m}) U_{ab} \; .
\end{equation}
Specifically, $U_{ab}$ are given by
\begin{subequations}
\begin{equation}
U_{12}  = 
\begin{pmatrix}
-1 & 0 & 0 & 0 & 0 \\
 0 & 1 & 0 & 0 & 0 \\
 0 & 0 & -1 & 0 & 0 \\
 0 & 0 & 0 & 1 & 0 \\
 0 & 0 & 0 & 0 & i
\end{pmatrix} \; , \;
 U_{25}  = 
\begin{pmatrix}
0 & 0 & 0 & -i & 0 \\
 0 & 0 & i & 0 & 0 \\
 0 & -i & 0 & 0 & 0 \\
 i & 0 & 0 & 0 & 0 \\
 0 & 0 & 0 & 0 & i
\end{pmatrix} 
\end{equation}
\begin{equation}
U_{34} = 
\begin{pmatrix}
-1 & 0 & 0 & 0 & 0 \\
 0 & 1 & 0 & 0 & 0 \\
 0 & 0 & 1 & 0 & 0 \\
 0 & 0 & 0 & -1 & 0 \\
 0 & 0 & 0 & 0 & i
\end{pmatrix} \; , \;
U_{45}  = 
\begin{pmatrix}
0 & 0 & -i & 0 & 0 \\
 0 & 0 & 0 & -i & 0 \\
 i & 0 & 0 & 0 & 0 \\
 0 & i & 0 & 0 & 0 \\
 0 & 0 & 0 & 0 & i
\end{pmatrix}
\end{equation}
\label{eqn:OSp14-U-matrices}
\end{subequations}
It is easy to check that these matrices satisfy 
\begin{equation}
\begin{split}
& (U_{12})^{2} = (U_{25})^{2} = (U_{34})^{2} = (U_{45})^{2} = \mathcal{P}_{1|4} \\
& U_{12}U_{25} = - \mathcal{P}_{1|4}U_{25}U_{12} \; , \;\; 
U_{34}U_{45} = - \mathcal{P}_{1|4}U_{45}U_{34} \; , \\
& U_{25}U_{45} = - \mathcal{P}_{1|4}U_{45}U_{25} \\
& U_{12}U_{34}=U_{34}U_{12} \; , \;\; U_{12}U_{45}=U_{45}U_{12} 
\; , \;\; U_{25}U_{34}=U_{34}U_{25} \; ,
\end{split}
\end{equation}
where 
\begin{equation}
\mathcal{P}_{1|4}=
\begin{pmatrix}
1_{4} & 0 \\ 0 & -1
 \end{pmatrix}
 \; .
 \end{equation}
By the general argument in section \ref{subsec:z2z2UOSp(1|4)}, one concludes that in some sectors {\em all} 
the entanglement levels are four$\times$(integer)-fold degenerate as is seen 
in Fig.~\ref{fig:ES-Osp14-adj}. 
\bibliographystyle{apsrev4-1}
\bibliography{./references/quantInfo,%
./references/topological_order,%
./references/misc,./references/field_theory,%
./references/2Dsystems,./references/haldane,./references/QHE-TI,%
./references/ref,./references/References}

\begin{thebibliography}{73}%
\makeatletter
\providecommand \@ifxundefined [1]{%
 \@ifx{#1\undefined}
}%
\providecommand \@ifnum [1]{%
 \ifnum #1\expandafter \@firstoftwo
 \else \expandafter \@secondoftwo
 \fi
}%
\providecommand \@ifx [1]{%
 \ifx #1\expandafter \@firstoftwo
 \else \expandafter \@secondoftwo
 \fi
}%
\providecommand \natexlab [1]{#1}%
\providecommand \enquote  [1]{``#1''}%
\providecommand \bibnamefont  [1]{#1}%
\providecommand \bibfnamefont [1]{#1}%
\providecommand \citenamefont [1]{#1}%
\providecommand \href@noop [0]{\@secondoftwo}%
\providecommand \href [0]{\begingroup \@sanitize@url \@href}%
\providecommand \@href[1]{\@@startlink{#1}\@@href}%
\providecommand \@@href[1]{\endgroup#1\@@endlink}%
\providecommand \@sanitize@url [0]{\catcode `\\12\catcode `\$12\catcode
  `\&12\catcode `\#12\catcode `\^12\catcode `\_12\catcode `\%12\relax}%
\providecommand \@@startlink[1]{}%
\providecommand \@@endlink[0]{}%
\providecommand \url  [0]{\begingroup\@sanitize@url \@url }%
\providecommand \@url [1]{\endgroup\@href {#1}{\urlprefix }}%
\providecommand \urlprefix  [0]{URL }%
\providecommand \Eprint [0]{\href }%
\providecommand \doibase [0]{http://dx.doi.org/}%
\providecommand \selectlanguage [0]{\@gobble}%
\providecommand \bibinfo  [0]{\@secondoftwo}%
\providecommand \bibfield  [0]{\@secondoftwo}%
\providecommand \translation [1]{[#1]}%
\providecommand \BibitemOpen [0]{}%
\providecommand \bibitemStop [0]{}%
\providecommand \bibitemNoStop [0]{.\EOS\space}%
\providecommand \EOS [0]{\spacefactor3000\relax}%
\providecommand \BibitemShut  [1]{\csname bibitem#1\endcsname}%
\let\auto@bib@innerbib\@empty
\bibitem [{\citenamefont {Affleck}\ \emph {et~al.}(1987)\citenamefont
  {Affleck}, \citenamefont {Kennedy}, \citenamefont {Lieb},\ and\ \citenamefont
  {Tasaki}}]{Affleck-K-L-T-87}%
  \BibitemOpen
  \bibfield  {author} {\bibinfo {author} {\bibfnamefont {I.}~\bibnamefont
  {Affleck}}, \bibinfo {author} {\bibfnamefont {T.}~\bibnamefont {Kennedy}},
  \bibinfo {author} {\bibfnamefont {E.~H.}\ \bibnamefont {Lieb}}, \ and\
  \bibinfo {author} {\bibfnamefont {H.}~\bibnamefont {Tasaki}},\ }\href
  {http://link.aps.org/doi/10.1103/PhysRevLett.59.799} {\bibfield  {journal}
  {\bibinfo  {journal} {Phys. Rev. Lett.}\ }\textbf {\bibinfo {volume} {59}},\
  \bibinfo {pages} {799} (\bibinfo {year} {1987})}\BibitemShut {NoStop}%
\bibitem [{\citenamefont {Affleck}\ \emph {et~al.}(1988)\citenamefont
  {Affleck}, \citenamefont {Kennedy}, \citenamefont {Lieb},\ and\ \citenamefont
  {Tasaki}}]{Affleck-K-L-T-88a}%
  \BibitemOpen
  \bibfield  {author} {\bibinfo {author} {\bibfnamefont {I.}~\bibnamefont
  {Affleck}}, \bibinfo {author} {\bibfnamefont {T.}~\bibnamefont {Kennedy}},
  \bibinfo {author} {\bibfnamefont {E.~H.}\ \bibnamefont {Lieb}}, \ and\
  \bibinfo {author} {\bibfnamefont {H.}~\bibnamefont {Tasaki}},\ }\href
  {http://dx.doi.org/10.1007/BF01218021} {\bibfield  {journal} {\bibinfo
  {journal} {Comm. Math. Phys.}\ }\textbf {\bibinfo {volume} {115}},\ \bibinfo
  {pages} {477} (\bibinfo {year} {1988})}\BibitemShut {NoStop}%
\bibitem [{\citenamefont {Haldane}(1983{\natexlab{a}})}]{Haldane-83a}%
  \BibitemOpen
  \bibfield  {author} {\bibinfo {author} {\bibfnamefont {F.~D.~M.}\
  \bibnamefont {Haldane}},\ }\href@noop {} {\bibfield  {journal} {\bibinfo
  {journal} {Phys. Lett.}\ }\textbf {\bibinfo {volume} {{\bf 93A}}},\ \bibinfo
  {pages} {464} (\bibinfo {year} {1983}{\natexlab{a}})}\BibitemShut {NoStop}%
\bibitem [{\citenamefont {Haldane}(1983{\natexlab{b}})}]{Haldane-83b}%
  \BibitemOpen
  \bibfield  {author} {\bibinfo {author} {\bibfnamefont {F.~D.~M.}\
  \bibnamefont {Haldane}},\ }\href@noop {} {\bibfield  {journal} {\bibinfo
  {journal} {Phys. Rev. Lett.}\ }\textbf {\bibinfo {volume} {{\bf 50}}},\
  \bibinfo {pages} {1153} (\bibinfo {year} {1983}{\natexlab{b}})}\BibitemShut
  {NoStop}%
\bibitem [{\citenamefont {Hagiwara}\ \emph {et~al.}(1990)\citenamefont
  {Hagiwara}, \citenamefont {Katsumata}, \citenamefont {Affleck}, \citenamefont
  {Halperin},\ and\ \citenamefont {Renard}}]{Hagiwara-K-A-H-R-90}%
  \BibitemOpen
  \bibfield  {author} {\bibinfo {author} {\bibfnamefont {M.}~\bibnamefont
  {Hagiwara}}, \bibinfo {author} {\bibfnamefont {K.}~\bibnamefont {Katsumata}},
  \bibinfo {author} {\bibfnamefont {I.}~\bibnamefont {Affleck}}, \bibinfo
  {author} {\bibfnamefont {B.}~\bibnamefont {Halperin}}, \ and\ \bibinfo
  {author} {\bibfnamefont {J.}~\bibnamefont {Renard}},\ }\href@noop {}
  {\bibfield  {journal} {\bibinfo  {journal} {Phys.Rev.Lett.}\ }\textbf
  {\bibinfo {volume} {{\bf 65}}},\ \bibinfo {pages} {3181} (\bibinfo {year}
  {1990})}\BibitemShut {NoStop}%
\bibitem [{\citenamefont {den Nijs}\ and\ \citenamefont
  {Rommelse}(1989)}]{denNijs-R-89}%
  \BibitemOpen
  \bibfield  {author} {\bibinfo {author} {\bibfnamefont {M.}~\bibnamefont {den
  Nijs}}\ and\ \bibinfo {author} {\bibfnamefont {K.}~\bibnamefont {Rommelse}},\
  }\href@noop {} {\bibfield  {journal} {\bibinfo  {journal} {Phys.Rev.}\
  }\textbf {\bibinfo {volume} {{\bf B40}}},\ \bibinfo {pages} {4709} (\bibinfo
  {year} {1989})}\BibitemShut {NoStop}%
\bibitem [{\citenamefont {Tasaki}(1991)}]{Tasaki-91}%
  \BibitemOpen
  \bibfield  {author} {\bibinfo {author} {\bibfnamefont {H.}~\bibnamefont
  {Tasaki}},\ }\href@noop {} {\bibfield  {journal} {\bibinfo  {journal}
  {Phys.Rev.Lett.}\ }\textbf {\bibinfo {volume} {{\bf 66}}},\ \bibinfo {pages}
  {798} (\bibinfo {year} {1991})}\BibitemShut {NoStop}%
\bibitem [{\citenamefont {Kennedy}\ and\ \citenamefont
  {Tasaki}(1992{\natexlab{a}})}]{Kennedy-T-92-PRB}%
  \BibitemOpen
  \bibfield  {author} {\bibinfo {author} {\bibfnamefont {T.}~\bibnamefont
  {Kennedy}}\ and\ \bibinfo {author} {\bibfnamefont {H.}~\bibnamefont
  {Tasaki}},\ }\href {http://link.aps.org/doi/10.1103/PhysRevB.45.304}
  {\bibfield  {journal} {\bibinfo  {journal} {Phys. Rev. B}\ }\textbf {\bibinfo
  {volume} {45}},\ \bibinfo {pages} {304} (\bibinfo {year}
  {1992}{\natexlab{a}})}\BibitemShut {NoStop}%
\bibitem [{\citenamefont {Kennedy}\ and\ \citenamefont
  {Tasaki}(1992{\natexlab{b}})}]{Kennedy-T-92-CMP}%
  \BibitemOpen
  \bibfield  {author} {\bibinfo {author} {\bibfnamefont {T.}~\bibnamefont
  {Kennedy}}\ and\ \bibinfo {author} {\bibfnamefont {H.}~\bibnamefont
  {Tasaki}},\ }\href {http://dx.doi.org/10.1007/BF02097239} {\bibfield
  {journal} {\bibinfo  {journal} {Comm. Math. Phys.}\ }\textbf {\bibinfo
  {volume} {147}},\ \bibinfo {pages} {431} (\bibinfo {year}
  {1992}{\natexlab{b}})},\ \bibinfo {note} {10.1007/BF02097239}\BibitemShut
  {NoStop}%
\bibitem [{\citenamefont {Hatsugai}(1992)}]{Hatsugai-92}%
  \BibitemOpen
  \bibfield  {author} {\bibinfo {author} {\bibfnamefont {Y.}~\bibnamefont
  {Hatsugai}},\ }\href@noop {} {\bibfield  {journal} {\bibinfo  {journal}
  {J.Phys.Soc.Jpn.}\ }\textbf {\bibinfo {volume} {{\bf 61}}},\ \bibinfo {pages}
  {3856} (\bibinfo {year} {1992})}\BibitemShut {NoStop}%
\bibitem [{\citenamefont {Oshikawa}(1992)}]{Oshikawa-92}%
  \BibitemOpen
  \bibfield  {author} {\bibinfo {author} {\bibfnamefont {M.}~\bibnamefont
  {Oshikawa}},\ }\href@noop {} {\bibfield  {journal} {\bibinfo  {journal}
  {J.Phys. Condens.Matter}\ }\textbf {\bibinfo {volume} {{\bf 4}}},\ \bibinfo
  {pages} {7469} (\bibinfo {year} {1992})}\BibitemShut {NoStop}%
\bibitem [{\citenamefont {Totsuka}\ and\ \citenamefont
  {Suzuki}(1995)}]{Totsuka-S-mpg-95}%
  \BibitemOpen
  \bibfield  {author} {\bibinfo {author} {\bibfnamefont {K.}~\bibnamefont
  {Totsuka}}\ and\ \bibinfo {author} {\bibfnamefont {M.}~\bibnamefont
  {Suzuki}},\ }\href@noop {} {\bibfield  {journal} {\bibinfo  {journal}
  {J.Phys.:condens.matter}\ }\textbf {\bibinfo {volume} {{\bf 7}}},\ \bibinfo
  {pages} {1639} (\bibinfo {year} {1995})}\BibitemShut {NoStop}%
\bibitem [{\citenamefont {Nishiyama}\ \emph {et~al.}(1995)\citenamefont
  {Nishiyama}, \citenamefont {Totsuka}, \citenamefont {Hatano},\ and\
  \citenamefont {Suzuki}}]{Nishiyama-T-H-S-95}%
  \BibitemOpen
  \bibfield  {author} {\bibinfo {author} {\bibfnamefont {Y.}~\bibnamefont
  {Nishiyama}}, \bibinfo {author} {\bibfnamefont {K.}~\bibnamefont {Totsuka}},
  \bibinfo {author} {\bibfnamefont {N.}~\bibnamefont {Hatano}}, \ and\ \bibinfo
  {author} {\bibfnamefont {M.}~\bibnamefont {Suzuki}},\ }\href@noop {}
  {\bibfield  {journal} {\bibinfo  {journal} {J.Phys.Soc.Jpn.}\ }\textbf
  {\bibinfo {volume} {{\bf 64}}},\ \bibinfo {pages} {414} (\bibinfo {year}
  {1995})}\BibitemShut {NoStop}%
\bibitem [{\citenamefont {Girvin}\ and\ \citenamefont
  {Arovas}(1989)}]{Girvin-A-89}%
  \BibitemOpen
  \bibfield  {author} {\bibinfo {author} {\bibfnamefont {S.~M.}\ \bibnamefont
  {Girvin}}\ and\ \bibinfo {author} {\bibfnamefont {D.~P.}\ \bibnamefont
  {Arovas}},\ }\href {http://stacks.iop.org/1402-4896/T27/156} {\bibfield
  {journal} {\bibinfo  {journal} {Physica Scripta}\ }\textbf {\bibinfo {volume}
  {T27}},\ \bibinfo {pages} {156} (\bibinfo {year} {1989})}\BibitemShut
  {NoStop}%
\bibitem [{\citenamefont {Arovas}\ \emph {et~al.}(1988)\citenamefont {Arovas},
  \citenamefont {Auerbach},\ and\ \citenamefont {Haldane}}]{Arovas-A-H-88}%
  \BibitemOpen
  \bibfield  {author} {\bibinfo {author} {\bibfnamefont {D.~P.}\ \bibnamefont
  {Arovas}}, \bibinfo {author} {\bibfnamefont {A.}~\bibnamefont {Auerbach}}, \
  and\ \bibinfo {author} {\bibfnamefont {F.~D.~M.}\ \bibnamefont {Haldane}},\
  }\href@noop {} {\bibfield  {journal} {\bibinfo  {journal} {Phys.Rev.Lett.}\
  }\textbf {\bibinfo {volume} {{\bf 60}}},\ \bibinfo {pages} {531} (\bibinfo
  {year} {1988})}\BibitemShut {NoStop}%
\bibitem [{\citenamefont {Levin}\ and\ \citenamefont {Wen}(2006)}]{Levin-W-06}%
  \BibitemOpen
  \bibfield  {author} {\bibinfo {author} {\bibfnamefont {M.}~\bibnamefont
  {Levin}}\ and\ \bibinfo {author} {\bibfnamefont {X.-G.}\ \bibnamefont
  {Wen}},\ }\href {http://link.aps.org/abstract/PRL/v96/e110405} {\bibfield
  {journal} {\bibinfo  {journal} {Phys. Rev. Lett.}\ }\textbf {\bibinfo
  {volume} {96}},\ \bibinfo {pages} {110405} (\bibinfo {year}
  {2006})}\BibitemShut {NoStop}%
\bibitem [{\citenamefont {Kitaev}\ and\ \citenamefont
  {Preskill}(2006)}]{Kitaev-P-06}%
  \BibitemOpen
  \bibfield  {author} {\bibinfo {author} {\bibfnamefont {A.}~\bibnamefont
  {Kitaev}}\ and\ \bibinfo {author} {\bibfnamefont {J.}~\bibnamefont
  {Preskill}},\ }\href {http://link.aps.org/abstract/PRL/v96/e110404}
  {\bibfield  {journal} {\bibinfo  {journal} {Phys. Rev. Lett.}\ }\textbf
  {\bibinfo {volume} {96}},\ \bibinfo {pages} {110404} (\bibinfo {year}
  {2006})}\BibitemShut {NoStop}%
\bibitem [{\citenamefont {Li}\ and\ \citenamefont {Haldane}(2008)}]{Li-H-08}%
  \BibitemOpen
  \bibfield  {author} {\bibinfo {author} {\bibfnamefont {H.}~\bibnamefont
  {Li}}\ and\ \bibinfo {author} {\bibfnamefont {F.~D.~M.}\ \bibnamefont
  {Haldane}},\ }\href {http://link.aps.org/doi/10.1103/PhysRevLett.101.010504}
  {\bibfield  {journal} {\bibinfo  {journal} {Phys. Rev. Lett.}\ }\textbf
  {\bibinfo {volume} {101}},\ \bibinfo {pages} {010504} (\bibinfo {year}
  {2008})}\BibitemShut {NoStop}%
\bibitem [{\citenamefont {Gu}\ and\ \citenamefont {Wen}(2009)}]{Gu-W-09}%
  \BibitemOpen
  \bibfield  {author} {\bibinfo {author} {\bibfnamefont {Z.-C.}\ \bibnamefont
  {Gu}}\ and\ \bibinfo {author} {\bibfnamefont {X.-G.}\ \bibnamefont {Wen}},\
  }\href {http://link.aps.org/abstract/PRB/v80/e155131} {\bibfield  {journal}
  {\bibinfo  {journal} {Phys. Rev. B}\ }\textbf {\bibinfo {volume} {80}},\
  \bibinfo {pages} {155131} (\bibinfo {year} {2009})}\BibitemShut {NoStop}%
\bibitem [{\citenamefont {Pollmann}\ \emph {et~al.}(2010)\citenamefont
  {Pollmann}, \citenamefont {Turner}, \citenamefont {Berg},\ and\ \citenamefont
  {Oshikawa}}]{Pollmann-T-B-O-10}%
  \BibitemOpen
  \bibfield  {author} {\bibinfo {author} {\bibfnamefont {F.}~\bibnamefont
  {Pollmann}}, \bibinfo {author} {\bibfnamefont {A.~M.}\ \bibnamefont
  {Turner}}, \bibinfo {author} {\bibfnamefont {E.}~\bibnamefont {Berg}}, \ and\
  \bibinfo {author} {\bibfnamefont {M.}~\bibnamefont {Oshikawa}},\ }\href
  {http://link.aps.org/doi/10.1103/PhysRevB.81.064439} {\bibfield  {journal}
  {\bibinfo  {journal} {Phys. Rev. B}\ }\textbf {\bibinfo {volume} {81}},\
  \bibinfo {pages} {064439} (\bibinfo {year} {2010})}\BibitemShut {NoStop}%
\bibitem [{\citenamefont {Pollmann}\ \emph {et~al.}(2012)\citenamefont
  {Pollmann}, \citenamefont {Berg}, \citenamefont {Turner},\ and\ \citenamefont
  {Oshikawa}}]{Pollmann-B-T-O-12}%
  \BibitemOpen
  \bibfield  {author} {\bibinfo {author} {\bibfnamefont {F.}~\bibnamefont
  {Pollmann}}, \bibinfo {author} {\bibfnamefont {E.}~\bibnamefont {Berg}},
  \bibinfo {author} {\bibfnamefont {A.~M.}\ \bibnamefont {Turner}}, \ and\
  \bibinfo {author} {\bibfnamefont {M.}~\bibnamefont {Oshikawa}},\ }\href
  {http://link.aps.org/doi/10.1103/PhysRevB.85.075125} {\bibfield  {journal}
  {\bibinfo  {journal} {Phys. Rev. B}\ }\textbf {\bibinfo {volume} {85}},\
  \bibinfo {pages} {075125} (\bibinfo {year} {2012})}\BibitemShut {NoStop}%
\bibitem [{\citenamefont {Chen}\ \emph
  {et~al.}(2011{\natexlab{a}})\citenamefont {Chen}, \citenamefont {Gu},\ and\
  \citenamefont {Wen}}]{Chen-G-W-11}%
  \BibitemOpen
  \bibfield  {author} {\bibinfo {author} {\bibfnamefont {X.}~\bibnamefont
  {Chen}}, \bibinfo {author} {\bibfnamefont {Z.-C.}\ \bibnamefont {Gu}}, \ and\
  \bibinfo {author} {\bibfnamefont {X.-G.}\ \bibnamefont {Wen}},\ }\href
  {http://link.aps.org/doi/10.1103/PhysRevB.83.035107} {\bibfield  {journal}
  {\bibinfo  {journal} {Phys. Rev. B}\ }\textbf {\bibinfo {volume} {83}},\
  \bibinfo {pages} {035107} (\bibinfo {year} {2011}{\natexlab{a}})}\BibitemShut
  {NoStop}%
\bibitem [{\citenamefont {Chen}\ \emph
  {et~al.}(2011{\natexlab{b}})\citenamefont {Chen}, \citenamefont {Gu},\ and\
  \citenamefont {Wen}}]{Chen-G-W-11b}%
  \BibitemOpen
  \bibfield  {author} {\bibinfo {author} {\bibfnamefont {X.}~\bibnamefont
  {Chen}}, \bibinfo {author} {\bibfnamefont {Z.-C.}\ \bibnamefont {Gu}}, \ and\
  \bibinfo {author} {\bibfnamefont {X.-G.}\ \bibnamefont {Wen}},\ }\href
  {http://link.aps.org/doi/10.1103/PhysRevB.84.235128} {\bibfield  {journal}
  {\bibinfo  {journal} {Phys. Rev. B}\ }\textbf {\bibinfo {volume} {84}},\
  \bibinfo {pages} {235128} (\bibinfo {year} {2011}{\natexlab{b}})}\BibitemShut
  {NoStop}%
\bibitem [{\citenamefont {Schuch}\ \emph {et~al.}(2011)\citenamefont {Schuch},
  \citenamefont {P\'{e}rez-Garc\'{i}a},\ and\ \citenamefont
  {Cirac}}]{Schuch-G-C-11}%
  \BibitemOpen
  \bibfield  {author} {\bibinfo {author} {\bibfnamefont {N.}~\bibnamefont
  {Schuch}}, \bibinfo {author} {\bibfnamefont {D.}~\bibnamefont
  {P\'{e}rez-Garc\'{i}a}}, \ and\ \bibinfo {author} {\bibfnamefont
  {I.}~\bibnamefont {Cirac}},\ }\href
  {http://link.aps.org/doi/10.1103/PhysRevB.84.165139} {\bibfield  {journal}
  {\bibinfo  {journal} {Phys. Rev. B}\ }\textbf {\bibinfo {volume} {84}},\
  \bibinfo {pages} {165139} (\bibinfo {year} {2011})}\BibitemShut {NoStop}%
\bibitem [{\citenamefont {Duivenvoorden}\ and\ \citenamefont
  {Quella}({\natexlab{a}})}]{Duiv-Quella-12a}%
  \BibitemOpen
  \bibfield  {author} {\bibinfo {author} {\bibfnamefont {K.}~\bibnamefont
  {Duivenvoorden}}\ and\ \bibinfo {author} {\bibfnamefont {T.}~\bibnamefont
  {Quella}},\ }\href@noop {} {\enquote {\bibinfo {title} {On topological phases
  of spin chains},}\ }  \bibinfo {note}
  {arXiv:1206.2462}\BibitemShut {NoStop}%
\bibitem [{\citenamefont {Duivenvoorden}\ and\ \citenamefont
  {Quella}({\natexlab{b}})}]{Duiv-Quella-12b}%
  \BibitemOpen
  \bibfield  {author} {\bibinfo {author} {\bibfnamefont {K.}~\bibnamefont
  {Duivenvoorden}}\ and\ \bibinfo {author} {\bibfnamefont {T.}~\bibnamefont
  {Quella}},\ }\href@noop {} {\enquote {\bibinfo {title} {A discriminating
  string order parameter for topological phases of gapped {SU(N)} spin
  chains},}\ } \bibinfo {note} {arXiv:1208.0697}\BibitemShut
  {NoStop}%
\bibitem [{\citenamefont {Zhang}\ and\ \citenamefont
  {Arovas}(1989)}]{Zhang-A-89}%
  \BibitemOpen
  \bibfield  {author} {\bibinfo {author} {\bibfnamefont {S.}~\bibnamefont
  {Zhang}}\ and\ \bibinfo {author} {\bibfnamefont {D.}~\bibnamefont {Arovas}},\
  }\href@noop {} {\bibfield  {journal} {\bibinfo  {journal} {Phys. Rev. B}\
  }\textbf {\bibinfo {volume} {40}},\ \bibinfo {pages} {2708} (\bibinfo {year}
  {1989})}\BibitemShut {NoStop}%
\bibitem [{\citenamefont {Penc}\ and\ \citenamefont {Shiba}(1995)}]{Penc-S-95}%
  \BibitemOpen
  \bibfield  {author} {\bibinfo {author} {\bibfnamefont {K.}~\bibnamefont
  {Penc}}\ and\ \bibinfo {author} {\bibfnamefont {H.}~\bibnamefont {Shiba}},\
  }\href {http://link.aps.org/doi/10.1103/PhysRevB.52.R715} {\bibfield
  {journal} {\bibinfo  {journal} {Phys. Rev. B}\ }\textbf {\bibinfo {volume}
  {52}},\ \bibinfo {pages} {R715} (\bibinfo {year} {1995})}\BibitemShut
  {NoStop}%
\bibitem [{\citenamefont {Xu}\ \emph {et~al.}(2000)\citenamefont {Xu},
  \citenamefont {Aeppli}, \citenamefont {Bisher}, \citenamefont {Broholm},
  \citenamefont {DiTusa}, \citenamefont {Frost}, \citenamefont {Ito},
  \citenamefont {Oka}, \citenamefont {Paul}, \citenamefont {Takagi} \emph
  {et~al.}}]{xu2000hqs}%
  \BibitemOpen
  \bibfield  {author} {\bibinfo {author} {\bibfnamefont {G.}~\bibnamefont
  {Xu}}, \bibinfo {author} {\bibfnamefont {G.}~\bibnamefont {Aeppli}}, \bibinfo
  {author} {\bibfnamefont {M.}~\bibnamefont {Bisher}}, \bibinfo {author}
  {\bibfnamefont {C.}~\bibnamefont {Broholm}}, \bibinfo {author} {\bibfnamefont
  {J.}~\bibnamefont {DiTusa}}, \bibinfo {author} {\bibfnamefont
  {C.}~\bibnamefont {Frost}}, \bibinfo {author} {\bibfnamefont
  {T.}~\bibnamefont {Ito}}, \bibinfo {author} {\bibfnamefont {K.}~\bibnamefont
  {Oka}}, \bibinfo {author} {\bibfnamefont {R.}~\bibnamefont {Paul}}, \bibinfo
  {author} {\bibfnamefont {H.}~\bibnamefont {Takagi}},  \emph {et~al.},\
  }\href@noop {} {\bibfield  {journal} {\bibinfo  {journal} {Science}\ }\textbf
  {\bibinfo {volume} {289}},\ \bibinfo {pages} {419} (\bibinfo {year}
  {2000})}\BibitemShut {NoStop}%
\bibitem [{\citenamefont {Hasebe}(2005)}]{hasebe2005PRL}%
  \BibitemOpen
  \bibfield  {author} {\bibinfo {author} {\bibfnamefont {K.}~\bibnamefont
  {Hasebe}},\ }\href@noop {} {\bibfield  {journal} {\bibinfo  {journal} {Phys.
  Rev. Lett.}\ }\textbf {\bibinfo {volume} {94}},\ \bibinfo {pages} {206802}
  (\bibinfo {year} {2005})}\BibitemShut {NoStop}%
\bibitem [{\citenamefont {Arovas}\ \emph {et~al.}(2009)\citenamefont {Arovas},
  \citenamefont {Hasebe}, \citenamefont {Qi},\ and\ \citenamefont
  {Zhang}}]{Arovas-H-Q-Z-09}%
  \BibitemOpen
  \bibfield  {author} {\bibinfo {author} {\bibfnamefont {D.~P.}\ \bibnamefont
  {Arovas}}, \bibinfo {author} {\bibfnamefont {K.}~\bibnamefont {Hasebe}},
  \bibinfo {author} {\bibfnamefont {X.-L.}\ \bibnamefont {Qi}}, \ and\ \bibinfo
  {author} {\bibfnamefont {S.-C.}\ \bibnamefont {Zhang}},\ }\href
  {http://link.aps.org/doi/10.1103/PhysRevB.79.224404} {\bibfield  {journal}
  {\bibinfo  {journal} {Phys. Rev. B}\ }\textbf {\bibinfo {volume} {79}},\
  \bibinfo {pages} {224404} (\bibinfo {year} {2009})}\BibitemShut {NoStop}%
\bibitem [{\citenamefont {Yu}\ and\ \citenamefont {Yang}(2008)}]{YuYang2008}%
  \BibitemOpen
  \bibfield  {author} {\bibinfo {author} {\bibfnamefont {Y.}~\bibnamefont
  {Yu}}\ and\ \bibinfo {author} {\bibfnamefont {K.}~\bibnamefont {Yang}},\
  }\href@noop {} {\bibfield  {journal} {\bibinfo  {journal} {Phys. Rev. Lett.}\
  }\textbf {\bibinfo {volume} {100}},\ \bibinfo {pages} {090404} (\bibinfo
  {year} {2008})}\BibitemShut {NoStop}%
\bibitem [{\citenamefont {Hasebe}\ and\ \citenamefont
  {Totsuka}(2011)}]{Hasebe-T-11}%
  \BibitemOpen
  \bibfield  {author} {\bibinfo {author} {\bibfnamefont {K.}~\bibnamefont
  {Hasebe}}\ and\ \bibinfo {author} {\bibfnamefont {K.}~\bibnamefont
  {Totsuka}},\ }\href {http://link.aps.org/doi/10.1103/PhysRevB.84.104426}
  {\bibfield  {journal} {\bibinfo  {journal} {Phys. Rev. B}\ }\textbf {\bibinfo
  {volume} {84}},\ \bibinfo {pages} {104426} (\bibinfo {year}
  {2011})}\BibitemShut {NoStop}%
\bibitem [{\citenamefont {Tu}\ \emph {et~al.}(2009)\citenamefont {Tu},
  \citenamefont {Zhang}, \citenamefont {Xiang}, \citenamefont {Liu},\ and\
  \citenamefont {Ng}}]{Tu-Z-X-X-N-09}%
  \BibitemOpen
  \bibfield  {author} {\bibinfo {author} {\bibfnamefont {H.-H.}\ \bibnamefont
  {Tu}}, \bibinfo {author} {\bibfnamefont {G.-M.}\ \bibnamefont {Zhang}},
  \bibinfo {author} {\bibfnamefont {T.}~\bibnamefont {Xiang}}, \bibinfo
  {author} {\bibfnamefont {Z.-X.}\ \bibnamefont {Liu}}, \ and\ \bibinfo
  {author} {\bibfnamefont {T.-K.}\ \bibnamefont {Ng}},\ }\href
  {http://link.aps.org/doi/10.1103/PhysRevB.80.014401} {\bibfield  {journal}
  {\bibinfo  {journal} {Phys. Rev. B}\ }\textbf {\bibinfo {volume} {80}},\
  \bibinfo {pages} {014401} (\bibinfo {year} {2009})}\BibitemShut {NoStop}%
\bibitem [{\citenamefont {Schuricht}\ and\ \citenamefont
  {Rachel}(2008)}]{Schuricht-R-08}%
  \BibitemOpen
  \bibfield  {author} {\bibinfo {author} {\bibfnamefont {D.}~\bibnamefont
  {Schuricht}}\ and\ \bibinfo {author} {\bibfnamefont {S.}~\bibnamefont
  {Rachel}},\ }\href {http://link.aps.org/doi/10.1103/PhysRevB.78.014430}
  {\bibfield  {journal} {\bibinfo  {journal} {Phys. Rev. B}\ }\textbf {\bibinfo
  {volume} {78}},\ \bibinfo {pages} {014430} (\bibinfo {year}
  {2008})}\BibitemShut {NoStop}%
\bibitem [{\citenamefont {Verstraete}\ and\ \citenamefont
  {Cirac}(2006)}]{Verstraete-C-06}%
  \BibitemOpen
  \bibfield  {author} {\bibinfo {author} {\bibfnamefont {F.}~\bibnamefont
  {Verstraete}}\ and\ \bibinfo {author} {\bibfnamefont {J.~I.}\ \bibnamefont
  {Cirac}},\ }\href {http://link.aps.org/doi/10.1103/PhysRevB.73.094423}
  {\bibfield  {journal} {\bibinfo  {journal} {Phys. Rev. B}\ }\textbf {\bibinfo
  {volume} {73}},\ \bibinfo {pages} {094423} (\bibinfo {year}
  {2006})}\BibitemShut {NoStop}%
\bibitem [{\citenamefont {Hastings}(2006)}]{Hastings-06}%
  \BibitemOpen
  \bibfield  {author} {\bibinfo {author} {\bibfnamefont {M.~B.}\ \bibnamefont
  {Hastings}},\ }\href {http://link.aps.org/doi/10.1103/PhysRevB.73.085115}
  {\bibfield  {journal} {\bibinfo  {journal} {Phys. Rev. B}\ }\textbf {\bibinfo
  {volume} {73}},\ \bibinfo {pages} {085115} (\bibinfo {year}
  {2006})}\BibitemShut {NoStop}%
\bibitem [{\citenamefont {Hastings}(2007)}]{Hastings-area-law-07}%
  \BibitemOpen
  \bibfield  {author} {\bibinfo {author} {\bibfnamefont {M.~B.}\ \bibnamefont
  {Hastings}},\ }\href {http://stacks.iop.org/1742-5468/2007/P08024} {\bibfield
   {journal} {\bibinfo  {journal} {J. Stat. Mech.: Theory and Experiment}\
  }\textbf {\bibinfo {volume} {2007}},\ \bibinfo {pages} {P08024} (\bibinfo
  {year} {2007})}\BibitemShut {NoStop}%
\bibitem [{\citenamefont {P{\'{e}}rez-Garc{\'{i}}a}\ \emph
  {et~al.}(2007)\citenamefont {P{\'{e}}rez-Garc{\'{i}}a}, \citenamefont
  {Verstraete}, \citenamefont {Wolf},\ and\ \citenamefont
  {Cirac}}]{Garcia-V-W-C-07}%
  \BibitemOpen
  \bibfield  {author} {\bibinfo {author} {\bibfnamefont {D.}~\bibnamefont
  {P{\'{e}}rez-Garc{\'{i}}a}}, \bibinfo {author} {\bibfnamefont
  {F.}~\bibnamefont {Verstraete}}, \bibinfo {author} {\bibfnamefont
  {M.}~\bibnamefont {Wolf}}, \ and\ \bibinfo {author} {\bibfnamefont
  {J.}~\bibnamefont {Cirac}},\ }\href {http://arxiv.org/abs/quant-ph/0608197}
  {\bibfield  {journal} {\bibinfo  {journal} {Quantum Inf. Comput.}\ }\textbf
  {\bibinfo {volume} {7}},\ \bibinfo {pages} {401} (\bibinfo {year}
  {2007})}\BibitemShut {NoStop}%
\bibitem [{\citenamefont {Frappat}\ \emph {et~al.}(2000)\citenamefont
  {Frappat}, \citenamefont {Sciarrino},\ and\ \citenamefont
  {Sorba}}]{SUSY-dictionary}%
  \BibitemOpen
  \bibfield  {author} {\bibinfo {author} {\bibfnamefont {L.}~\bibnamefont
  {Frappat}}, \bibinfo {author} {\bibfnamefont {A.}~\bibnamefont {Sciarrino}},
  \ and\ \bibinfo {author} {\bibfnamefont {P.}~\bibnamefont {Sorba}},\
  }\href@noop {} {\emph {\bibinfo {title} {Dictionary on Lie Algebras and
  Superalgebras}}}\ (\bibinfo  {publisher} {Academic Press},\ \bibinfo {year}
  {2000})\BibitemShut {NoStop}%
\bibitem [{\citenamefont {Hasebe}(2011)}]{hasebe-2011}%
  \BibitemOpen
  \bibfield  {author} {\bibinfo {author} {\bibfnamefont {K.}~\bibnamefont
  {Hasebe}},\ }\href@noop {} {\bibfield  {journal} {\bibinfo  {journal}
  {Nucl.Phys. B}\ }\textbf {\bibinfo {volume} {853}},\ \bibinfo {pages} {777}
  (\bibinfo {year} {2011})}\BibitemShut {NoStop}%
\bibitem [{\citenamefont {Hasebe}\ and\ \citenamefont
  {Totsuka}()}]{Hasebe-T-unpub-12}%
  \BibitemOpen
  \bibfield  {author} {\bibinfo {author} {\bibfnamefont {K.}~\bibnamefont
  {Hasebe}}\ and\ \bibinfo {author} {\bibfnamefont {K.}~\bibnamefont
  {Totsuka}},\ }\href@noop {} {}\bibinfo {note} {Unpublished}\BibitemShut
  {NoStop}%
\bibitem [{\citenamefont {Scalapino}\ \emph {et~al.}(1998)\citenamefont
  {Scalapino}, \citenamefont {Zhang},\ and\ \citenamefont
  {Hanke}}]{Scalapino-Z-H-98}%
  \BibitemOpen
  \bibfield  {author} {\bibinfo {author} {\bibfnamefont {D.}~\bibnamefont
  {Scalapino}}, \bibinfo {author} {\bibfnamefont {S.-C.}\ \bibnamefont
  {Zhang}}, \ and\ \bibinfo {author} {\bibfnamefont {W.}~\bibnamefont
  {Hanke}},\ }\href {http://link.aps.org/abstract/PRB/v58/p443} {\bibfield
  {journal} {\bibinfo  {journal} {Phys. Rev. B}\ }\textbf {\bibinfo {volume}
  {58}},\ \bibinfo {pages} {443} (\bibinfo {year} {1998})}\BibitemShut
  {NoStop}%
\bibitem [{\citenamefont {Tu}\ \emph {et~al.}(2008)\citenamefont {Tu},
  \citenamefont {Zhang},\ and\ \citenamefont {Xiang}}]{Tu-Z-X-08}%
  \BibitemOpen
  \bibfield  {author} {\bibinfo {author} {\bibfnamefont {H.-H.}\ \bibnamefont
  {Tu}}, \bibinfo {author} {\bibfnamefont {G.-M.}\ \bibnamefont {Zhang}}, \
  and\ \bibinfo {author} {\bibfnamefont {T.}~\bibnamefont {Xiang}},\ }\href
  {http://link.aps.org/doi/10.1103/PhysRevB.78.094404} {\bibfield  {journal}
  {\bibinfo  {journal} {Phys. Rev. B}\ }\textbf {\bibinfo {volume} {78}},\
  \bibinfo {pages} {094404} (\bibinfo {year} {2008})}\BibitemShut {NoStop}%
\bibitem [{\citenamefont {Majumdar}\ and\ \citenamefont
  {D.K.Ghosh}(1969)}]{Majumdar-G-69}%
  \BibitemOpen
  \bibfield  {author} {\bibinfo {author} {\bibfnamefont {C.}~\bibnamefont
  {Majumdar}}\ and\ \bibinfo {author} {\bibnamefont {D.K.Ghosh}},\ }\href@noop
  {} {\bibfield  {journal} {\bibinfo  {journal} {J.Math.Phys.}\ }\textbf
  {\bibinfo {volume} {{\bf 10}}},\ \bibinfo {pages} {1388,1399} (\bibinfo
  {year} {1969})}\BibitemShut {NoStop}%
\bibitem [{\citenamefont {Majumdar}(1970)}]{Majumdar-70}%
  \BibitemOpen
  \bibfield  {author} {\bibinfo {author} {\bibfnamefont {C.}~\bibnamefont
  {Majumdar}},\ }\href@noop {} {\bibfield  {journal} {\bibinfo  {journal}
  {J.Phys.}\ }\textbf {\bibinfo {volume} {{\bf C3}}},\ \bibinfo {pages} {911}
  (\bibinfo {year} {1970})}\BibitemShut {NoStop}%
\bibitem [{\citenamefont {Hida}(1992)}]{Hida-92}%
  \BibitemOpen
  \bibfield  {author} {\bibinfo {author} {\bibfnamefont {K.}~\bibnamefont
  {Hida}},\ }\href {http://link.aps.org/doi/10.1103/PhysRevB.45.2207}
  {\bibfield  {journal} {\bibinfo  {journal} {Phys. Rev. B}\ }\textbf {\bibinfo
  {volume} {45}},\ \bibinfo {pages} {2207} (\bibinfo {year}
  {1992})}\BibitemShut {NoStop}%
\bibitem [{\citenamefont {Thomale}\ \emph
  {et~al.}(2010{\natexlab{a}})\citenamefont {Thomale}, \citenamefont
  {Sterdyniak}, \citenamefont {Regnault},\ and\ \citenamefont
  {Bernevig}}]{Thomale-S-R-B-10}%
  \BibitemOpen
  \bibfield  {author} {\bibinfo {author} {\bibfnamefont {R.}~\bibnamefont
  {Thomale}}, \bibinfo {author} {\bibfnamefont {A.}~\bibnamefont {Sterdyniak}},
  \bibinfo {author} {\bibfnamefont {N.}~\bibnamefont {Regnault}}, \ and\
  \bibinfo {author} {\bibfnamefont {B.~A.}\ \bibnamefont {Bernevig}},\ }\href
  {http://link.aps.org/doi/10.1103/PhysRevLett.104.180502} {\bibfield
  {journal} {\bibinfo  {journal} {Phys. Rev. Lett.}\ }\textbf {\bibinfo
  {volume} {104}},\ \bibinfo {pages} {180502} (\bibinfo {year}
  {2010}{\natexlab{a}})}\BibitemShut {NoStop}%
\bibitem [{\citenamefont {Regnault}\ \emph {et~al.}(2009)\citenamefont
  {Regnault}, \citenamefont {Bernevig},\ and\ \citenamefont
  {Haldane}}]{Regnault-B-H-09}%
  \BibitemOpen
  \bibfield  {author} {\bibinfo {author} {\bibfnamefont {N.}~\bibnamefont
  {Regnault}}, \bibinfo {author} {\bibfnamefont {B.~A.}\ \bibnamefont
  {Bernevig}}, \ and\ \bibinfo {author} {\bibfnamefont {F.~D.~M.}\ \bibnamefont
  {Haldane}},\ }\href {http://link.aps.org/doi/10.1103/PhysRevLett.103.016801}
  {\bibfield  {journal} {\bibinfo  {journal} {Phys. Rev. Lett.}\ }\textbf
  {\bibinfo {volume} {103}},\ \bibinfo {pages} {016801} (\bibinfo {year}
  {2009})}\BibitemShut {NoStop}%
\bibitem [{\citenamefont {L\"{a}uchli}\ \emph {et~al.}(2010)\citenamefont
  {L\"{a}uchli}, \citenamefont {Bergholtz}, \citenamefont {Suorsa},\ and\
  \citenamefont {Haque}}]{Lauchli-B-S-H-10}%
  \BibitemOpen
  \bibfield  {author} {\bibinfo {author} {\bibfnamefont {A.~M.}\ \bibnamefont
  {L\"{a}uchli}}, \bibinfo {author} {\bibfnamefont {E.~J.}\ \bibnamefont
  {Bergholtz}}, \bibinfo {author} {\bibfnamefont {J.}~\bibnamefont {Suorsa}}, \
  and\ \bibinfo {author} {\bibfnamefont {M.}~\bibnamefont {Haque}},\ }\href
  {http://link.aps.org/doi/10.1103/PhysRevLett.104.156404} {\bibfield
  {journal} {\bibinfo  {journal} {Phys. Rev. Lett.}\ }\textbf {\bibinfo
  {volume} {104}},\ \bibinfo {pages} {156404} (\bibinfo {year}
  {2010})}\BibitemShut {NoStop}%
\bibitem [{\citenamefont {Fidkowski}(2010)}]{Fidkowski-10}%
  \BibitemOpen
  \bibfield  {author} {\bibinfo {author} {\bibfnamefont {L.}~\bibnamefont
  {Fidkowski}},\ }\href {\doibase 10.1103/PhysRevLett.104.130502} {\bibfield
  {journal} {\bibinfo  {journal} {Phys. Rev. Lett.}\ }\textbf {\bibinfo
  {volume} {104}},\ \bibinfo {pages} {130502} (\bibinfo {year}
  {2010})}\BibitemShut {NoStop}%
\bibitem [{\citenamefont {Prodan}\ \emph {et~al.}(2010)\citenamefont {Prodan},
  \citenamefont {Hughes},\ and\ \citenamefont {Bernevig}}]{Prodan-H-B-10}%
  \BibitemOpen
  \bibfield  {author} {\bibinfo {author} {\bibfnamefont {E.}~\bibnamefont
  {Prodan}}, \bibinfo {author} {\bibfnamefont {T.~L.}\ \bibnamefont {Hughes}},
  \ and\ \bibinfo {author} {\bibfnamefont {B.~A.}\ \bibnamefont {Bernevig}},\
  }\href {http://link.aps.org/doi/10.1103/PhysRevLett.105.115501} {\bibfield
  {journal} {\bibinfo  {journal} {Phys. Rev. Lett.}\ }\textbf {\bibinfo
  {volume} {105}},\ \bibinfo {pages} {115501} (\bibinfo {year}
  {2010})}\BibitemShut {NoStop}%
\bibitem [{\citenamefont {Turner}\ \emph {et~al.}(2010)\citenamefont {Turner},
  \citenamefont {Zhang},\ and\ \citenamefont {Vishwanath}}]{Turner-Z-V-10}%
  \BibitemOpen
  \bibfield  {author} {\bibinfo {author} {\bibfnamefont {A.~M.}\ \bibnamefont
  {Turner}}, \bibinfo {author} {\bibfnamefont {Y.}~\bibnamefont {Zhang}}, \
  and\ \bibinfo {author} {\bibfnamefont {A.}~\bibnamefont {Vishwanath}},\
  }\href {http://link.aps.org/doi/10.1103/PhysRevB.82.241102} {\bibfield
  {journal} {\bibinfo  {journal} {Phys. Rev. B}\ }\textbf {\bibinfo {volume}
  {82}},\ \bibinfo {pages} {241102} (\bibinfo {year} {2010})}\BibitemShut
  {NoStop}%
\bibitem [{\citenamefont {Thomale}\ \emph
  {et~al.}(2010{\natexlab{b}})\citenamefont {Thomale}, \citenamefont {Arovas},\
  and\ \citenamefont {Bernevig}}]{Thomale-A-B-10}%
  \BibitemOpen
  \bibfield  {author} {\bibinfo {author} {\bibfnamefont {R.}~\bibnamefont
  {Thomale}}, \bibinfo {author} {\bibfnamefont {D.~P.}\ \bibnamefont {Arovas}},
  \ and\ \bibinfo {author} {\bibfnamefont {B.~A.}\ \bibnamefont {Bernevig}},\
  }\href {http://link.aps.org/doi/10.1103/PhysRevLett.105.116805} {\bibfield
  {journal} {\bibinfo  {journal} {Phys. Rev. Lett.}\ }\textbf {\bibinfo
  {volume} {105}},\ \bibinfo {pages} {116805} (\bibinfo {year}
  {2010}{\natexlab{b}})}\BibitemShut {NoStop}%
\bibitem [{\citenamefont {Vidal}(2003)}]{Vidal-03}%
  \BibitemOpen
  \bibfield  {author} {\bibinfo {author} {\bibfnamefont {G.}~\bibnamefont
  {Vidal}},\ }\href {http://link.aps.org/doi/10.1103/PhysRevLett.91.147902}
  {\bibfield  {journal} {\bibinfo  {journal} {Phys. Rev. Lett.}\ }\textbf
  {\bibinfo {volume} {91}},\ \bibinfo {pages} {147902} (\bibinfo {year}
  {2003})}\BibitemShut {NoStop}%
\bibitem [{\citenamefont {Or{\'{u}}s}\ and\ \citenamefont
  {Vidal}(2008)}]{Orus-V-08}%
  \BibitemOpen
  \bibfield  {author} {\bibinfo {author} {\bibfnamefont {R.}~\bibnamefont
  {Or{\'{u}}s}}\ and\ \bibinfo {author} {\bibfnamefont {G.}~\bibnamefont
  {Vidal}},\ }\href {http://link.aps.org/doi/10.1103/PhysRevB.78.155117}
  {\bibfield  {journal} {\bibinfo  {journal} {Phys. Rev. B}\ }\textbf {\bibinfo
  {volume} {78}},\ \bibinfo {pages} {155117} (\bibinfo {year}
  {2008})}\BibitemShut {NoStop}%
\bibitem [{\citenamefont {Fan}\ \emph {et~al.}(2004)\citenamefont {Fan},
  \citenamefont {Korepin},\ and\ \citenamefont {Roychowdhury}}]{Fan-K-R-04}%
  \BibitemOpen
  \bibfield  {author} {\bibinfo {author} {\bibfnamefont {H.}~\bibnamefont
  {Fan}}, \bibinfo {author} {\bibfnamefont {V.}~\bibnamefont {Korepin}}, \ and\
  \bibinfo {author} {\bibfnamefont {V.}~\bibnamefont {Roychowdhury}},\ }\href
  {http://link.aps.org/doi/10.1103/PhysRevLett.93.227203} {\bibfield  {journal}
  {\bibinfo  {journal} {Phys. Rev. Lett.}\ }\textbf {\bibinfo {volume} {93}},\
  \bibinfo {pages} {227203} (\bibinfo {year} {2004})}\BibitemShut {NoStop}%
\bibitem [{\citenamefont {Katsura}\ \emph {et~al.}(2007)\citenamefont
  {Katsura}, \citenamefont {Hirano},\ and\ \citenamefont
  {Hatsugai}}]{katsura-2007-76}%
  \BibitemOpen
  \bibfield  {author} {\bibinfo {author} {\bibfnamefont {H.}~\bibnamefont
  {Katsura}}, \bibinfo {author} {\bibfnamefont {T.}~\bibnamefont {Hirano}}, \
  and\ \bibinfo {author} {\bibfnamefont {Y.}~\bibnamefont {Hatsugai}},\
  }\href@noop {} {\bibfield  {journal} {\bibinfo  {journal} {Phys. Rev. B}\
  }\textbf {\bibinfo {volume} {76}},\ \bibinfo {pages} {012401} (\bibinfo
  {year} {2007})}\BibitemShut {NoStop}%
\bibitem [{\citenamefont {Katsura}\ \emph {et~al.}(2008)\citenamefont
  {Katsura}, \citenamefont {Hirano},\ and\ \citenamefont
  {Korepin}}]{katsura-2008-41}%
  \BibitemOpen
  \bibfield  {author} {\bibinfo {author} {\bibfnamefont {H.}~\bibnamefont
  {Katsura}}, \bibinfo {author} {\bibfnamefont {T.}~\bibnamefont {Hirano}}, \
  and\ \bibinfo {author} {\bibfnamefont {V.~E.}\ \bibnamefont {Korepin}},\
  }\href@noop {} {\bibfield  {journal} {\bibinfo  {journal} {J. Phys. A:Math
  and Theor.}\ }\textbf {\bibinfo {volume} {41}},\ \bibinfo {pages} {135304}
  (\bibinfo {year} {2008})}\BibitemShut {NoStop}%
\bibitem [{sup()}]{suppl}%
  \BibitemOpen
  \href@noop {} {\bibinfo  {journal} {See Supplemental Material at
  http://link.aps.org/supplemental/ 10.1103/PhysRevB.00.000000 for entanglement
  of superqudit pairs.}\ }\BibitemShut {NoStop}%
\bibitem [{\citenamefont {P{\'{e}}rez-Garc{\'{i}}a}\ \emph
  {et~al.}(2008)\citenamefont {P{\'{e}}rez-Garc{\'{i}}a}, \citenamefont {Wolf},
  \citenamefont {Sanz}, \citenamefont {Verstraete},\ and\ \citenamefont
  {Cirac}}]{Garcia-W-S-V-C-08}%
  \BibitemOpen
\bibfield  {journal} {  }\bibfield  {author} {\bibinfo {author} {\bibfnamefont
  {D.}~\bibnamefont {P{\'{e}}rez-Garc{\'{i}}a}}, \bibinfo {author}
  {\bibfnamefont {M.~M.}\ \bibnamefont {Wolf}}, \bibinfo {author}
  {\bibfnamefont {M.}~\bibnamefont {Sanz}}, \bibinfo {author} {\bibfnamefont
  {F.}~\bibnamefont {Verstraete}}, \ and\ \bibinfo {author} {\bibfnamefont
  {J.~I.}\ \bibnamefont {Cirac}},\ }\href
  {http://link.aps.org/doi/10.1103/PhysRevLett.100.167202} {\bibfield
  {journal} {\bibinfo  {journal} {Phys. Rev. Lett.}\ }\textbf {\bibinfo
  {volume} {100}},\ \bibinfo {pages} {167202} (\bibinfo {year}
  {2008})}\BibitemShut {NoStop}%
\bibitem [{Note1()}]{Note1}%
  \BibitemOpen
  \bibinfo {note} {When $\protect \mathcal {S}$ is half-odd-integer, $\protect
  \mathcal {T}^{2} =-P$ which generalizes $\protect \mathcal {T}^{2}=-\protect
  \mathbf {1}$ for the SU(2) case.}\BibitemShut {Stop}%
\bibitem [{Note2()}]{Note2}%
  \BibitemOpen
  \bibinfo {note} {Specifically, $(\protect \mathbb {Z}_2 \times \protect
  \mathbb {Z}_2 )^2$-symmetry can be defined for the SO(5) states where all the
  allowed weights at each site are integers (e.g. the vector- and the adjoint
  representations).}\BibitemShut {Stop}%
\bibitem [{\citenamefont {Tu}\ and\ \citenamefont
  {Or\'us}(2011)}]{TuandOrusPRB2011}%
  \BibitemOpen
  \bibfield  {author} {\bibinfo {author} {\bibfnamefont {H.-H.}\ \bibnamefont
  {Tu}}\ and\ \bibinfo {author} {\bibfnamefont {R.}~\bibnamefont {Or\'us}},\
  }\href@noop {} {\bibfield  {journal} {\bibinfo  {journal} {Phys. Rev. B}\
  }\textbf {\bibinfo {volume} {84}},\ \bibinfo {pages} {140407} (\bibinfo
  {year} {2011})}\BibitemShut {NoStop}%
\bibitem [{Note3()}]{Note3}%
  \BibitemOpen
  \bibinfo {note} {This is the case for the class of UOSp(1$|$4) states
  discussed here. For the vector representation, for instance, we have a
  slightly different form of $R^{ab}$.}\BibitemShut {Stop}%
\bibitem [{\citenamefont {Haegeman}\ \emph {et~al.}(2012)\citenamefont
  {Haegeman}, \citenamefont {P{\'{e}}rez-Garc{\'{i}}a}, \citenamefont {Cirac},\
  and\ \citenamefont {Schuch}}]{Haegeman-G-C-S-12}%
  \BibitemOpen
  \bibfield  {author} {\bibinfo {author} {\bibfnamefont {J.}~\bibnamefont
  {Haegeman}}, \bibinfo {author} {\bibfnamefont {D.}~\bibnamefont
  {P{\'{e}}rez-Garc{\'{i}}a}}, \bibinfo {author} {\bibfnamefont
  {I.}~\bibnamefont {Cirac}}, \ and\ \bibinfo {author} {\bibfnamefont
  {N.}~\bibnamefont {Schuch}},\ }\href
  {http://link.aps.org/doi/10.1103/PhysRevLett.109.050402} {\bibfield
  {journal} {\bibinfo  {journal} {Phys. Rev. Lett.}\ }\textbf {\bibinfo
  {volume} {109}},\ \bibinfo {pages} {050402} (\bibinfo {year}
  {2012})}\BibitemShut {NoStop}%
\bibitem [{\citenamefont {Pollmann}\ and\ \citenamefont
  {Turner}(2012)}]{Pollmann-T-12}%
  \BibitemOpen
  \bibfield  {author} {\bibinfo {author} {\bibfnamefont {F.}~\bibnamefont
  {Pollmann}}\ and\ \bibinfo {author} {\bibfnamefont {A.~M.}\ \bibnamefont
  {Turner}},\ }\href {http://link.aps.org/doi/10.1103/PhysRevB.86.125441}
  {\bibfield  {journal} {\bibinfo  {journal} {Phys. Rev. B}\ }\textbf {\bibinfo
  {volume} {86}},\ \bibinfo {pages} {125441} (\bibinfo {year}
  {2012})}\BibitemShut {NoStop}%
\bibitem [{\citenamefont {Kl$\ddot{\mbox{u}}$mper}\ \emph
  {et~al.}(1992)\citenamefont {Kl$\ddot{\mbox{u}}$mper}, \citenamefont
  {Schadschneider},\ and\ \citenamefont {J.Zittartz}}]{Klumper-S-Z-92}%
  \BibitemOpen
  \bibfield  {author} {\bibinfo {author} {\bibfnamefont {A.}~\bibnamefont
  {Kl$\ddot{\mbox{u}}$mper}}, \bibinfo {author} {\bibfnamefont
  {A.}~\bibnamefont {Schadschneider}}, \ and\ \bibinfo {author} {\bibnamefont
  {J.Zittartz}},\ }\href@noop {} {\bibfield  {journal} {\bibinfo  {journal}
  {Z.Phys.}\ }\textbf {\bibinfo {volume} {{\bf B87}}},\ \bibinfo {pages} {281}
  (\bibinfo {year} {1992})}\BibitemShut {NoStop}%
\bibitem [{\citenamefont {Totsuka}\ and\ \citenamefont
  {Suzuki}(1994)}]{Totsuka-S-94}%
  \BibitemOpen
  \bibfield  {author} {\bibinfo {author} {\bibfnamefont {K.}~\bibnamefont
  {Totsuka}}\ and\ \bibinfo {author} {\bibfnamefont {M.}~\bibnamefont
  {Suzuki}},\ }\href {\doibase 10.1088/0305-4470/27/19/017} {\bibfield
  {journal} {\bibinfo  {journal} {J. Phys. A: Math. Gen.}\ }\textbf {\bibinfo
  {volume} {27}},\ \bibinfo {pages} {6443} (\bibinfo {year}
  {1994})}\BibitemShut {NoStop}%
\bibitem [{\citenamefont {Anderson}(1973)}]{Anderson-73}%
  \BibitemOpen
  \bibfield  {author} {\bibinfo {author} {\bibfnamefont {P.}~\bibnamefont
  {Anderson}},\ }\href
  {http://www.sciencedirect.com/science/article/B6TXC-48DYHGJ-BY/2/c8776a1d5fd%
4d78e8e2c4e37a53a0c73} {\bibfield  {journal} {\bibinfo  {journal} {Mater. Res.
  Bull.}\ }\textbf {\bibinfo {volume} {8}},\ \bibinfo {pages} {153} (\bibinfo
  {year} {1973})}\BibitemShut {NoStop}%
\bibitem [{\citenamefont {Fazekas}\ and\ \citenamefont
  {Anderson}(1974)}]{Fazekas-A-74}%
  \BibitemOpen
  \bibfield  {author} {\bibinfo {author} {\bibfnamefont {P.}~\bibnamefont
  {Fazekas}}\ and\ \bibinfo {author} {\bibfnamefont {P.}~\bibnamefont
  {Anderson}},\ }\href@noop {} {\bibfield  {journal} {\bibinfo  {journal}
  {Phil. Mag.}\ }\textbf {\bibinfo {volume} {30}},\ \bibinfo {pages} {423}
  (\bibinfo {year} {1974})}\BibitemShut {NoStop}%
\bibitem [{\citenamefont {Rokhsar}\ and\ \citenamefont
  {Kivelson}(1988)}]{Rokhsar-K-88}%
  \BibitemOpen
  \bibfield  {author} {\bibinfo {author} {\bibfnamefont {D.~S.}\ \bibnamefont
  {Rokhsar}}\ and\ \bibinfo {author} {\bibfnamefont {S.~A.}\ \bibnamefont
  {Kivelson}},\ }\href {http://link.aps.org/abstract/PRL/v61/p2376} {\bibfield
  {journal} {\bibinfo  {journal} {Phys. Rev. Lett.}\ }\textbf {\bibinfo
  {volume} {61}},\ \bibinfo {pages} {2376} (\bibinfo {year}
  {1988})}\BibitemShut {NoStop}%
\bibitem [{\citenamefont {Schuch}\ \emph {et~al.}(2012)\citenamefont {Schuch},
  \citenamefont {Poilblanc}, \citenamefont {Cirac},\ and\ \citenamefont
  {P\'{e}rez-Garc\'{i}a}}]{Schuch-P-C-G-12}%
  \BibitemOpen
  \bibfield  {author} {\bibinfo {author} {\bibfnamefont {N.}~\bibnamefont
  {Schuch}}, \bibinfo {author} {\bibfnamefont {D.}~\bibnamefont {Poilblanc}},
  \bibinfo {author} {\bibfnamefont {J.~I.}\ \bibnamefont {Cirac}}, \ and\
  \bibinfo {author} {\bibfnamefont {D.}~\bibnamefont {P\'{e}rez-Garc\'{i}a}},\
  }\href {http://link.aps.org/doi/10.1103/PhysRevB.86.115108} {\bibfield
  {journal} {\bibinfo  {journal} {Phys. Rev. B}\ }\textbf {\bibinfo {volume}
  {86}},\ \bibinfo {pages} {115108} (\bibinfo {year} {2012})}\BibitemShut
  {NoStop}%
\end{thebibliography}%
\end{document}